\documentclass[a4paper,onecolumn,11pt,unpublished]{quantumarticle}
\pdfoutput=1

\usepackage[utf8]{inputenc}
\usepackage[english]{babel}
\usepackage[T1]{fontenc}
\usepackage{amsmath,amssymb,amsthm}
\usepackage{graphicx}
\usepackage{microtype}
\usepackage{booktabs,array}
\usepackage{arydshln}
\usepackage[switch]{lineno}
\usepackage{bm}
\usepackage{dcolumn}
\usepackage{multirow}
\usepackage{algorithm}
\usepackage{algpseudocode}
\usepackage{qcircuit}
\usepackage{braket}
\usepackage{tikz}
\usetikzlibrary{arrows.meta}
\usepackage[font={small,sf}]{caption}
\usepackage{subcaption}
\usepackage[numbers,sort&compress]{natbib}
\usepackage{hyperref}
\hypersetup{linkcolor=quantumviolet,citecolor=quantumviolet,urlcolor=quantumviolet}

\numberwithin{equation}{section}

\newcommand{\artifactlink}{\url{https://github.com/kupl/qmiep-artifact}}

\DeclareMathOperator*{\argmax}{arg\,max}
\DeclareMathOperator{\tr}{tr}
\DeclareMathOperator{\spec}{spec}

\DeclareRobustCommand{\cnot}{\ensuremath{\mathit{CNOT}}}
\DeclareRobustCommand{\cz}{\ensuremath{\mathit{CZ}}}

\newcommand{\mux}{\gate{}}
\newcommand{\qcunitarygate}[1]{\gate{%
  \makebox[\ifdim\width<1.15em 1.15em\else\width\fi][c]{%
    $\vphantom{V^\dagger U_1}\smash{#1}$}}}

\newcommand{\rgate}{R}
\expandafter\newcommand\csname ucg_rz\endcsname{\mathrm{UCG}(R_z)}
\newcommand{\ucgrz}{\csname ucg_rz\endcsname}
\newcommand{\ucgrzgate}[1]{\gate{R_z#1}}
\newcommand{\ucgrzctrl}{\mux\qwx[-1]}

\newcommand{\qcbundle}[1]{%
  \push{\rule{1.6em}{0pt}}\qw
  \POS[0,0]+<-0.8em,0em>;[0,0]+<0.8em,0em>**\dir{-},%
  [0,0]+<0.5em,-0.35em>;[0,0]+<0.8em,0.35em>**\dir{-},%
  [0,0]+<0.25em,0.65em>*!R{\scriptstyle #1}%
}

\newcommand{\qcC}{1.15em}
\newcommand{\qcR}{1.50em}
\newcommand{\qcCm}{1.05em}
\newcommand{\qcRm}{1.35em}

\newcommand{\qcRw}{1.45em}

\newcommand{\qcRd}{1.10em}

\newcommand{\statementcircuit}[1]{%
  \vcenter{\hbox{\fontsize{10.2}{12}\selectfont$\Qcircuit @C=0.67em @R=0.88em {#1}$}}%
}

\makeatletter
\newcommand{\framedotwidth}[1]{%
  \def\framezerodot@@{\hbox{\kern-.5#1\vrule width#1 height.5#1 depth.5#1}}}
\makeatother
\framedotwidth{1.6pt}

\newcommand{\conu}{C} 
 
\newcommand{\diagonal}{D} 

\newcommand{\costX}{\textsf{\upshape cost}_{X}}
\newcommand{\costU}{\textsf{\upshape cost}_{U}}

\newcommand{\costucgrz}{\textsf{\upshape cost}_{\ucgrz}}
\newcommand{\costR}{\textsf{\upshape cost}_{\rgate}}
\newcommand{\costmethodlabel}[1]{\scriptscriptstyle\text{\upshape #1}}
\newcommand{\costheuristic}[1]{\costU^{\costmethodlabel{\oursheuristic{}}}(#1)}
\newcommand{\costheuristicopt}[1]{\costU^{\costmethodlabel{\oursheuristicopt{}}}(#1)}
\newcommand{\costDheuristicopt}[1]{\costUD^{\costmethodlabel{\oursheuristicopt{}}}(#1)}
\newcommand{\costanalytic}[1]{\costU^{\costmethodlabel{\oursanalytic{}}}(#1)}
\newcommand{\costPDheuristic}[1]{\costUPD^{\costmethodlabel{\oursheuristic{}}}(#1)}

\newcommand{\costPDanalytic}[1]{\costUPD^{\costmethodlabel{\oursanalytic{}}}(#1)}
\newcommand{\costUD}{\textsf{\upshape cost}_{U_{\diagonal}}}
\newcommand{\costUBlockZXZ}{\costU^{\costmethodlabel{Block-ZXZ}}}
\newcommand{\costUDBlockZXZ}{\costUD^{\costmethodlabel{Block-ZXZ}}}
\newcommand{\costUPD}{\textsf{\upshape cost}_{U_{P\diagonal}}}
\newcommand{\costSP}{\textsf{\upshape cost}_{sp}}
\newcommand{\costIso}{\textsf{\upshape cost}_{\mathrm{Iso}}}
\newcommand{\costIsoPD}{\textsf{\upshape cost}_{\mathrm{Iso}_{\mathrm{PD}}}}
\newcommand{\costBE}{\textsf{\upshape cost}_{\mathrm{BE}}}

\DeclareRobustCommand{\oursanalytic}{\texorpdfstring{\ensuremath{\mathrm{Ours}_{\mathrm{Analytic}}}}{Ours Analytic}}
\DeclareRobustCommand{\oursheuristic}{\texorpdfstring{\ensuremath{\mathrm{Ours}_{\mathrm{Heur}}}}{Ours Heur}}
\DeclareRobustCommand{\oursheuristicopt}{\texorpdfstring{\ensuremath{\mathrm{Ours}_{\mathrm{Heur\text{-}Opt}}}}{Ours Heur-Opt}}
\newcommand{\OurAnalyticCost}{%
  \frac{85}{192}4^n-\frac{3}{2}2^n+\frac{8}{3}}
\newcommand{\OurHeuristicCost}{%
  \frac{41}{96}4^n-\frac{9}{8}2^n+\frac{2}{3}}
\newcommand{\OurHeuristicOptCost}{%
  \frac{5}{12}4^n-\frac{3}{4}2^n-n+\frac{4}{3}}

\makeatletter
\ExplSyntaxOn
\newcommand{\benchmarkCNOT}[2]{
  \group_begin:
    \tl_set:Nn \@currentlabel {#2}
    \tl_remove_all:Nn \@currentlabel {,}
    \label{cnot:#1}
  \group_end:
  #2
}
\ExplSyntaxOff
\makeatother

\newcommand{\siableBEcost}{\frac{11}{48}4^n-2^n+\frac{7}{3}}
\newcommand{\OurAnalyticBECost}{%
  \frac{85}{384}4^n-2^n+\frac{10}{3}}
\newcommand{\OurHeuristicBECost}{%
  \frac{41}{192}4^n-\frac{3}{4}2^n+\frac{4}{3}}
\newcommand{\OurHeuristicOptBECost}{%
  \frac{5}{24}4^n-2^{n-1}-n+\frac{8}{3}}

\newcommand{\ItenIsoCost}{%
  \frac{23}{64}4^n-\frac{5}{4}2^n+1}
\newcommand{\LiIsoCost}{%
  \frac{11}{32}4^n-\frac{5}{4}2^n+2}
\newcommand{\OurAnalyticIsoCost}{%
  \frac{85}{256}4^n-\frac{5}{4}2^n+3}
\newcommand{\OurHeuristicIsoCost}{%
  \frac{41}{128}4^n-\frac{15}{16}2^n+1}
\newcommand{\OurHeuristicOptIsoCost}{%
  \frac{5}{16}4^n-\frac{5}{8}2^n-n+2}

\theoremstyle{plain}
\newtheorem{theorem}{Theorem}[section]
\newtheorem{proposition}[theorem]{Proposition}
\newtheorem{fact}[theorem]{Fact}
\newtheorem{corollary}[theorem]{Corollary}
\newtheorem{claim}[theorem]{Claim}
\newtheorem{conjecture}[theorem]{Conjecture}
\newtheorem*{claim*}{Claim}

\theoremstyle{definition}
\newtheorem{definition}[theorem]{Definition}
\newtheorem{example}[theorem]{Example}
\newtheorem*{problem}{Problem}
\newtheorem{remark}[theorem]{Remark}

\begin{document}

\title{Quantum Circuit Synthesis for Unitaries via the Quantum Multiplicative Inverse Eigenvalue Problem}

\author{Chan-Gu Kang}
\email{changukang@korea.ac.kr}
\affiliation{Department of Computer Science, Korea University, 145 Anam-ro, Seongbuk-gu, Seoul 02841, Republic of Korea}

\author{Hakjoo Oh}
\email{hakjoo\_oh@korea.ac.kr}
\affiliation{Department of Computer Science, Korea University, 145 Anam-ro, Seongbuk-gu, Seoul 02841, Republic of Korea}

\date{November 27, 2026}

\maketitle

\begin{abstract}
In this work, we present an ancilla-free method for synthesizing quantum circuits that implement a given unitary with fewer CNOT gates.
We introduce the quantum multiplicative inverse eigenvalue problem (QMIEP)
as a quantum variant of the classical multiplicative inverse eigenvalue problem.
Given a unitary $U$, the QMIEP asks for a permutation matrix $P$ and a diagonal unitary $D$ such that $DPU$ has an antipodal spectrum.
We exploit this spectral structure to reduce the CNOT count during the synthesis.
Our fully analytic construction applies to every unitary and reduces the asymptotic CNOT count by $3.4\%$
relative to the state-of-the-art optimized Block-ZXZ decomposition.
A second construction increases this reduction to $6.8\%$ when a numerical heuristic solves the required instances
of the QMIEP, which we conjecture is solvable for every unitary.
We also describe an optimization that can reduce the CNOT count further.
We provide software implementations of all our methods and code to reproduce our experiments.
\end{abstract}

\section{Introduction}
\label{sec:introduction}

Quantum computing is a promising next-generation computing technology.
Its importance stems from the promise of quantum advantage over classical
computers in fields such as quantum chemistry and
physics~\cite{feynman1982,lloyd1996,berry2015,peruzzo2014},
cryptography~\cite{shor1997}, and
optimization~\cite{grover1996,farhi2014,cerezo2021,abbas2024}.

\paragraph{Problem}
In this paper, we study the ancilla-free unitary synthesis problem: constructing an
$n$-qubit circuit that implements a given unitary $U\in U(2^n)$.
Here, we consider circuits composed of arbitrary 1-qubit gates
and \cnot{} gates, with the goal of minimizing the \cnot{} count.
Reducing this count is crucial for reliably executing quantum programs
on noisy quantum computers.
See Section~\ref{sec:background} for further context on the problem formulation.

\paragraph{Motivation}
Unitary synthesis is one of the most fundamental tasks in quantum programming.
It is directly required by many quantum applications, including POVM
implementation through unitary dilation~\cite{ivashkov2024,davies1970} and
Hamiltonian simulation through Trotterization~\cite{lloyd1996,koh2026}.

Unitary synthesis also underlies several other tasks in quantum programming.
These include state preparation~\cite{plesch2011,araujo2024,li2026}, which
encodes a classical vector in the amplitudes of a quantum state;
isometry synthesis~\cite{iten2016}, which implements a given isometric map
as a quantum circuit; and block encoding~\cite{low2019,gilyen2019,li2026},
which constructs a circuit implementing a unitary that contains a given
matrix as a block.
These constructions serve as building blocks for many other quantum applications~\cite{zhang2026}.
Unitary synthesis is a subroutine in the best-known algorithms for these
tasks; therefore, reducing its \cnot{} count also lowers their implementation costs.
Section~\ref{sec:application} shows how our unitary synthesis method and its variants reduce their \cnot{} counts,
improving on the state-of-the-art.

\paragraph{Existing Works}
Unitary synthesis has been studied for more than three
decades~\cite{barenco1995,knill1995,tucci1999,cybenko2001,aho2003,vartiainen2004,mottonen2004,mottonen2006,shende2006,devos2016,krol2022,mansky2023}, and much of
this work has sought to close the gap between constructive \cnot{} upper
bounds and the parameter-counting lower bound $\frac{1}{4}(4^n - 3n - 1)$~\cite{shende2006}.
Some representative works are as follows.
The first general construction, given in 1995~\cite{barenco1995},
required $O(n^3 4^n)$ gates.
A breakthrough came in 2006 with the Quantum Shannon Decomposition
(QSD)~\cite{shende2006}, which established the previous state-of-the-art with a \cnot{} count of
$\frac{23}{48}4^n - \frac{3}{2}2^n + \frac{4}{3}$,
bringing the asymptotic ratio to the lower bound below~2.
The current state-of-the-art, established in 2024~\cite{krol2024} by
building on the Block-ZXZ decomposition~\cite{devos2016}, attains
$\frac{22}{48}4^n - \frac{3}{2}2^n + \frac{5}{3}$.
This is an asymptotic improvement of $4.3\%$ over QSD in 18 years,
maybe reflecting the difficulty of the problem.

\begin{table*}[t]
    \caption{\cnot{} counts for synthesis of a generic $n$-qubit unitary
    for $n \geq 3$ ($n=1$ is trivial, and optimal synthesis is known for
    $n=2$~\cite{shende2004,vatan2004,vidal2004}).
    Note that \oursanalytic{} is analytic with guaranteed universality, as are
    the comparable QSD and Block-ZXZ methods.
    \oursheuristic{} instead relies on a numerical heuristic and on a
    conjecture for universal solvability that is supported by numerical
    evidence, while offering a further reduction in \cnot{} count.
    For \oursheuristicopt{}, the \cnot{} counts for $4\leq n\leq 8$ are averages over 100 Haar-random unitaries
    per size (Table~\ref{tab:synthesis-runtime}), and the cost formula gives an achievable lower bound.}
    \label{tab:cnot-count-comparison}
    \centering
    \begingroup
        \footnotesize
        \renewcommand{\arraystretch}{1.55}
        \setlength{\tabcolsep}{10pt}
        \setlength{\arrayrulewidth}{0.35pt}
        \noindent\resizebox{\linewidth}{!}{%
        \begin{tabular}{lccccccc}
            \toprule
            \multirow{2}{*}{\textbf{Method}}
                & \multicolumn{6}{c}{\textbf{Qubits, }$\bm{n}$}
                & \multirow{2}{*}{\textbf{\boldmath\cnot{} cost }$\bm{\costU(n)}$} \\
            \cmidrule(lr){2-7}
                & $\mathbf{3}$ & $\mathbf{4}$ & $\mathbf{5}$
                & $\mathbf{6}$ & $\mathbf{7}$ & $\mathbf{8}$ & \\
            \midrule
            QSD (optimized)~\cite{shende2006}
                & $20$ & $100$ & $444$ & $1868$ & $7660$ & $31020$
                & $\frac{23}{48}4^n-\frac{3}{2}2^n+\frac{4}{3}$ \\
            Block-ZXZ (optimized)~\cite{krol2024}
                & $19$ & $\ref*{cnot:blockzxz-4}$ & $\ref*{cnot:blockzxz-5}$ & $\ref*{cnot:blockzxz-6}$ & $\ref*{cnot:blockzxz-7}$ & $\ref*{cnot:blockzxz-8}$
                & $\frac{22}{48}4^n-\frac{3}{2}2^n+\frac{5}{3}$ \\
            \textbf{\oursanalytic{}}
                & $\mathbf{19}$ & $\mathbf{\ref*{cnot:analytic-4}}$ & $\mathbf{\ref*{cnot:analytic-5}}$ & $\mathbf{\ref*{cnot:analytic-6}}$
                & $\mathbf{\ref*{cnot:analytic-7}}$ & $\mathbf{\ref*{cnot:analytic-8}}$
                & $\bm{\OurAnalyticCost}$ \\
            \addlinespace[1.5pt]
            \cdashline{1-8}[2pt/2.5pt]
            \textbf{\oursheuristic{}}
                & $\mathbf{19}$ & $\mathbf{\ref*{cnot:heuristic-4}}$ & $\mathbf{\ref*{cnot:heuristic-5}}$ & $\mathbf{\ref*{cnot:heuristic-6}}$
                & $\mathbf{\ref*{cnot:heuristic-7}}$ & $\mathbf{\ref*{cnot:heuristic-8}}$
                & $\bm{\OurHeuristicCost}$ \\
            \textbf{\oursheuristicopt{}}
                & $\mathbf{19}$ & $\mathbf{\ref*{cnot:heuristic-opt-4}}$ & $\mathbf{\ref*{cnot:heuristic-opt-5}}$ & $\mathbf{\ref*{cnot:heuristic-opt-6}}$
                & $\mathbf{\ref*{cnot:heuristic-opt-7}}$ & $\mathbf{\ref*{cnot:heuristic-opt-8}}$
                & $\bm{\geq \OurHeuristicOptCost}$ \\
            \midrule
            Theoretical lower bound~\cite{shende2006}
                & $14$ & $61$ & $252$ & $1020$ & $4091$ & $16378$
                & $\left\lceil\frac{1}{4}\left(4^n-3n-1\right)\right\rceil$ \\
            \bottomrule
        \end{tabular}
        }%
        \par
    \endgroup
\end{table*}

\paragraph{Key Results}
We present a unitary synthesis method that improves on the state-of-the-art
in \cnot{} count, as summarized in Table~\ref{tab:cnot-count-comparison}.
We propose three modes of synthesis:
\begin{itemize}
    \item \oursanalytic{}: We give a fully analytic construction with a
    \cnot{} count of $\OurAnalyticCost$, an asymptotic reduction of $3.4\%$ relative to the
    state-of-the-art.
    \item \oursheuristic{}: We further give a method with a \cnot{} count
    of $\OurHeuristicCost$, an asymptotic reduction of $6.8\%$ relative to the
    state-of-the-art.
    We prove that this improvement holds on a subset of $U(2^n)$ with
    non-zero probability (i.e., non-zero Haar measure), although universality remains unproven.
    \begin{itemize}
        \item[$\circ$] We conjecture that \oursheuristic{} is universal,
        achieving the stated \cnot{} reduction for every unitary.
        Our numerical study provides strong support for the universal solvability
        of the underlying problem (see Section~\ref{sec:general-case}).
    \end{itemize}

    \item \oursheuristicopt{}: We can further reduce the \cnot{} count by refining \oursheuristic{}.
    The savings depend on the generic subunitaries encountered during synthesis,
    so the resulting cost can be viewed probabilistically.
    It is bounded above by the cost of \oursheuristic{} and can be as low as $\OurHeuristicOptCost$.
\end{itemize}
We present \oursanalytic{} and \oursheuristic{} in Section~\ref{sec:synthesis}
and describe \oursheuristicopt{} in Section~\ref{sec:further-optimal-d-equivalent}.

Our improved unitary synthesis method and its underlying technique also apply to
state preparation, isometry synthesis, and block encoding. As shown in
Section~\ref{sec:application}, we obtain lower \cnot{} counts than the state-of-the-art
for all three tasks.

Implementations of the proposed synthesis algorithms and code for reproducing
all numerical experiments in this paper are available in our software
artifact\footnote{\artifactlink}.

\paragraph{Our Approach}

Our method constructs quantum circuits that implement a given unitary up to
permutation and diagonal factors, a relation we call $PD$-equivalence.
Synthesizing a $PD$-equivalent unitary can require fewer \cnot{} gates than
synthesizing the target unitary exactly.
We then compensate for these residual factors during synthesis using a
technique we call $PD$-migration, yielding an exact implementation whose
overall \cnot{} count is lower than that of existing techniques.
To find such $PD$-equivalent unitaries, we formulate a quantum variant of
the multiplicative inverse eigenvalue problem (MIEP), a classical problem
in matrix analysis, which we call QMIEP.
Section~\ref{sec:synthesis} describes the method in detail.

\begin{table}[!t]
  \caption[Notation for quantum gates used throughout this paper.]{Notation
    for quantum gates used throughout this paper. Gate parameters (e.g.,
    $\theta$ in $R_z(\theta)$) are often omitted when clear from context.}
  \label{tab:notation}
  \centering
  \begingroup
  \small
  \setlength{\tabcolsep}{0pt}
  \renewcommand{\arraystretch}{1.15}
  \newcommand{\notationcircuit}[1]{%
    \ensuremath{\vcenter{\hbox{\small$\Qcircuit @C=0.55em @R=0.75em {#1}$}}}%
  }
  \let\notationbundle\qcbundle
  \begin{tabular}{@{}>{\centering\arraybackslash}m{.12\linewidth}%
    @{\hspace{.015\linewidth}}>{\centering\arraybackslash}m{.18\linewidth}%
    @{\hspace{.035\linewidth}}>{\raggedright\arraybackslash}m{.65\linewidth}@{}}
    \toprule
    \textbf{Symbol} & \textbf{Circuit} & \textbf{Description} \\
    \midrule
    \addlinespace[6pt]
    $H$
    & \notationcircuit{& \gate{H} & \qw}
    & Hadamard gate:
      $H=\frac{1}{\sqrt{2}}
        \left(\begin{smallmatrix}1&1\\1&-1\end{smallmatrix}\right).$ \\
    \addlinespace[12pt]
    $S$
    & \notationcircuit{& \gate{S} & \qw}
    & Phase gate:
      $S=\left(\begin{smallmatrix}1&0\\0&i\end{smallmatrix}\right).$ \\
    \addlinespace[12pt]
    $U$
    & \notationcircuit{& \notationbundle{n} & \gate{U} & \qw}
    & Generic $n$-qubit unitary; $L,M,N,O,V,W,A,B,C$ are also used throughout this paper. \\
    \addlinespace[12pt]
    $U_D$
    & \notationcircuit{& \notationbundle{n} & \gate{U_D} & \qw}
    & $D$-equivalent unitary to $U$; see Definition~\ref{def:D-equivalence}. \\
    \addlinespace[12pt]
    $U_{PD}$
    & \notationcircuit{& \notationbundle{n} & \gate{U_{PD}} & \qw}
    & a unitary $PD$-equivalent to $U$; see Definition~\ref{def:PD-equivalence}. \\
    \addlinespace[12pt]
    $D$
    & \notationcircuit{& \notationbundle{n} & \gate{D} & \qw}
    & $n$-qubit diagonal unitary: $D=\operatorname{diag}(e^{i\theta_0},\ldots,e^{i\theta_{2^n-1}}).$ \\
    \addlinespace[12pt]
    $P$
    & \notationcircuit{& \notationbundle{n} & \gate{P} & \qw}
    & $n$-qubit permutation matrix: $P=\sum_{i}|\sigma(i)\rangle\langle i|$, $\sigma\in S_{2^n}.$ \\
    \addlinespace[12pt]
    $\cnot{}$
    & \notationcircuit{& \ctrl{1} & \qw \\ & \targ & \qw}
    & Controlled-NOT gate:
      $|b_1b_2\rangle\mapsto|b_1,\,b_1\oplus b_2\rangle.$ \\
    \addlinespace[12pt]
    $\cz{}$
    & \notationcircuit{& \ctrl{1} & \qw \\ & \control \qw & \qw}
    & Controlled-$Z$ gate:
      $|b_1b_2\rangle\mapsto(-1)^{b_1b_2}|b_1b_2\rangle.$ \\
    \addlinespace[12pt]
    $R_z$
    & \notationcircuit{& \gate{R_z(\theta)} & \qw}
    & $z$-axis rotation:
      $R_z(\theta)=e^{-i\theta Z/2}
        =\operatorname{diag}(e^{-i\theta/2},e^{i\theta/2}).$ \\
    \addlinespace[12pt]
    $CU$
    & \notationcircuit{
      & \qw & \qw & \ctrl{1} & \qw \\
      & \notationbundle{n-1} & \qw & \gate{U} & \qw
    }
    & Controlled $(n-1)$-qubit unitary:
      \newline
      $CU=|0\rangle\langle 0|\otimes I+|1\rangle\langle 1|\otimes U.$ \\
    \addlinespace[12pt]
    $\displaystyle\bigoplus_{j=0}^{2^m-1}U_j$
    & \notationcircuit{
      & \notationbundle{m} & \qw & \gate{} \qwx[1] & \qw \\
      & \notationbundle{k} & \qw & \gate{\mathcal{U}} & \qw
    }
    & Multiplexor with $m$ controls and $k$-qubit unitaries $U_j$:
      \newline
      $\bigoplus_{j=0}^{2^m-1} U_j=\sum_{j=0}^{2^m-1} |j\rangle\langle j|\otimes U_j$,
      $\mathcal{U}=(U_0,\ldots,U_{2^m-1})$. \\
    \addlinespace[12pt]
    $\rgate(\vec\theta)$
    & \notationcircuit{& \notationbundle{n} & \gate{\rgate(\vec\theta)} & \qw}
    & UCG($R_z$) with rotation angles $\vec\theta\in\mathbb{R}^{2^{n-1}}$ and one boundary \cnot{} removed
      (Section~\ref{sec:prelim:diagonal}). \\
    \addlinespace[6pt]
    \bottomrule
  \end{tabular}
  \endgroup
\end{table}

\section{Preliminaries} \label{sec:preliminary}
In this section, we introduce the background and notation used throughout
the paper, together with circuit identities from prior work that serve as
building blocks for our synthesis method.

\paragraph{Notation}

Table~\ref{tab:notation} summarizes the gate and circuit notation used
throughout this paper. We measure circuit cost by the number of \cnot{}
gates. For a class $X$ of gates, we write
\[
    \costX(n)
    := \text{the \cnot{} count of the construction for a generic $n$-qubit gate in $X$}.
\]
Each cost expression is derived from the corresponding circuit
implementation presented throughout this paper.

\subsection{Background on the Unitary Synthesis Problem} \label{sec:background}

Here, we state the problem and its context more precisely.
Given a unitary matrix $U\in U(2^n)$, the ancilla-free unitary synthesis problem is to
construct an $n$-qubit circuit implementing $U$ over a prescribed gate set.
We consider arbitrary 1-qubit gates and \cnot{} gates. Together, these gates
form a universal gate set, so every unitary can be represented exactly by a
finite sequence of them~\cite{barenco1995,nielsen2010}.

Our objective is to construct such circuits with as few \cnot{} gates as
possible. \cnot{}-count minimization is a standard objective in unitary and
related circuit synthesis~\cite{shende2006,iten2016,krol2024,vidal2004,vatan2004,shende2004b,mottonen2004,bergholm2005,plesch2011,araujo2024,malvetti2021,li2026}.
This objective reflects the physical cost of 2-qubit operations, which
require controls for interactions between qubits and typically have higher
error rates and longer execution times than 1-qubit operations.

For instance, consider IBM's recent Fez quantum processor, whose native 2-qubit
gate is \cz{} (equivalent to \cnot{} up to 1-qubit gates, since
$\cz{} = (I \otimes H)\,\cnot{}\,(I \otimes H)$).
On this device, the median error rates of \cz{} gates and 1-qubit gates are
$0.330\%$ and $0.0258\%$, respectively, and their durations are
$84\,\mathrm{ns}$ and $24\,\mathrm{ns}$~\cite{ibmfez2025}.
Thus, a \cz{} gate has approximately $12.8$ times the error and $3.5$ times
the duration of a 1-qubit gate,
which highlights the importance of reducing \cnot{} counts.

Minimizing the \cnot{} count is particularly important for near-term
(NISQ) devices.
Other gate sets and computational models lead to different synthesis
objectives. For instance, another major objective may be $T$-gate optimization
because of the higher cost of implementing $T$ gates in fault-tolerant quantum
computing.

Synthesis with ancillary qubits is another major research direction.
In this work, we focus on ancilla-free synthesis, implementing $n$-qubit
unitaries using $n$-qubit circuits.
This setting remains valuable given the space complexity of quantum computers (e.g., more qubit interactions may induce noise).
See Section~\ref{sec:related} for related references and further discussion.

We also note that our synthesis algorithm targets Haar-generic unitaries, i.e.,
unitaries sampled from the Haar distribution over the unitary group, and the
\cnot~costs we derive hold for such unitaries.
Although our methods remain applicable to non-generic unitaries, such as
sparse unitaries~\cite{malvetti2021}, specialized methods that exploit such
structure may achieve lower \cnot~costs.
Accordingly, our work, along with comparable works, should be understood as
improving the worst-case \cnot~cost over the unitary group.

The problem we call ``unitary synthesis'' should be considered distinct from the
problem posed in quantum circuit complexity~\cite{aaronson2007}. There, the problem
asks for a polynomial-time quantum algorithm that can approximate any $n$-qubit unitary
$U$ by querying a Boolean oracle chosen for $U$. In this setting, the resources
of interest are the oracle queries together with the elementary gates needed to achieve a
given accuracy. The unitary synthesis problem of interest here is to construct a
circuit that implements $U$ exactly, with the explicit goal of minimizing the
number of \cnot{} gates. (Although the two problems have largely been studied
separately, possible connections between them may be interesting.)

\subsection{Controlled-Diagonal and Uniformly Controlled \texorpdfstring{$R_z$}{Rz}} \label{sec:prelim:diagonal}

An $n$-qubit diagonal gate has the form
$\diagonal = \operatorname{diag}(e^{i\theta_0}, \ldots, e^{i\theta_{2^n-1}})$,
where $\theta_j \in \mathbb{R}$.
The decomposition of its controlled version
uses a \textit{uniformly controlled $R_z$ gate}, denoted by $\ucgrz$.
With a specific qubit as the target and the remaining $n-1$ qubits as controls,
this gate is defined for $\vec\theta\in\mathbb{R}^{2^{n-1}}$ by
\[
\begin{aligned}
\ucgrz(\vec{\theta})
&= \sum_{y \in \{0,1\}^{n-1}} R_z(\theta_y) \otimes \ket{y}\bra{y}
&= \sum_{\substack{b \in \{0,1\} \\ y \in \{0,1\}^{n-1}}}
e^{-i(-1)^b\theta_y/2}\ket{b,y}\bra{b,y}.
\end{aligned}
\]
The definition shows that $\ucgrz$ is a multiplexor controlled by an
$(n-1)$-qubit register. We therefore use multiplexor notation for this gate
in circuit diagrams.
Controlled diagonal gates admit the following decomposition.
\begin{proposition}[Controlled-diagonal implementation~{\cite[Theorem~7]{shende2006}}]\label{prop:shende_theorem_7}
For any diagonal gate $\diagonal\in U(2^{n-1})$, there exists a diagonal gate
$\diagonal'\in U(2^{n-1})$ such that the following circuit identity holds:
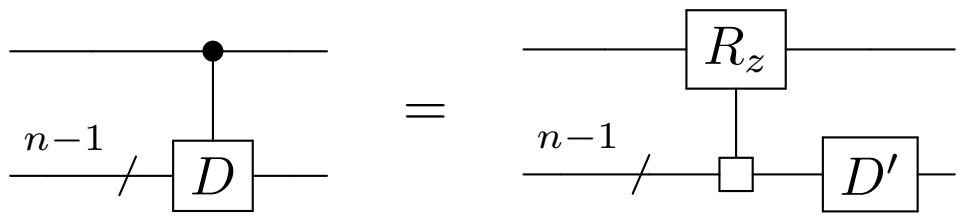
\begin{equation*}
\statementcircuit{
  & \qw           & \ctrl{1}         & \push{\vphantom{\boxed{R_z}}}\qw & \qw \\
  & \qcbundle{n-1} & \gate{\diagonal} & \qw & \qw
}
\quad = \quad
\statementcircuit{
  & \qw           & \ucgrzgate{} & \qw              & \qw \\
  & \qcbundle{n-1} & \ucgrzctrl & \gate{\diagonal'} & \qw
}
\end{equation*}
\end{proposition}

An $n$-qubit $\ucgrz$ gate can be implemented as an alternating
sequence of $R_z$ rotations and \cnot{} gates~\cite{mottonen2004}.
\begin{proposition}[$\ucgrz$ implementation~\cite{mottonen2004}] \label{prop:ucgrz}
For $n\geq 2$ and $\vec\theta\in\mathbb{R}^{2^{n-1}}$, there exists
$\vec\theta'\in\mathbb{R}^{2^{n-1}}$ such that the following circuit identities hold:
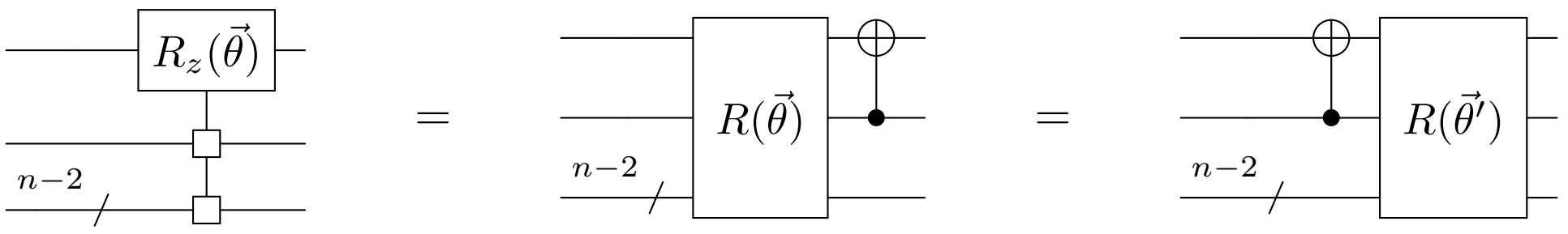
\begin{equation}\label{qc:ucg-rz-boundary-cnot}
\statementcircuit{
  & \qw           & \ucgrzgate{(\vec\theta)} & \qw \\
  & \qw           & \ucgrzctrl   & \qw \\
  & \qcbundle{n-2} & \ucgrzctrl   & \qw
}
\qquad = \qquad
\statementcircuit{
  & \qw           & \multigate{2}{\rgate(\vec\theta)} & \targ     & \qw \\
  & \qw           & \ghost{\rgate(\vec\theta)}        & \ctrl{-1} & \qw \\
  & \qcbundle{n-2} & \ghost{\rgate(\vec\theta)}        & \qw       & \qw
}
\qquad = \qquad
\statementcircuit{
  & \qw           & \targ     & \multigate{2}{\rgate(\vec\theta')} & \qw \\
  & \qw           & \ctrl{-1} & \ghost{\rgate(\vec\theta')}        & \qw \\
  & \qcbundle{n-2} & \qw       & \ghost{\rgate(\vec\theta')}        & \qw
}
\end{equation}
Here, $\rgate(\vec\theta)$\footnote{We adopted the notation $\rgate$ from~\cite{li2026}.}
is defined by
\begin{equation}\label{eq:r-gate-definition}
\rgate(\vec\theta)=
\sum_{\substack{b\in\{0,1\} \\ y=(y_0,\ldots,y_{n-2})\in\{0,1\}^{n-1}}}
e^{-i(-1)^b\theta_y/2}\ket{b\oplus y_0,y}\bra{b,y}.
\end{equation}
The implementation of $\rgate$ contains $2^{n-1}-1$ \cnot{} gates interleaved with
$2^{n-1}$ $R_z$ rotations.
The resulting \cnot{} counts are
\begin{equation}\label{eq:ucgrz-r-cost}
  \costucgrz(n)=2^{n-1},
  \qquad
  \costR(n)=\costucgrz(n)-1=2^{n-1}-1.
\end{equation}
\end{proposition}

Note that there are two choices for this implementation: the boundary
\cnot{} gate can be placed either at the beginning or at the end of the circuit.
This flexibility underlies the \cnot{}-reduction technique
of~\cite{krol2024}, which we also reproduce in Section~\ref{sec:prelim:block_zxz}.

\subsection{Controlled and Multiplexed Unitaries} \label{sec:prelim:controlled_u}

For $U \in U(2^{n-1})$, the $2^n$-dimensional operator
$\conu U := I_{2^{n-1}} \oplus U$ is called a \textit{controlled unitary}.
The implementation of $\conu U$ uses the \textit{spectral decomposition}: for a $k$-dimensional unitary $U$, there exist a $k$-dimensional unitary $V$ and a
diagonal unitary $\diagonal$ such that $U=V\diagonal V^\dagger$.
The \textit{spectrum} $\spec(U)$ is the multiset of eigenvalues of $U$, counted with algebraic multiplicity.
For $\diagonal = \operatorname{diag}(e^{i\theta_0}, \ldots, e^{i\theta_{k-1}})$ with $\theta_j\in\mathbb{R}$,
we have $\spec(U) = \{e^{i\theta_0}, \ldots, e^{i\theta_{k-1}}\}$.
Using this decomposition, $\conu U$ can be implemented by applying the
control only to $\diagonal$:
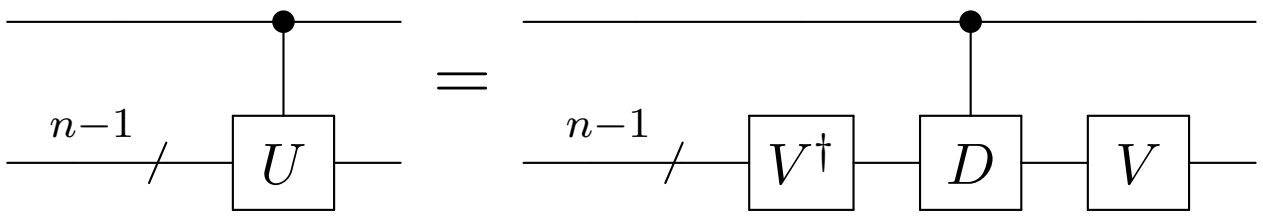
\begin{equation} \label{eq:controlled-unitary}
\begin{array}{c}
\Qcircuit @C=\qcC @R=\qcR {
  & \qw & \ctrl{1} & \qw \\
  & \qcbundle{n-1} & \qcunitarygate{U} & \qw
}
\;\;\raisebox{-1.1em}{\Large $=$}\;\;
\Qcircuit @C=\qcC @R=\qcR {
  &\qw & \qw      & \ctrl{1} & \qw      & \qw \\
  & \qcbundle{n-1}  & \qcunitarygate{V^\dagger} & \qcunitarygate{\diagonal} & \qcunitarygate{V} & \qw
}
\end{array}.
\end{equation}

This construction generalizes to a multiplexor. For
$U_0, U_1 \in U(2^{n-1})$, the $2^n$-dimensional operator
$U := U_0 \oplus U_1$ acts as
$U\ket{0}\ket{\psi} = \ket{0}U_0\ket{\psi}$ and
$U\ket{1}\ket{\psi} = \ket{1}U_1\ket{\psi}$ for
$\ket{\psi} \in \mathbb{C}^{2^{n-1}}$.
As with $\conu U$, a multiplexor can be implemented using two
$(n-1)$-qubit unitaries and one $\conu\diagonal$ gate:

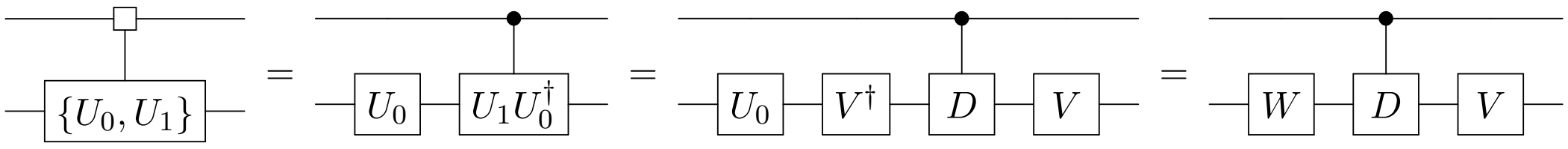
\begin{equation} \label{eq:multiplexor}
\resizebox{0.92\linewidth}{!}{$\displaystyle
\vcenter{\hbox{$
\Qcircuit @C=\qcCm @R=\qcRm {
  & \mux \qwx[1]     & \qw \\
  & \qcunitarygate{\{U_0,U_1\}} & \qw
}
 $}}
\;=\;
\vcenter{\hbox{$
\Qcircuit @C=\qcCm @R=\qcRm {
  & \qw      & \ctrl{1}            & \qw \\
  & \qcunitarygate{U_0} & \qcunitarygate{U_1U_0^{\dagger}} & \qw
}
 $}}
\;=\;
\vcenter{\hbox{$
\Qcircuit @C=\qcCm @R=\qcRm {
  & \qw      & \qw                & \ctrl{1} & \qw      & \qw \\
  & \qcunitarygate{U_0} & \qcunitarygate{V^{\dagger}} & \qcunitarygate{\diagonal} & \qcunitarygate{V} & \qw
}
 $}}
\;=\;
\vcenter{\hbox{$
\Qcircuit @C=\qcCm @R=\qcRm {
   & \qw                & \ctrl{1} & \qw      & \qw \\
 & \qcunitarygate{W} & \qcunitarygate{\diagonal} & \qcunitarygate{V} & \qw
}
 $}}
$}
\end{equation}
Here, $V$ and $\diagonal$ are obtained from the spectral decomposition
$U_1U_0^\dagger = V\diagonal V^\dagger$.
The last equality follows by absorbing $U_0$ into $V^\dagger$ and defining
$W := V^\dagger U_0$.
This construction is a basic building block of most unitary synthesis methods,
particularly those that can be viewed as instances of recursive Cartan
decomposition~\cite{wierichs2025}.
These include the Block-ZXZ decomposition, which serves as our starting point
for unitary synthesis.

Note that the cost of this implementation depends on the cost of $\conu\diagonal$,
which in turn depends on the spectrum of the unitary.
In general,
$\diagonal$ is fully entangled across the qubit register.
Its controlled implementation is therefore generically costly in this
synthesis approach.
Later, we present a method that constructs a controlled unitary
with a separable spectrum at the expense of some residual factors.
These residual factors can be mitigated as part of the overall unitary synthesis,
reducing the total synthesis cost.

\subsection{Baseline Unitary Synthesis} \label{sec:prelim:block_zxz}
Our technique is compatible with most unitary synthesis methods based on
recursive Cartan decompositions~\cite{wierichs2025}.
To make the presentation concrete, we use the state-of-the-art Block-ZXZ method
as our baseline.
We first review this method before showing how our technique further reduces
its synthesis cost.

\subsubsection{Block-ZXZ Decomposition}
The Block-ZXZ decomposition~\cite{devos2014,fuhr2015}
provides a recursive decomposition of a unitary.
We state it below, together with a representation for later use:
\begin{proposition}[Block-ZXZ decomposition in a quantum circuit implementation~\cite{devos2016}]\label{prop:blockzxz}
For any $n$-qubit unitary $U$ with $n\geq 3$, the following circuit identities hold:
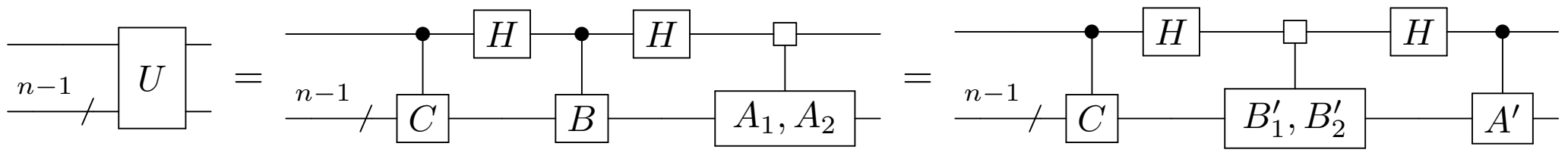
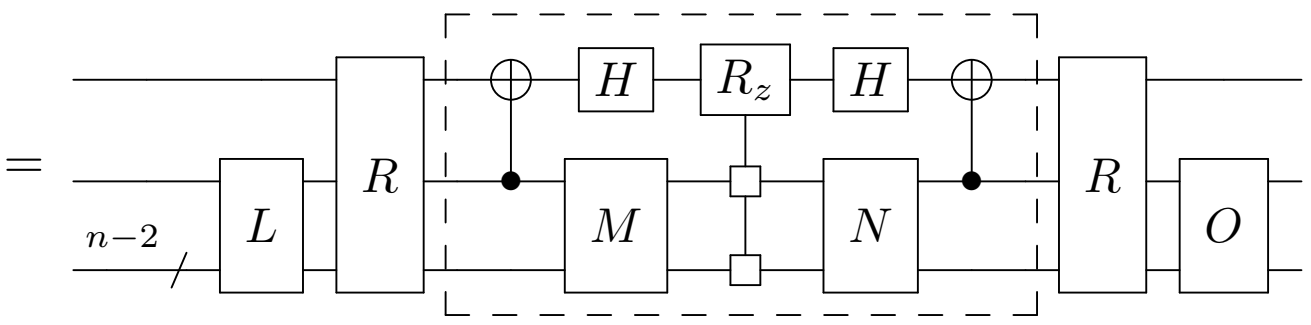
\begin{subequations}\label{qc:block-zxz}
\begin{gather}
\resizebox{0.90\linewidth}{!}{$\displaystyle
\statementcircuit{
& \qw           & \multigate{1}{U} & \qw \\
& \qcbundle{n-1} & \ghost{U}        & \qw
}
\;=\;
\statementcircuit{
& \qw           & \ctrl{1} & \gate{H} & \ctrl{1} & \gate{H} & \mux \qwx[1]   & \qw \\
& \qcbundle{n-1} & \gate{C} & \qw      & \gate{B} & \qw      & \gate{\,A_1,A_2\,} & \qw
}
\;=\;
\statementcircuit{
  & \qw           & \ctrl{1} & \gate{H} & \mux \qwx[1]       & \gate{H} & \ctrl{1} & \qw \\
  & \qcbundle{n-1} & \gate{C} & \qw      & \gate{\,B'_1,B'_2\,} & \qw      & \gate{A'} & \qw
}$}\label{qc:block-zxz-ucg:block}
\\[1em]
\scalebox{0.90}{$\displaystyle
=\;
\statementcircuit{
  & \qw           & \qw              & \multigate{2}{\rgate} & \qw & \targ     & \gate{H}         & \ucgrzgate{} & \gate{H}         & \targ                                     & \qw & \multigate{2}{\rgate} & \qw              & \qw \\
  & \qw           & \multigate{1}{L} & \ghost{\rgate}        & \qw & \ctrl{-1} & \multigate{1}{M} & \ucgrzctrl         & \multigate{1}{N} & \ctrl{-1}                                 & \qw & \ghost{\rgate}        & \multigate{1}{O} & \qw \\
  & \qcbundle{n-2} & \ghost{L}        & \ghost{\rgate}        & \qw & \qw       & \ghost{M}        & \ucgrzctrl                 & \ghost{N}        & \qw \gategroup{1}{6}{3}{10}{1.8em}{--}       & \qw & \ghost{\rgate}        & \ghost{O}        & \qw
}$}\label{qc:block-zxz-ucg:expanded}
\end{gather}
\end{subequations}
\end{proposition}
\begin{proof}[Proof sketch]
    The first circuit identity follows directly from~\cite{devos2016}.
    For the second identity, we decompose the rightmost multiplexor
    $A_1 \oplus A_2$ according to Eq.~\eqref{eq:multiplexor}.
    We then absorb the resulting adjacent $(n-1)$-qubit unitary into
    the controlled-$B$ gate to form a multiplexor.
    For the final identity, we decompose each controlled and multiplexed
    unitary and expand the resulting controlled diagonal gates using
    Propositions~\ref{prop:shende_theorem_7} and~\ref{prop:ucgrz},
    absorbing the remaining diagonal factors into adjacent $(n-1)$-qubit unitaries.
\end{proof}
The Block-ZXZ decomposition is applied recursively until the remaining
unitaries act on 2 qubits. A generic 2-qubit unitary can be implemented
using 3 \cnot{} gates, which is optimal~\cite{vidal2004,vatan2004,shende2004}.
In our notation, we write $\costU(2)=3$.
As stated in Proposition~\ref{prop:ucgrz}, the $\rgate$ and $\ucgrz$ gates
at each stage can be decomposed into \cnot{} and 1-qubit gates.
Together, these steps yield a complete synthesis of an arbitrary unitary
using only \cnot{} and 1-qubit gates.

\subsubsection{\cnot{} Absorption}
A further optimization eliminates the two boundary \cnot{} gates adjacent
to the $\rgate$ gates~\cite{krol2024}.
The following circuit identity holds:
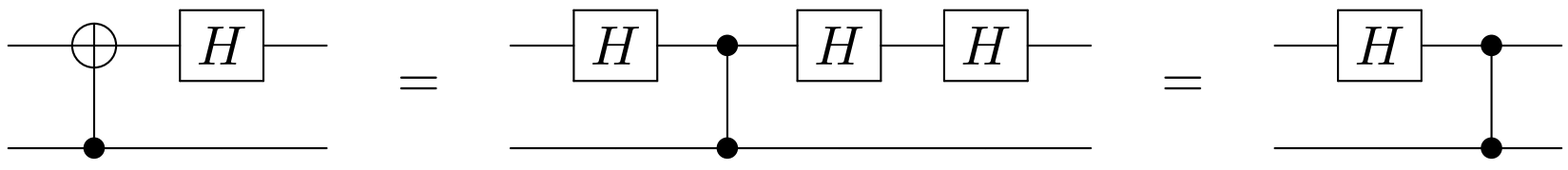
\begin{equation*}
\vcenter{\hbox{$
\Qcircuit @C=\qcC @R=\qcRd {
 & \targ     & \gate{H} & \qw \\
 & \ctrl{-1} & \qw      & \qw
}
$}}
\quad = \quad
\vcenter{\hbox{$
\Qcircuit @C=\qcC @R=\qcRd {
 & \gate{H}  & \ctrl{1}     & \gate{H} & \gate{H} & \qw \\
 & \qw       & \control \qw & \qw      & \qw      & \qw
}
$}}
\quad = \quad
\vcenter{\hbox{$
\Qcircuit @C=\qcC @R=\qcRd {
& \gate{H} & \ctrl{1}     & \qw \\
& \qw      & \control \qw & \qw
}
$}}
\end{equation*}

In the dashed subcircuit of~\eqref{qc:block-zxz-ucg:expanded}, the identity
converts the two \cnot{} gates into \cz{} gates, which are absorbed into a
single multiplexed unitary as follows:

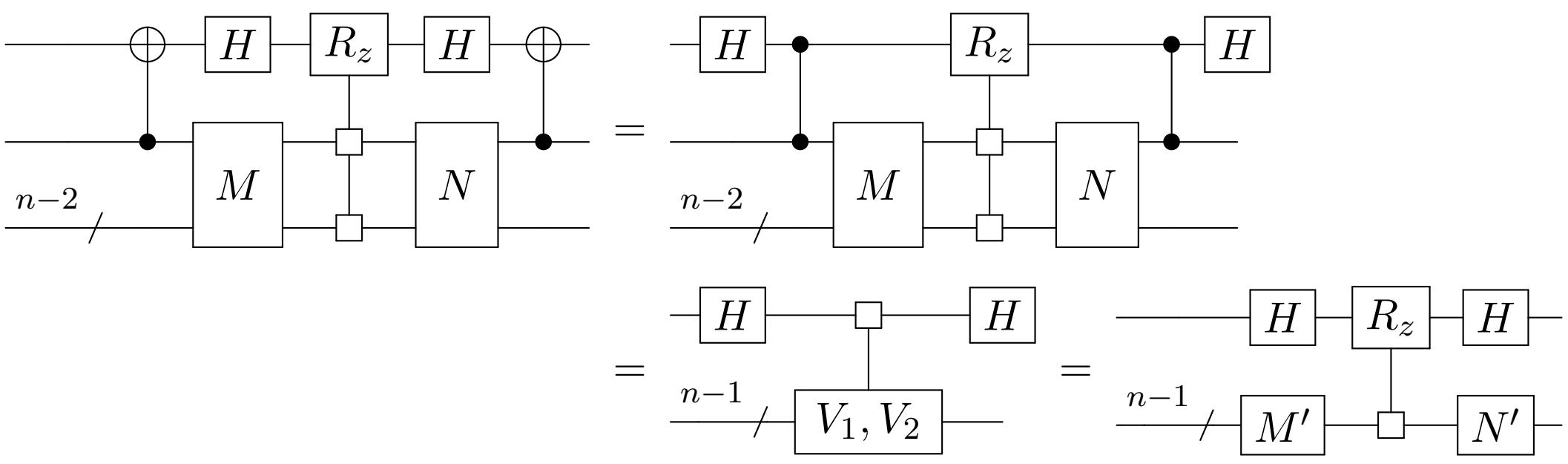
\begin{equation} \label{qc:block-zxz-cz}
\begin{aligned}
\vcenter{\hbox{$
\Qcircuit @C=0.65em @R=\qcRd {
  & \qw           & \targ     & \gate{H}         & \ucgrzgate{} & \gate{H}         & \targ     & \qw \\
  & \qw           & \ctrl{-1} & \multigate{1}{M} & \ucgrzctrl         & \multigate{1}{N} & \ctrl{-1} & \qw \\
  & \qcbundle{n-2} & \qw       & \ghost{M}        & \ucgrzctrl                 & \ghost{N}        & \qw       & \qw
}
$}}
&\;=\;
\vcenter{\hbox{$
\Qcircuit @C=0.65em @R=\qcRd {
  & \gate{H}      & \ctrl{1}     & \qw               & \ucgrzgate{} & \qw               & \ctrl{1}     & \gate{H} \\
  & \qw           & \control \qw & \multigate{1}{M}  & \ucgrzctrl         & \multigate{1}{N}  & \control \qw & \qw \\
  & \qcbundle{n-2} & \qw          & \ghost{M}         & \ucgrzctrl                 & \ghost{N}         & \qw          & \qw
}
$}}
\\[6pt]
&\;=\;
\vcenter{\hbox{$
\Qcircuit @C=0.65em @R=\qcRd {
  & \gate{H}      & \mux \qwx[1]       & \gate{H} \\
  & \qcbundle{n-1} & \gate{\,V_1,V_2\,} & \qw
}
$}}
\;=\;
\vcenter{\hbox{$
\Qcircuit @C=0.65em @R=\qcRd {
  & \qw           & \gate{H}  & \ucgrzgate{} & \gate{H}  & \qw \\
  & \qcbundle{n-1} & \gate{M'} & \ucgrzctrl                  & \gate{N'} & \qw
}
$}}
\end{aligned}
\end{equation}
Applying this identity to the Block-ZXZ decomposition gives the following
decomposition for an arbitrary unitary.

\begin{proposition}[\cnot{} absorption in Block-ZXZ~\cite{krol2024}]\label{prop:block-zxz-cx-absorbed}
The following circuit identity holds:
\begingroup
\setlength{\abovedisplayskip}{12pt plus 2pt minus 2pt}
\setlength{\abovedisplayshortskip}{12pt plus 2pt minus 2pt}
\setlength{\belowdisplayskip}{12pt plus 2pt minus 2pt}
\setlength{\belowdisplayshortskip}{12pt plus 2pt minus 2pt}
\begin{equation}\label{qc:block-zxz-cx-absorbed}
\statementcircuit{
  & \qw           & \multigate{1}{U} & \qw \\
  & \qcbundle{n-1} & \ghost{U}        & \qw
}
\quad = \quad
\statementcircuit{
  & \qw           & \qw              & \multigate{2}{\rgate} & \gate{H}          & \ucgrzgate{} & \gate{H}          & \multigate{2}{\rgate} & \qw              & \qw \\
  & \qw           & \multigate{1}{L} & \ghost{\rgate}        & \multigate{1}{M} & \ucgrzctrl   & \multigate{1}{N} & \ghost{\rgate}        & \multigate{1}{O} & \qw \\
  & \qcbundle{n-2} & \ghost{L}        & \ghost{\rgate}        & \ghost{M}        & \ucgrzctrl   & \ghost{N}        & \ghost{\rgate}        & \ghost{O}        & \qw
}
\end{equation}
\endgroup
\end{proposition}

\subsection{\texorpdfstring{$D$}{D}-Equivalence and \texorpdfstring{$D$}{D}-migration} \label{sec:prelim:block_zxz:diagonal_migrate}
To match the state-of-the-art \cnot{} counts reported in Table~\ref{tab:cnot-count-comparison},
we need an additional optimization called \textit{diagonal migration}, or $D$-migration for short.

\begin{definition} \label{def:D-equivalence}
For $U\in U(2^n)$, we call $U_D$ \textit{equivalent to $U$ up to a diagonal unitary},
or \textit{$D$-equivalent} for short,\footnote{The $D$ in
``$D$-equivalent'' is a label; it does not specify a particular diagonal matrix.}
if there exists a diagonal unitary $D$ such that $U=D U_D$.
We denote the implementation cost of $U_D$ by $\costUD(n)$.
\end{definition}
Taking the adjoint of the left-multiplier form $U^\dagger=DV$ gives the right-multiplier form $U=V^\dagger D^\dagger$.

In the recursive step of the Block-ZXZ synthesis of an $n$-qubit unitary,
assume inductively that we have a $D$-equivalent implementation of each
$(n-1)$-qubit unitary.
We then move the residual diagonal gates through the $\rgate$ and $\ucgrz$ gates.
The following proposition gives the required commutation identities, which follow by direct computation.
\begin{proposition}[Diagonal commutation]\label{prop:diagonal-commutation}
For $n\geq 2$, the following circuit identities hold:
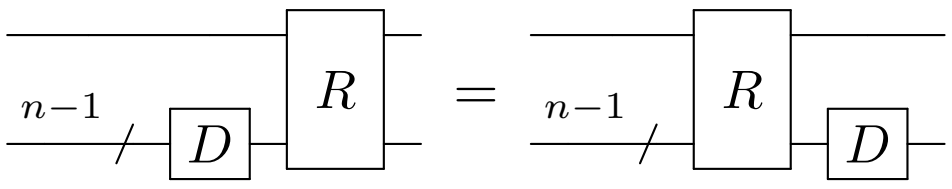
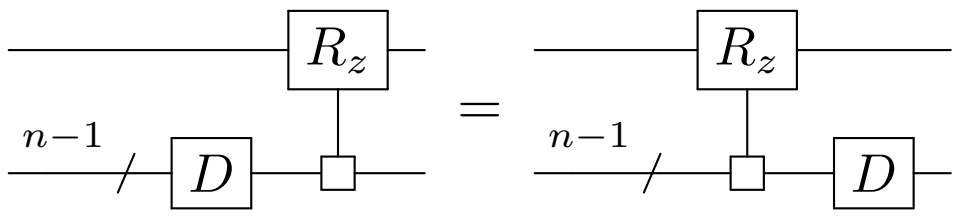
\begin{equation*}
\statementcircuit{
  & \qw           & \qw              & \multigate{1}{\rgate} & \qw \\
  & \qcbundle{n-1} & \gate{\diagonal} & \ghost{\rgate}        & \qw
}
\;=\;
\statementcircuit{
  & \qw           & \multigate{1}{\rgate} & \qw              & \qw \\
  & \qcbundle{n-1} & \ghost{\rgate}        & \gate{\diagonal} & \qw
}
\quad\text{and}\quad
\statementcircuit{
  & \qw           & \qw              & \ucgrzgate{} & \qw \\
  & \qcbundle{n-1} & \gate{\diagonal} & \ucgrzctrl   & \qw
}
\;=\;
\statementcircuit{
  & \qw           & \ucgrzgate{} & \qw              & \qw \\
  & \qcbundle{n-1} & \ucgrzctrl   & \gate{\diagonal} & \qw
}
\end{equation*}
\end{proposition}

For $U,V\in U(2^{n-1})$ with $U=\diagonal U_{\diagonal}$, the commutation relation above
yields the following circuit identity, which we call $D$-migration:
\begin{equation}\label{qc:diagonal-migration}
\begin{aligned}
&
\vcenter{\hbox{$
\Qcircuit @C=0.65em @R=\qcRw {
  & \qw           & \qw      & \ucgrzgate{} & \qw      & \qw \\
  & \qcbundle{n-1} & \gate{U} & \ucgrzctrl   & \gate{V} & \qw
}
$}}
\;=\;
\vcenter{\hbox{$
\Qcircuit @C=0.65em @R=\qcRw {
  & \qw           & \qw                  & \qw              & \ucgrzgate{} & \qw      & \qw \\
  & \qcbundle{n-1} & \gate{U_{\diagonal}} & \gate{\diagonal} & \ucgrzctrl   & \gate{V} & \qw
}
$}}
\\[1.5em]
&\;=\;
\vcenter{\hbox{$
\Qcircuit @C=0.65em @R=\qcRw {
  & \qw           & \qw                  & \ucgrzgate{} & \qw              & \qw      & \qw \\
  & \qcbundle{n-1} & \gate{U_{\diagonal}} & \ucgrzctrl   & \gate{\diagonal} & \gate{V} & \qw
}
$}}
\;=\;
\vcenter{\hbox{$
\Qcircuit @C=0.65em @R=\qcRw {
  & \qw           & \qw                  & \ucgrzgate{} & \qw       & \qw \\
  & \qcbundle{n-1} & \gate{U_{\diagonal}} & \ucgrzctrl   & \gate{V'} & \qw
}
$}}
\end{aligned}
\end{equation}
where $V'=V\diagonal$ can again be treated as a generic $(n-1)$-qubit unitary.
The same identity holds with $\rgate$ in place of
$\ucgrz$.

Using these identities, we obtain the final exact and $D$-equivalent forms
of the state-of-the-art unitary synthesis baseline.
The 2-qubit case serves as the base case:
every $U\in U(4)$ can be factored as $U=D U_D$, where
two \cnot{} gates suffice to implement $U_D$~\cite{shende2004}.
In our notation, this gives $\costUD(2)=2$.
Applying $D$-migration to the $(n-1)$-qubit unitaries on either side
of each $\ucgrz$ or $\rgate$ gate in Equation~\eqref{qc:block-zxz-cx-absorbed}
then yields the following exact and $D$-equivalent implementations.

\begin{proposition}\label{prop:diagonal-migrated-synthesis}
For $n\geq 3$ and any $n$-qubit unitary $U$, the following circuit identities hold.
\begin{itemize}
\item \textit{Exact synthesis}~\cite{krol2024}:
\begin{equation*}
\statementcircuit{
  & \qw           & \multigate{1}{U} & \qw \\
  & \qcbundle{n-1} & \ghost{U}        & \qw
}
\quad = \quad
\statementcircuit{
  & \qw           & \qw              & \multigate{2}{\rgate} & \gate{H}          & \ucgrzgate{} & \gate{H}          & \multigate{2}{\rgate} & \qw              & \qw \\
  & \qw           & \multigate{1}{L_D} & \ghost{\rgate}        & \multigate{1}{M_D} & \ucgrzctrl   & \multigate{1}{N_D} & \ghost{\rgate}        & \multigate{1}{O} & \qw \\
  & \qcbundle{n-2} & \ghost{L_D}        & \ghost{\rgate}        & \ghost{M_D}        & \ucgrzctrl   & \ghost{N_D}        & \ghost{\rgate}        & \ghost{O}        & \qw
}
\end{equation*}
The \cnot{} count is
\begin{equation}\label{eq:cost-block-zxz-exact}
\begin{aligned}
  \costUBlockZXZ(n)
  &= \costUBlockZXZ(n-1)+3\costUDBlockZXZ(n-1)\\
  &\quad +2\costR(n)+\costucgrz(n).
\end{aligned}
\end{equation}
\item \textit{$D$-equivalent synthesis}~\cite{krol2024,shende2006}:
\begin{equation}\label{qc:diagonal-migrated-modulus}
\statementcircuit{
  & \qw           & \multigate{1}{U_D} & \qw \\
  & \qcbundle{n-1} & \ghost{U_D}        & \qw
}
\quad = \quad
\statementcircuit{
  & \qw           & \qw              & \multigate{2}{\rgate} & \gate{H}          & \ucgrzgate{} & \gate{H}          & \multigate{2}{\rgate} & \qw              & \qw \\
  & \qw           & \multigate{1}{L_D} & \ghost{\rgate}        & \multigate{1}{M_D} & \ucgrzctrl   & \multigate{1}{N_D} & \ghost{\rgate}        & \multigate{1}{O_D} & \qw \\
  & \qcbundle{n-2} & \ghost{L_D}        & \ghost{\rgate}        & \ghost{M_D}        & \ucgrzctrl   & \ghost{N_D}        & \ghost{\rgate}        & \ghost{O_D}        & \qw
}
\end{equation}
The \cnot{} count is
\begin{equation*}
  \costUDBlockZXZ(n)
  = 4\costUDBlockZXZ(n-1)+2\costR(n)+\costucgrz(n).
\end{equation*}
\end{itemize}
\end{proposition}
$U_D$ in Eq.~\eqref{qc:diagonal-migrated-modulus} is obtained
by replacing $O$ with $D\cdot O_D$ and then discarding the diagonal factor $D$.

Substituting the expressions for $\costR(n)$ and $\costucgrz(n)$
from Eq.~\eqref{eq:ucgrz-r-cost} into the recurrences above
and solving them with the base cases $\costUBlockZXZ(2)=3$ and
$\costUDBlockZXZ(2)=2$ yields
\begin{equation*}
  \costUBlockZXZ(n)=\frac{22}{48}4^n-\frac{3}{2}2^n+\frac{5}{3},
  \qquad
  \costUDBlockZXZ(n)=\frac{22}{48}4^n-\frac{3}{2}2^n+\frac{2}{3}.
\end{equation*}
The exact synthesis count matches the optimized Block-ZXZ count
reported in Table~\ref{tab:cnot-count-comparison}.

\begin{remark}\label{rem:diagonal-migration-savings}
The described $D$-equivalent synthesis method saves only one \cnot{} gate
compared with the corresponding exact implementation
(i.e., $\costUBlockZXZ(n)-\costUDBlockZXZ(n)=1$).
Let $D_{2^n}$ denote the group of $2^n\times 2^n$ diagonal unitaries.
The space of $D$-equivalence classes $D_{2^n}\backslash U(2^n)$ has real dimension
$4^n-2^n$, a reduction of $2^n$ parameters from $U(2^n)$. This parameter argument motivates the search for
$D$-equivalent synthesis methods with savings that grow with $n$.

Although we propose another equivalence relation between unitaries later,
developing more efficient $D$-equivalent synthesis methods remains important
for improving quantum circuit synthesis. Our approach is related to $D$-equivalence and provides a method for
efficient $D$-equivalent synthesis and, in turn, exact unitary synthesis,
although it has been shown not to apply to
all unitaries. See Section~\ref{sec:numerical_heuristic} for further details.

\end{remark}

\section{Proposed Method: \texorpdfstring{$PD$}{PD}-migration }~\label{sec:synthesis}
In this section, we describe our novel method for unitary synthesis.
Our method can be understood as a direct expansion of the $D$-migration depicted in Section~\ref{sec:prelim:block_zxz:diagonal_migrate}, namely,
$PD$-migration. Here, $D$ again stands for a diagonal unitary, while $P$ stands for
a \textit{permutation} matrix, which is a Boolean unitary matrix.

\subsection{\texorpdfstring{$PD$}{PD}-Equivalence and \texorpdfstring{$PD$}{PD}-migration}
We extend $D$-equivalence in Definition~\ref{def:D-equivalence} to
$PD$-equivalence, where $P$ stands for a permutation matrix
and $D$ for a diagonal unitary.
\begin{definition}\label{def:PD-equivalence}
  For $U\in U(2^n)$, we call $U_{PD}$ \textit{equivalent to $U$ up to a diagonal unitary
  and a permutation}, or \textit{$PD$-equivalent} for short,
  if there exist a diagonal unitary $D$ and a multiplexed permutation matrix $P=P_1\oplus P_2$, with
  $(n-1)$-qubit permutation matrices $P_1,P_2$, such that
  $U=PDU_{PD}$.
  We denote the implementation cost of $U_{PD}$ by $\costUPD(n)$.
\end{definition}
Likewise, taking the adjoint of $U^\dagger=PDV$ gives the right multiplier relation $U=V^\dagger D^\dagger P^\dagger$,
with $P^\dagger=P_1^\dagger\oplus P_2^\dagger$.
We use both forms of $D$- and $PD$-equivalence.

Here, the diagonal factor remains unrestricted, as in $D$-equivalence.
However, we emphasize that the permutation factor is restricted to the
multiplexed form $P=P_1\oplus P_2$ for our application to unitary synthesis.
Also, note that the order of the residual factors $P$ and $D$ need not be fixed,
since $PD=D'P$, where $D'=PDP^\dagger$ is diagonal.

Analogously to $D$-migration, we can also consider $PD$-migration: migrating the $P$ and $D$ factors
through the $R$ and $\ucgrz$ gates.
In this section, we assume that an implementation of $U_{PD}$ is available and show that $PD$-migration is possible.
Just as the diagonal $D$ commutes through $\rgate$ and $\ucgrz$
(Proposition~\ref{prop:diagonal-commutation}), the permutation factor $P$
can also be propagated through them, up to a reordering of parameters.
This requires $P$ to have the multiplexed form $P = P_1 \oplus P_2$
imposed in Definition~\ref{def:PD-equivalence}.

\begin{proposition}[$P = P_1 \oplus P_2 $ propagation through $\ucgrz$ and $\rgate$] \label{prop:pd-thru-r}
Let $P=P_1\oplus P_2$, where $P_1,P_2$ are $(n-2)$-qubit permutation matrices.
For parameters $\vec\theta,\vec\phi\in\mathbb{R}^{2^{n-1}}$, there exist reordered parameter vectors
$\vec\theta_{\pi}$ and $\vec\phi_{\pi}$ such that the following
circuit identities hold:
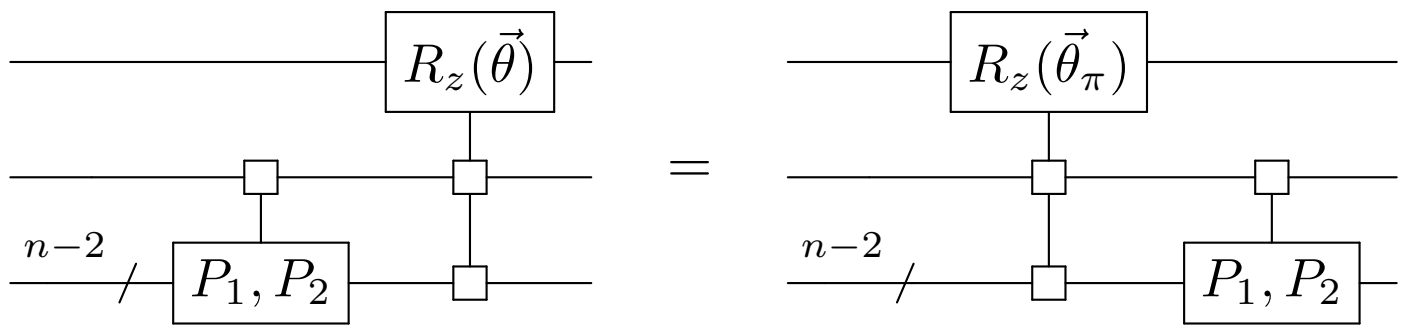
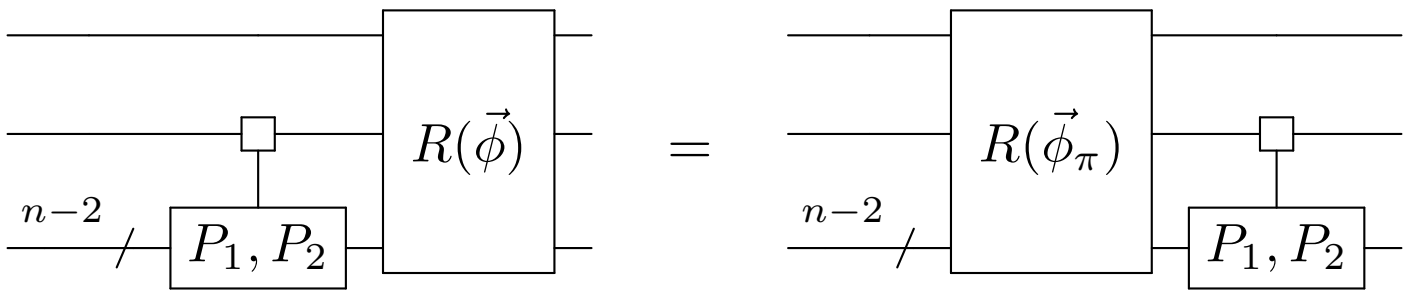
\begin{gather*}
\statementcircuit{
  & \qw           & \qw                 & \ucgrzgate{(\vec\theta)} & \qw \\
  & \qw           & \mux \qwx[1]        & \ucgrzctrl              & \qw \\
  & \qcbundle{n-2} & \gate{P_1,P_2}      & \ucgrzctrl              & \qw
}
\quad = \quad
\statementcircuit{
  & \qw           & \ucgrzgate{(\vec\theta_{\pi})} & \qw            & \qw \\
  & \qw           & \ucgrzctrl                     & \mux \qwx[1]   & \qw \\
  & \qcbundle{n-2} & \ucgrzctrl                     & \gate{P_1,P_2} & \qw
}
\\[0.4em]
\text{and}
\\[0.4em]
\statementcircuit{
  & \qw           & \qw            & \multigate{2}{\rgate(\vec\phi)} & \qw \\
  & \qw           & \mux \qwx[1]   & \ghost{\rgate(\vec\phi)}        & \qw \\
  & \qcbundle{n-2} & \gate{P_1,P_2} & \ghost{\rgate(\vec\phi)}        & \qw
}
\quad = \quad
\statementcircuit{
  & \qw           & \multigate{2}{\rgate(\vec\phi_{\pi})} & \qw            & \qw \\
  & \qw           & \ghost{\rgate(\vec\phi_{\pi})}        & \mux \qwx[1]   & \qw \\
  & \qcbundle{n-2} & \ghost{\rgate(\vec\phi_{\pi})}        & \gate{P_1,P_2} & \qw
}
\end{gather*}
\end{proposition}
\begin{proof}
For $\ucgrz$, $P$ actually can be any control-register permutation, not necessarily multiplexed.
Write $P=\sum_i\ket{\sigma(i)}\bra{i}$, where $\sigma\in S_{2^{n-1}}$,
and define $(\vec\theta_\pi)_i=\theta_{\sigma(i)}$.
Then
\begin{equation}\label{eq:ucgrz-permutation-propagation}
\begin{aligned}
\ucgrz(\vec\theta)(I\otimes P)
&=\sum_{\substack{b\in\{0,1\}\\0\leq i<2^{n-1}}}e^{-i(-1)^b\theta_{\sigma(i)}/2}\ket{b,\sigma(i)}\bra{b,i}\\
&=(I\otimes P)\sum_{\substack{b\in\{0,1\}\\0\leq i<2^{n-1}}}e^{-i(-1)^b\theta_{\sigma(i)}/2}\ket{b,i}\bra{b,i}
=(I\otimes P)\ucgrz(\vec\theta_\pi).
\end{aligned}
\end{equation}
For $\rgate$, with $(\vec\phi_\pi)_i=\phi_{\sigma(i)}$, the result follows
by direct computation using Eq.~\eqref{eq:r-gate-definition}, or equivalently,
from the following circuit identities:
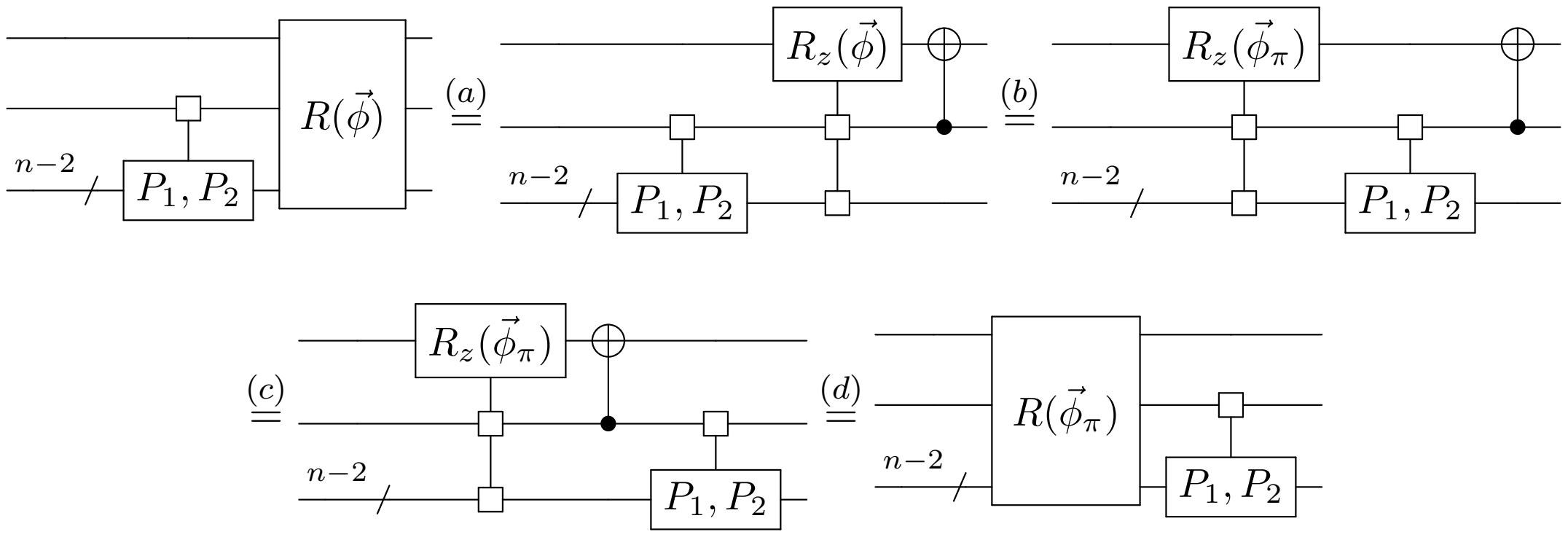
\begin{equation*}
\resizebox{0.94\linewidth}{!}{$\displaystyle
\begin{gathered}
\vcenter{\hbox{\small$
\Qcircuit @C=0.65em @R=0.85em {
  & \qw           & \qw            & \multigate{2}{\rgate(\vec\phi)} & \qw \\
  & \qw           & \mux \qwx[1]   & \ghost{\rgate(\vec\phi)}        & \qw \\
  & \qcbundle{n-2} & \gate{P_1,P_2} & \ghost{\rgate(\vec\phi)}        & \qw
}
$}}
\overset{(a)}{=}\vcenter{\hbox{\small$
\Qcircuit @C=0.65em @R=0.85em {
  & \qw           & \qw            & \ucgrzgate{(\vec\phi)} & \targ     & \qw \\
  & \qw           & \mux \qwx[1]   & \ucgrzctrl             & \ctrl{-1} & \qw \\
  & \qcbundle{n-2} & \gate{P_1,P_2} & \ucgrzctrl             & \qw       & \qw
}
$}}
\overset{(b)}{=}\vcenter{\hbox{\small$
\Qcircuit @C=0.65em @R=0.85em {
  & \qw           & \ucgrzgate{(\vec\phi_\pi)} & \qw            & \targ     & \qw \\
  & \qw           & \ucgrzctrl                 & \mux \qwx[1]   & \ctrl{-1} & \qw \\
  & \qcbundle{n-2} & \ucgrzctrl                 & \gate{P_1,P_2} & \qw       & \qw
}
$}}
\\[1.25em]
\overset{(c)}{=}\vcenter{\hbox{\small$
\Qcircuit @C=0.65em @R=0.85em {
  & \qw           & \ucgrzgate{(\vec\phi_\pi)} & \targ     & \qw            & \qw \\
  & \qw           & \ucgrzctrl                 & \ctrl{-1} & \mux \qwx[1]   & \qw \\
  & \qcbundle{n-2} & \ucgrzctrl                 & \qw       & \gate{P_1,P_2} & \qw
}
$}}
\overset{(d)}{=}\vcenter{\hbox{\small$
\Qcircuit @C=0.65em @R=0.85em {
  & \qw           & \multigate{2}{\rgate(\vec\phi_\pi)} & \qw            & \qw \\
  & \qw           & \ghost{\rgate(\vec\phi_\pi)}        & \mux \qwx[1]   & \qw \\
  & \qcbundle{n-2} & \ghost{\rgate(\vec\phi_\pi)}        & \gate{P_1,P_2} & \qw
}
$}}
\end{gathered}
$}
\end{equation*}
Here, (a) follows by applying a \cnot{} gate to both sides of Eq.~\eqref{qc:ucg-rz-boundary-cnot} and cancelling the adjacent \cnot{} gates;
(b) applies the commutation rule for $\ucgrz$ in Eq.~\eqref{eq:ucgrz-permutation-propagation};
(c) holds because $P$ preserves the control bit of the boundary \cnot{};
and (d) recombines the gates into $\rgate$ using its definition and \cnot{} cancellation.
\end{proof}

Note that the propagation through $\rgate$ requires the multiplexed form of $P$,
which motivates the restriction in
Definition~\ref{def:PD-equivalence}.
Applying Proposition~\ref{prop:pd-thru-r} to the Block-ZXZ decomposition
allows the multiplexed permutation and diagonal factors from a
$PD$-equivalent implementation to be migrated to the next subunitary,
analogously to $D$-migration in Eq.~\eqref{qc:diagonal-migration}.
In circuit form, for $U=PDU_{PD}$, this gives
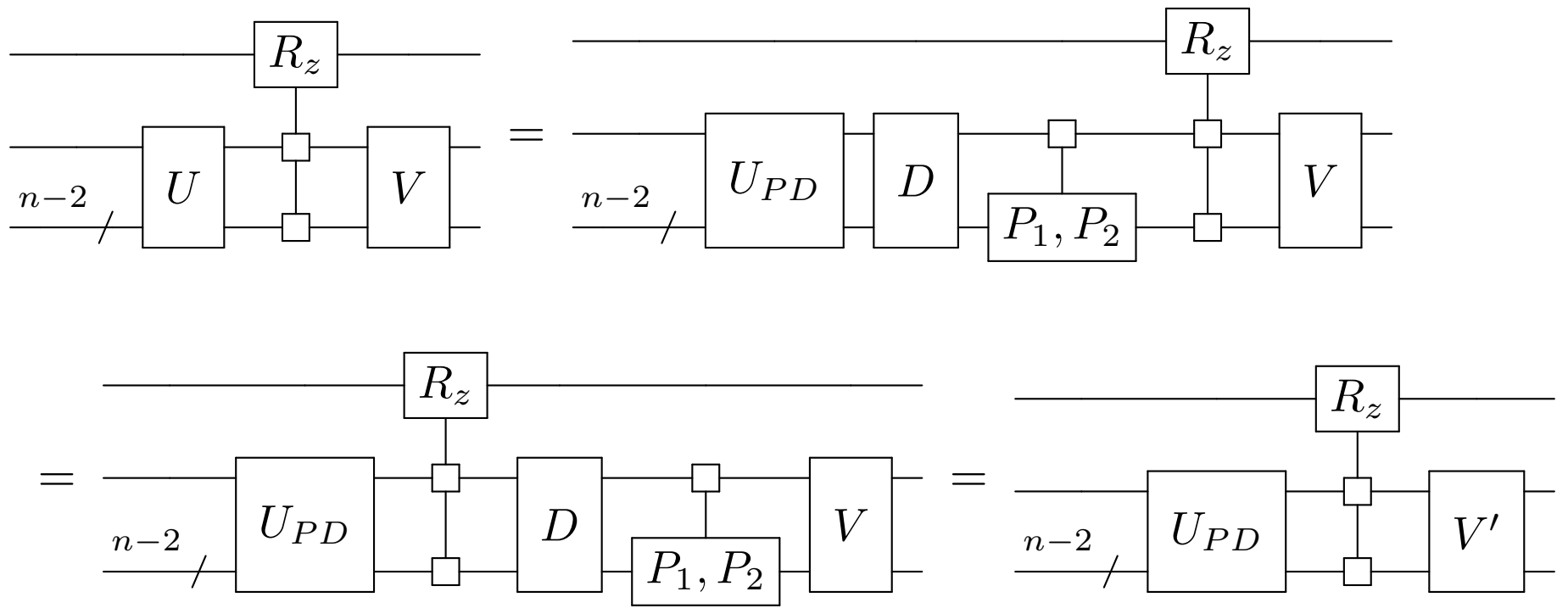
\begin{equation}\label{qc:pd-migration}
\begin{aligned}
&\statementcircuit{
  & \qw           & \qw      & \ucgrzgate{} & \qw      & \qw \\
  & \qw           & \multigate{1}{U} & \ucgrzctrl & \multigate{1}{V} & \qw \\
  & \qcbundle{n-2} & \ghost{U}        & \ucgrzctrl & \ghost{V}        & \qw
}
\;=\;
\statementcircuit{
  & \qw           & \qw           & \qw      & \qw      & \ucgrzgate{} & \qw      & \qw \\
  & \qw           & \multigate{1}{U_{PD}} & \multigate{1}{D} & \mux \qwx[1]  & \ucgrzctrl & \multigate{1}{V} & \qw \\
  & \qcbundle{n-2} & \ghost{U_{PD}}        & \ghost{D}        & \gate{P_1,P_2} & \ucgrzctrl & \ghost{V}        & \qw
}
\\[1.5em]
&\;=\;
\statementcircuit{
  & \qw           & \qw           & \ucgrzgate{} & \qw      & \qw      & \qw      & \qw \\
  & \qw           & \multigate{1}{U_{PD}} & \ucgrzctrl & \multigate{1}{D} & \mux \qwx[1]  & \multigate{1}{V} & \qw \\
  & \qcbundle{n-2} & \ghost{U_{PD}}        & \ucgrzctrl & \ghost{D}        & \gate{P_1,P_2} & \ghost{V}        & \qw
}
\;=\;
\statementcircuit{
  & \qw           & \qw           & \ucgrzgate{} & \qw       & \qw \\
  & \qw           & \multigate{1}{U_{PD}} & \ucgrzctrl & \multigate{1}{V'} & \qw \\
  & \qcbundle{n-2} & \ghost{U_{PD}}        & \ucgrzctrl & \ghost{V'}        & \qw
}
\end{aligned}
\end{equation}
where $V'=VPD$ can again be treated as a generic $(n-1)$-qubit unitary.
The same identity holds with $\rgate$ in place of $\ucgrz$.
Applying this identity successively to the circuit in Eq.~\eqref{qc:block-zxz-cx-absorbed} gives the following circuit identity:
\begin{theorem}\label{thm:pd-migrated-n-q-circuit}
  For an $n$-qubit unitary $U$ with $n\geq 3$, assume that $PD$-equivalent implementations
  of the $(n-1)$-qubit subunitaries are available. Then the following
  circuit identity holds:
  \begin{equation*}
  \statementcircuit{
    & \qw           & \multigate{1}{U} & \qw & \ar@{}[d]|{=} & & \qw           & \qw           & \multigate{1}{\rgate} & \gate{H}      & \ucgrzgate{} & \gate{H}      & \multigate{1}{\rgate} & \qw      & \qw \\
    & \qcbundle{n-1} & \ghost{U}        & \qw &               & & \qcbundle{n-1} & \gate{L_{PD}} & \ghost{\rgate}         & \gate{M_{PD}} & \ucgrzctrl                  & \gate{N_{PD}} & \ghost{\rgate}         & \gate{O} & \qw
}
  \end{equation*}
  which gives the following cost relation:
  \begin{equation} \label{eq:cost:U_by_PD_migrate}
      \costU (n) = 2\costR(n) + \costucgrz(n) + 3\costUPD(n-1) + \costU(n-1).
  \end{equation}
\end{theorem}

As in the case of $D$-migration in Proposition~\ref{prop:diagonal-migrated-synthesis},
we will show that each $PD$-equivalent quantum circuit implementation $L_{PD}, M_{PD}, N_{PD}$ has a lower implementation cost
than its exact counterpart or the $D$-equivalent implementation in~\eqref{qc:diagonal-migrated-modulus} (for $n\geq 4$),
yielding further cost reductions relative to the state of the art.

Note that a $D$-equivalent implementation is still obtained by the naive procedure of recursively replacing $O$ with $O_D$
and discarding the diagonal part until a 2-qubit unitary is reached, yielding the cost relation $\costUD(n) = \costU(n) - 1$.

\subsection{Quantum Multiplicative Inverse Eigenvalue Problem (QMIEP)}

It remains to show how to obtain a lower-cost $PD$-equivalent implementation.
To this end, we propose a method based on manipulating the spectrum of a
certain unitary matrix.

For general complex matrices, this task is known as the \textit{multiplicative
inverse eigenvalue problem (MIEP)}~\cite{HADELER196965, FRIEDLAND1975127}.
Given a complex matrix $A\in\mathbb{C}^{k\times k}$ and a target spectrum
$\mathcal{D}=\{e_1,\ldots,e_k\}$, the problem asks
whether there exists a complex diagonal matrix $\diagonal$ such that
$\spec(A\diagonal)=\mathcal{D}$.
It is known that if all principal minors of $A$ are nonzero, such a diagonal multiplier
$\diagonal$ exists for any target spectrum $\mathcal{D}$~\cite{FRIEDLAND1975127}.

Since the quantum circuits considered here consist of unitary operations, the
natural quantum setting requires the input matrix $A$ and the diagonal modifier
$\diagonal$ to be unitary and all target eigenvalues to have unit modulus.
We refer to MIEPs
subject to these constraints as \emph{quantum multiplicative inverse eigenvalue
problems} (QMIEPs). However, universal solvability does not extend to this setting.
For example, let $A$ be a non-diagonal unitary matrix
(whether or not all its principal minors are nonzero) and choose the target spectrum $\spec(I)$. If a
diagonal unitary $\diagonal$ satisfied
$\spec(\diagonal A)=\spec(I)$, unitarity would imply $\diagonal A=I$ and hence
$\diagonal=A^\dagger$, contradicting diagonality.

Thus, multiplication by a
diagonal unitary alone is insufficient.
To address this limitation, we consider a variant of the MIEP with two
modifications. First, we allow left multiplication by $\diagonal P$, where
$\diagonal$ is a diagonal unitary and $P$ is a permutation matrix. Second, we require the
resulting spectrum to have a specified structure. For the synthesis
considered here, we require this spectrum to be \textit{antipodal}.
The resulting problem is stated as follows.

\begin{problem}[QMIEP for an antipodal spectrum]
    Given $U \in U(2^{n})$, find a diagonal unitary matrix $\diagonal$ and a
    $2^{n}$-dimensional permutation matrix $P$\footnote{The unrestricted $P$ here is not to be confused with the multiplexed $P$ in Definition~\ref{def:PD-equivalence}.} such that
    $\diagonal P U$ has an \emph{antipodal} spectrum, meaning that each eigenvalue
    $\lambda$ and its negative $-\lambda$ occur with the same multiplicity.
    Equivalently, the spectrum can be enumerated as
    \begin{equation*}
        \spec(\diagonal P U) =
        \{
            e^{i\theta_{1}},\ldots,e^{i\theta_{2^{n-1}}},\;
            -e^{i\theta_{1}},\ldots,-e^{i\theta_{2^{n-1}}}
        \},
    \end{equation*}
    for some angles $\theta_{1},\ldots,\theta_{2^{n-1}} \in \mathbb{R}$.
\end{problem}

Throughout this paper, QMIEP refers to the antipodal-spectrum problem stated above.
Other forms of spectral simplification are left for future work; see
Section~\ref{sec:conclusion} for potential directions.
The problem is stated in terms of left multiplication by $DP$.
Solving it for $U^\dagger$ and taking the conjugate transpose gives
the corresponding right-multiplication formulation with $UP^\dagger D^\dagger$.
We therefore make no explicit distinction between left and right multiplication.

An antipodal spectrum is of particular interest because it is separable.
Given a QMIEP solution $D,P$, an appropriate ordering of the eigenvectors yields
the spectral decomposition $DPU=V(D'\otimes Z)V^\dagger$,
where $D'$ is a $2^{n-1}$-dimensional diagonal unitary,
$Z=\operatorname{diag}(1,-1)$ is the 1-qubit Pauli-$Z$ gate, and $V$ is unitary.
As we will show, this enables $PD$-equivalent implementations with fewer \cnot{} gates.

In Section~\ref{sec:theory}, we prove the following solvability results for
the QMIEP with an antipodal spectrum.

\begin{claim} \label{claim:two-qubit-case}
For every $U\in U(2^2)$, the QMIEP is solvable:
there exist $D,P$ such that $\spec(DPU)$ is antipodal.
\end{claim}
\begin{claim} \label{claim:general-case}
For each $n\geq 3$, the QMIEP is solvable on a set of non-zero Haar measure
in $U(2^n)$: each $U$ in the set admits $D,P$ such that $\spec(DPU)$ is antipodal.
\end{claim}
Claim~\ref{claim:general-case} provides only a weak guarantee of solvability
for $n$-qubit unitaries with $n>2$.
However, strong numerical and empirical evidence supports universal solvability,
motivating the conjecture that the QMIEP is solvable for all unitaries
and its use in unitary synthesis, as described below.
Proofs, solution methods, and further analysis of the problem are given in Section~\ref{sec:theory}.

\subsection{Obtaining a \texorpdfstring{$P\diagonal$}{PD}-equivalent implementation through QMIEP} \label{sec:implement_pd_modulus}

In this section, we show that using a QMIEP solution yields a $PD$-equivalent implementation
with fewer \cnot{} gates than the corresponding exact implementation.
Our starting point is that a QMIEP solution reduces the cost of
implementing controlled unitaries (up to $P,D$).

\begin{proposition} \label{prop:controlled_A_by_QMIEP}
  Suppose the QMIEP is solvable for $U\in U(2^{n-1})$.
  Then the following circuit identity holds:

  \begin{equation*}
  \statementcircuit{
    & \qw           & \ctrl{1}         & \qw \\
    & \qcbundle{n-2} & \multigate{1}{U} & \qw \\
    & \qw           & \ghost{U}        & \qw
}
  \quad = \quad
  \statementcircuit{
    & \qw           & \qw              & \ctrl{1}        & \ctrl{2} & \qw                         & \ctrl{1}          & \ctrl{1}          & \qw \\
    & \qcbundle{n-2}  & \multigate{1}{V} & \gate{D'} & \qw      & \multigate{1}{V^{\dagger}} & \multigate{1}{P} & \multigate{1}{D} & \qw \\
    & \qw           & \ghost{V}        & \qw             & \control\qw & \ghost{V^{\dagger}}         & \ghost{P}        & \ghost{D}        & \qw
}
  \end{equation*}
\end{proposition}
\begin{proof}

Let $(D_U,P_U)$ be a QMIEP solution for $U$, so that $W=D_U P_U U$ has an antipodal spectrum, 
giving the following identity:
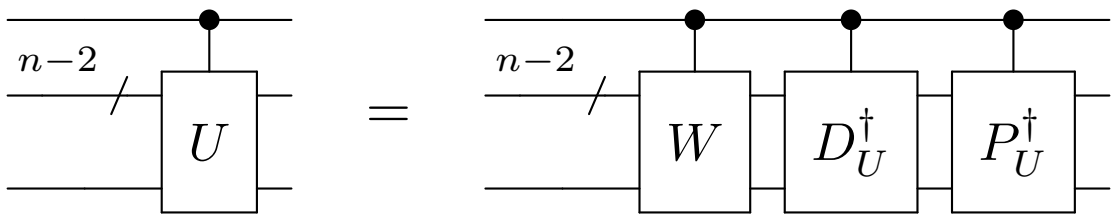
\begin{equation*}
\vcenter{\hbox{\small$
\Qcircuit @C=0.65em @R=0.85em {
  & \qw           & \ctrl{1}         & \qw \\
  & \qcbundle{n-2}  & \multigate{1}{U} & \qw \\
  & \qw           & \ghost{U}        & \qw
}
$}}
\quad = \quad\vcenter{\hbox{\small$
\Qcircuit @C=0.65em @R=0.85em {
  & \qw           & \ctrl{1}              & \ctrl{1}                   & \ctrl{1}                   & \qw \\
  & \qcbundle{n-2}  & \multigate{1}{W}      & \multigate{1}{D_U^\dagger} & \multigate{1}{P_U^\dagger} & \qw \\
  & \qw           & \ghost{W}             & \ghost{D_U^\dagger}        & \ghost{P_U^\dagger}        & \qw
}
$}}
\end{equation*}

The antipodal pairing gives the spectral decomposition $W=V^\dagger(D'\otimes Z)V$.
As noted in Eq.~\eqref{eq:controlled-unitary}, its controlled version
$CW$ can be implemented using $(n-1)$-qubit unitaries and a
controlled-$(D'\otimes Z)$ gate, which consists of a controlled-$D'$ gate
on the $(n-2)$-qubit register and a $CZ$ gate.
Setting $P=P_U^\dagger$ and $D=P_U^\dagger D_U^\dagger P_U$ gives $DP=P_U^\dagger D_U^\dagger$, as required.

\end{proof}

Applying Proposition~\ref{prop:controlled_A_by_QMIEP} yields the following
construction of a $PD$-equivalent implementation of a unitary recursively.

\begin{theorem}~\label{thm:pd_modulus_from_qmiep}
Let $U$ be an $n$-qubit unitary with $n\geq 3$.
Assuming that the QMIEP is solvable for $(n-1)$-qubit unitaries,
the following circuit identity holds:
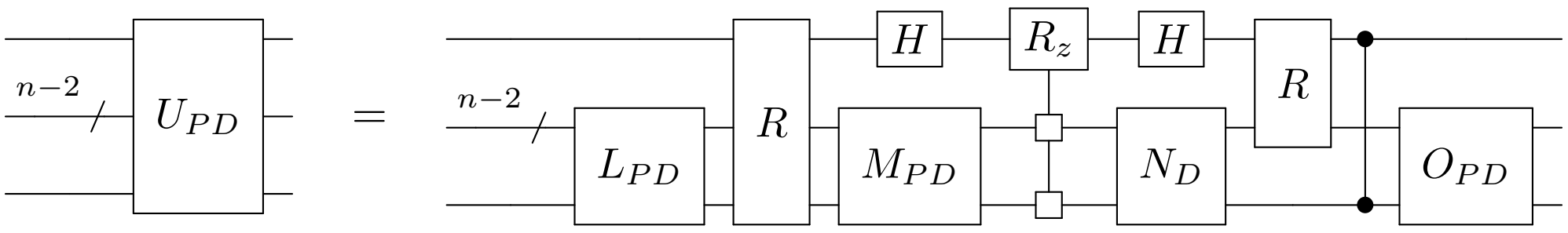
\begin{equation}\label{qc:pd-modulus-from-qmiep}
  \statementcircuit{
    & \qw          & \multigate{2}{U_{PD}} & \qw \\
    & \qcbundle{n-2} & \ghost{U_{PD}}        & \qw \\
    & \qw          & \ghost{U_{PD}}        & \qw
}
  \quad = \quad
  \statementcircuit{
    & \qw          & \qw                   & \multigate{2}{\rgate} & \gate{H}              & \ucgrzgate{} & \gate{H}             & \multigate{1}{\rgate} & \ctrl{2}    & \qw                   & \qw \\
    & \qcbundle{n-2} & \multigate{1}{L_{PD}} & \ghost{\rgate}        & \multigate{1}{M_{PD}} & \ucgrzctrl   & \multigate{1}{N_{D}} & \ghost{\rgate}        & \qw         & \multigate{1}{O_{PD}} & \qw \\
    & \qw          & \ghost{L_{PD}}        & \ghost{\rgate}        & \ghost{M_{PD}}        & \ucgrzctrl   & \ghost{N_{D}}        & \qw                    & \control\qw & \ghost{O_{PD}}        & \qw
}
\end{equation}
which gives the following cost relation:
\begin{equation}\label{eq:cost:UPD_by_QMIEP}
\begin{aligned}
  \costUPD(n)
  &= \costR(n)+\costR(n-1)+\costucgrz(n)\\
  &\quad +3\costUPD(n-1)+\costUD(n-1)+1.
\end{aligned}
\end{equation}
\end{theorem}

\begin{proof}
\begingroup
\footnotesize
\patchcmd{\multigate}{<1em,.9em>}{<0.25em,0.65em>}{}{}
\patchcmd{\ghost}{<1em,.9em>}{<0.25em,0.65em>}{}{}
\patchcmd{\gate}{<.6em>}{<0.35em>}{}{}
\newcommand{\pdproofentry}[2]{%
  \makebox[0.57\linewidth][l]{$\displaystyle #1$}\quad
  \parbox[c]{0.30\linewidth}{\raggedright #2}%
}
\allowdisplaybreaks[1]
\setlength{\jot}{5pt}
\begin{subequations}\label{qc:pd-modulus-derivation}
\begin{align}
& \pdproofentry{
\vcenter{\hbox{$
\Qcircuit @C=0.40em @R=0.85em {
   & \qw & \multigate{3}{U} & \qw \\
   & \qw & \ghost{U} & \qw \\
   & \qcbundle{n-3} & \ghost{U} & \qw \\
   & \qw & \ghost{U} & \qw
}
$}}
\quad =\quad \vcenter{\hbox{$
\Qcircuit @C=0.40em @R=0.85em {
   & \qw & \ctrl{1} & \gate{H} & \mux\qwx[1] & \gate{H} & \ctrl{1} & \qw \\
   & \qw & \multigate{2}{C} & \qw & \multigate{2}{B_1,B_2} & \qw & \multigate{2}{A} & \qw \\
   & \qcbundle{n-3} & \ghost{C} & \qw & \ghost{B_1,B_2} & \qw & \ghost{A} & \qw \\
   & \qw & \ghost{C} & \qw & \ghost{B_1,B_2} & \qw & \ghost{A} & \qw
}
$}}
}{Apply the Block-ZXZ decomposition as in Eq.~\eqref{qc:block-zxz-ucg:block}.}
\label{qc:pd-modulus-derivation:a} \\
& \pdproofentry{
=\quad \vcenter{\hbox{$
\Qcircuit @C=0.40em @R=0.85em {
   & \qw & \ctrl{1} & \gate{H} & \mux\qwx[1] & \gate{H} & \qw & \ctrl{1} & \ctrl{3} & \qw & \ctrl{1} & \ctrl{1} & \qw \\
   & \qw & \multigate{2}{C} & \qw & \multigate{2}{B_1,B_2} & \qw & \multigate{2}{V} & \multigate{1}{D'} & \qw & \multigate{2}{V^\dagger} & \multigate{2}{P} & \multigate{2}{D} & \qw \\
   & \qcbundle{n-3} & \ghost{C} & \qw & \ghost{B_1,B_2} & \qw & \ghost{V} & \ghost{D'} & \qw & \ghost{V^\dagger} & \ghost{P} & \ghost{D} & \qw \\
   & \qw & \ghost{C} & \qw & \ghost{B_1,B_2} & \qw & \ghost{V} & \qw & \control\qw & \ghost{V^\dagger} & \ghost{P} & \ghost{D} & \qw
}
$}}
}{Apply Proposition~\ref{prop:controlled_A_by_QMIEP} to the controlled-$A$ gate, which involves solving the QMIEP for the $(n-1)$-qubit unitary $A$.}
\label{qc:pd-modulus-derivation:b} \\
& \pdproofentry{
=\quad \vcenter{\hbox{$
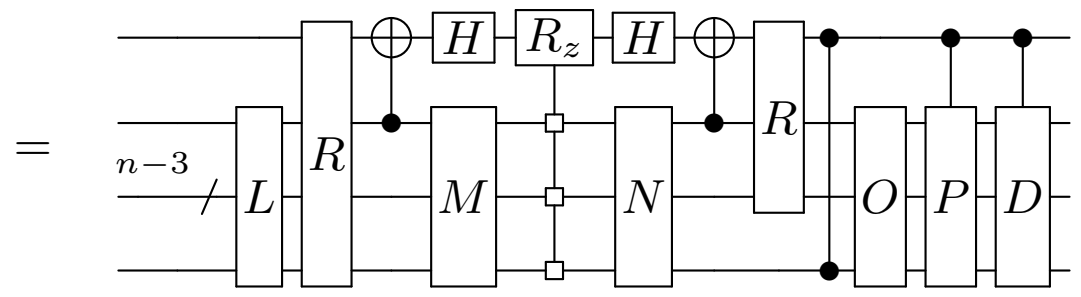
\Qcircuit @C=0.40em @R=0.85em {
   & \qw & \qw & \multigate{3}{\rgate} & \targ & \gate{H} & \ucgrzgate{} & \gate{H} & \targ & \multigate{2}{\rgate} & \ctrl{3} & \qw & \ctrl{1} & \ctrl{1} & \qw \\
   & \qw & \multigate{2}{L} & \ghost{\rgate} & \ctrl{-1} & \multigate{2}{M} & \ucgrzctrl & \multigate{2}{N} & \ctrl{-1} & \ghost{\rgate} & \qw & \multigate{2}{O} & \multigate{2}{P} & \multigate{2}{D} & \qw \\
   & \qcbundle{n-3} & \ghost{L} & \ghost{\rgate} & \qw & \ghost{M} & \ucgrzctrl & \ghost{N} & \qw & \ghost{\rgate} & \qw & \ghost{O} & \ghost{P} & \ghost{D} & \qw \\
   & \qw & \ghost{L} & \ghost{\rgate} & \qw & \ghost{M} & \ucgrzctrl & \ghost{N} & \qw & \qw & \control\qw & \ghost{O} & \ghost{P} & \ghost{D} & \qw
}
$}}
}{Decompose the controlled-$C$ gate and the $B_1\oplus B_2$ multiplexor using
Eqs.~\eqref{eq:controlled-unitary} and~\eqref{eq:multiplexor}, respectively.
Expand the controlled diagonal gates using
Propositions~\ref{prop:shende_theorem_7} and~\ref{prop:ucgrz},
and absorb adjacent $(n-1)$-qubit unitaries.
}
\label{qc:pd-modulus-derivation:c} \\
& \pdproofentry{
=\quad \vcenter{\hbox{$
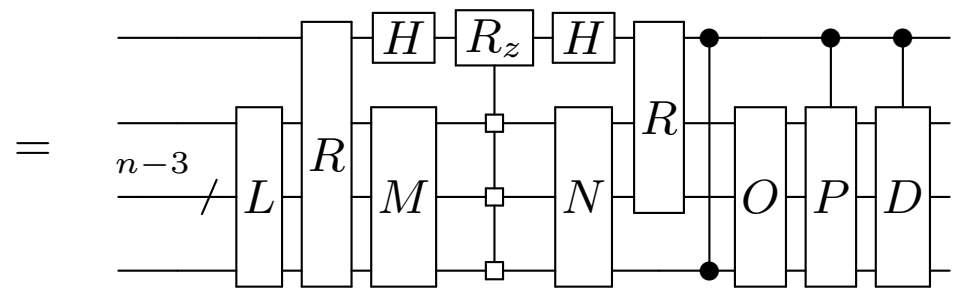
\Qcircuit @C=0.40em @R=0.85em {
   & \qw & \qw & \multigate{3}{\rgate} & \gate{H} & \ucgrzgate{} & \gate{H} & \multigate{2}{\rgate} & \ctrl{3} & \qw & \ctrl{1} & \ctrl{1} & \qw \\
   & \qw & \multigate{2}{L} & \ghost{\rgate} & \multigate{2}{M} & \ucgrzctrl & \multigate{2}{N} & \ghost{\rgate} & \qw & \multigate{2}{O} & \multigate{2}{P} & \multigate{2}{D} & \qw \\
   & \qcbundle{n-3} & \ghost{L} & \ghost{\rgate} & \ghost{M} & \ucgrzctrl & \ghost{N} & \ghost{\rgate} & \qw & \ghost{O} & \ghost{P} & \ghost{D} & \qw \\
   & \qw & \ghost{L} & \ghost{\rgate} & \ghost{M} & \ucgrzctrl & \ghost{N} & \qw & \control\qw & \ghost{O} & \ghost{P} & \ghost{D} & \qw
}
$}}
}{Absorb the boundary \cnot{} gates as in Eq.~\eqref{qc:block-zxz-cx-absorbed} and relabel the blocks $M$ and~$N$.}
\label{qc:pd-modulus-derivation:d} \\
& \pdproofentry{
=\quad \vcenter{\hbox{$
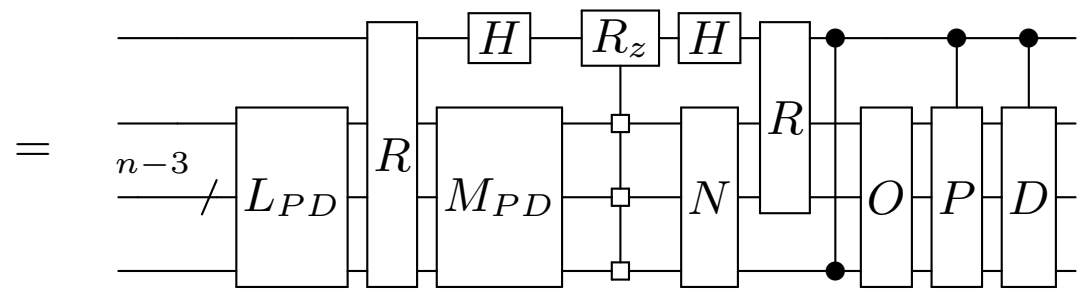
\Qcircuit @C=0.40em @R=0.85em {
   & \qw & \qw & \multigate{3}{\rgate} & \gate{H} & \ucgrzgate{} & \gate{H} & \multigate{2}{\rgate} & \ctrl{3} & \qw & \ctrl{1} & \ctrl{1} & \qw \\
   & \qw & \multigate{2}{L_{PD}} & \ghost{\rgate} & \multigate{2}{M_{PD}} & \ucgrzctrl & \multigate{2}{N} & \ghost{\rgate} & \qw & \multigate{2}{O} & \multigate{2}{P} & \multigate{2}{D} & \qw \\
   & \qcbundle{n-3} & \ghost{L_{PD}} & \ghost{\rgate} & \ghost{M_{PD}} & \ucgrzctrl & \ghost{N} & \ghost{\rgate} & \qw & \ghost{O} & \ghost{P} & \ghost{D} & \qw \\
   & \qw & \ghost{L_{PD}} & \ghost{\rgate} & \ghost{M_{PD}} & \ucgrzctrl & \ghost{N} & \qw & \control\qw & \ghost{O} & \ghost{P} & \ghost{D} & \qw
}
$}}
}{Apply $PD$-migration along $L\to M\to N$ as in Eq.~\eqref{qc:pd-migration} and relabel the updated block $N$.}
\label{qc:pd-modulus-derivation:e} \\
& \pdproofentry{
=\quad \vcenter{\hbox{$
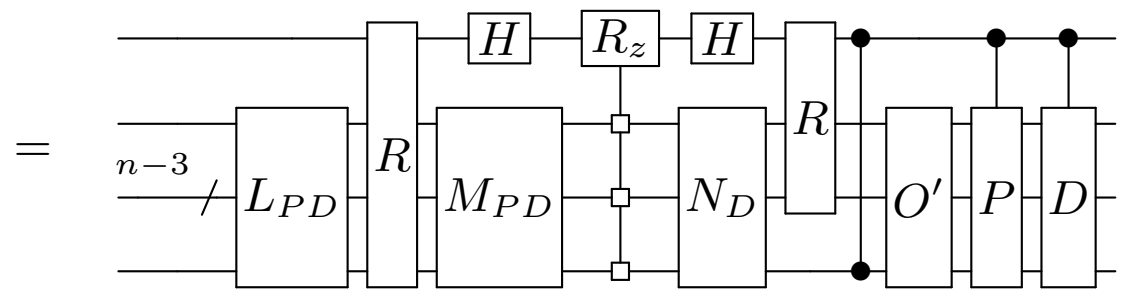
\Qcircuit @C=0.40em @R=0.85em {
   & \qw & \qw & \multigate{3}{\rgate} & \gate{H} & \ucgrzgate{} & \gate{H} & \multigate{2}{\rgate} & \ctrl{3} & \qw & \ctrl{1} & \ctrl{1} & \qw \\
   & \qw & \multigate{2}{L_{PD}} & \ghost{\rgate} & \multigate{2}{M_{PD}} & \ucgrzctrl & \multigate{2}{N_D} & \ghost{\rgate} & \qw & \multigate{2}{O'} & \multigate{2}{P} & \multigate{2}{D} & \qw \\
   & \qcbundle{n-3} & \ghost{L_{PD}} & \ghost{\rgate} & \ghost{M_{PD}} & \ucgrzctrl & \ghost{N_D} & \ghost{\rgate} & \qw & \ghost{O'} & \ghost{P} & \ghost{D} & \qw \\
   & \qw & \ghost{L_{PD}} & \ghost{\rgate} & \ghost{M_{PD}} & \ucgrzctrl & \ghost{N_D} & \qw & \control\qw & \ghost{O'} & \ghost{P} & \ghost{D} & \qw
}
$}}
}{Write $N=D_N N_D$ and commute $D_N$ through the $(n-1)$-qubit $\rgate$ block and the $CZ$ gate to obtain $O'=O D_N$.}
\label{qc:pd-modulus-derivation:f} \\
& \pdproofentry{
=\quad \vcenter{\hbox{$
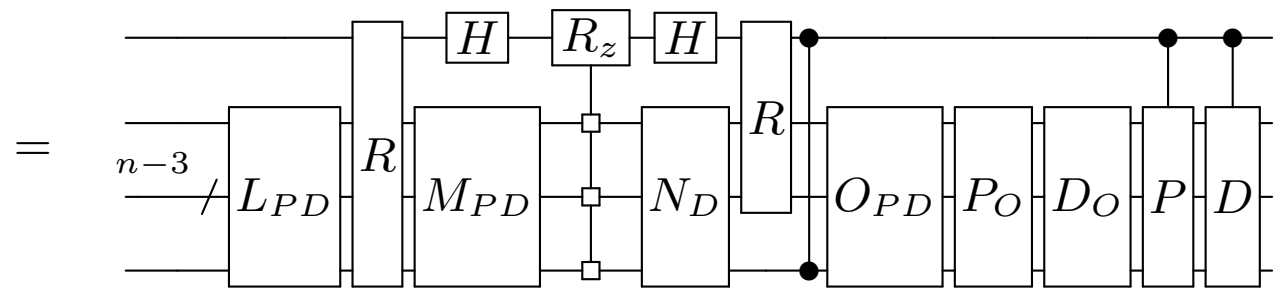
\Qcircuit @C=0.25em @R=0.85em {
   & \qw & \qw & \multigate{3}{\rgate} & \gate{H} & \ucgrzgate{} & \gate{H} & \multigate{2}{\rgate} & \ctrl{3} & \qw & \qw & \qw & \ctrl{1} & \ctrl{1} & \qw \\
   & \qw & \multigate{2}{L_{PD}} & \ghost{\rgate} & \multigate{2}{M_{PD}} & \ucgrzctrl & \multigate{2}{N_D} & \ghost{\rgate} & \qw & \multigate{2}{O_{PD}} & \multigate{2}{P_O} & \multigate{2}{D_O} & \multigate{2}{P} & \multigate{2}{D} & \qw \\
   & \qcbundle{n-3} & \ghost{L_{PD}} & \ghost{\rgate} & \ghost{M_{PD}} & \ucgrzctrl & \ghost{N_D} & \ghost{\rgate} & \qw & \ghost{O_{PD}} & \ghost{P_O} & \ghost{D_O} & \ghost{P} & \ghost{D} & \qw \\
   & \qw & \ghost{L_{PD}} & \ghost{\rgate} & \ghost{M_{PD}} & \ucgrzctrl & \ghost{N_D} & \qw & \control\qw & \ghost{O_{PD}} & \ghost{P_O} & \ghost{D_O} & \ghost{P} & \ghost{D} & \qw
}
$}}
}{Factor $O'=D_O P_O O_{PD}$.}
\label{qc:pd-modulus-derivation:g} \\
& \pdproofentry{
=\quad \vcenter{\hbox{$
\Qcircuit @C=0.25em @R=0.85em {
   & \qw & \qw & \multigate{3}{\rgate} & \gate{H} & \ucgrzgate{} & \gate{H} & \multigate{2}{\rgate} & \ctrl{3} & \qw & \mux\qwx[1] & \mux\qwx[1] & \qw \\
   & \qw & \multigate{2}{L_{PD}} & \ghost{\rgate} & \multigate{2}{M_{PD}} & \ucgrzctrl & \multigate{2}{N_D} & \ghost{\rgate} & \qw & \multigate{2}{O_{PD}} & \multigate{2}{P_1,P_2} & \multigate{2}{D_1,D_2} & \qw \\
   & \qcbundle{n-3} & \ghost{L_{PD}} & \ghost{\rgate} & \ghost{M_{PD}} & \ucgrzctrl & \ghost{N_D} & \ghost{\rgate} & \qw & \ghost{O_{PD}} & \ghost{P_1,P_2} & \ghost{D_1,D_2} & \qw \\
   & \qw & \ghost{L_{PD}} & \ghost{\rgate} & \ghost{M_{PD}} & \ucgrzctrl & \ghost{N_D} & \qw & \control\qw & \ghost{O_{PD}} & \ghost{P_1,P_2} & \ghost{D_1,D_2} & \qw
}
$}}
}{Absorb the residual factors $P_O$ and $D_O$ into the controlled permutation
and diagonal gates, yielding permutation and diagonal multiplexors.}
\label{qc:pd-modulus-derivation:h}
\end{align}
\end{subequations}
\endgroup
Discarding the rightmost permutation and diagonal multiplexors gives the desired
$PD$-equivalent implementation.
Note that the residual permutation has the multiplexed form required by
Definition~\ref{def:PD-equivalence}.

The cost relation follows directly from the circuit, counting the $CZ$ gate as
one \cnot{}, since $CZ=(I\otimes H)\,\mathrm{CNOT}\,(I\otimes H)$.
\end{proof}

\begin{remark}
Note that $N_D$ in the circuit in Eq.~\eqref{qc:pd-modulus-derivation:f} is essential,
meaning that it cannot in general be replaced by $N_{PD}$ via $PD$-migration through the
$R$ gate which is on the upper $n-1$ qubits and the $CZ$ gate
while preserving the gate sequence (up to a reordering of the parameters).
\end{remark}

\subsection{Two Modes of Synthesis}
In this section, we finalize the proposed synthesis method and derive the
costs presented in Table~\ref{tab:cnot-count-comparison}.
The construction in Theorem~\ref{thm:pd_modulus_from_qmiep}, and hence the synthesis
in Theorem~\ref{thm:pd-migrated-n-q-circuit}, depends on QMIEP solvability.
We distinguish two synthesis modes by the QMIEP instances they require:
\begin{itemize}
    \item {\oursanalytic{}~:}
    This mode relies only on analytic solutions to the 2-qubit QMIEP,
    whose solvability is guaranteed by Claim~\ref{claim:two-qubit-case}.
    \item {\oursheuristic{}~:}
    This mode relies on QMIEP solvers for unitaries of general size.
    Beyond 2 qubits, QMIEP solvability is only partially guaranteed by
    Claim~\ref{claim:general-case}, and universal solvability is conjectured.
    Numerical heuristics are used to find solutions.
\end{itemize}

The two modes offer a trade-off between guaranteed success and \cnot{} count.
\oursanalytic{} formally guarantees success for every input but requires more \cnot{} gates.
\oursheuristic{} requires fewer \cnot{} gates, but its success is empirically verified
rather than formally guaranteed.

Both modes use a 3-qubit $PD$-equivalent implementation as their base case.
For $n=3$, Theorem~\ref{thm:pd_modulus_from_qmiep} requires solving the QMIEP
only for 2-qubit unitaries, for which an analytic solution always exists
by Claim~\ref{claim:two-qubit-case}.
For the 2-qubit subunitaries, we set $P=I$ and take their $PD$-equivalent
implementations to be the corresponding $D$-equivalent implementations, giving
$\costUPD(2)=\costUD(2)=2$.
Substituting these costs into Eq.~\eqref{eq:cost:UPD_by_QMIEP}
yields the following common base case:
\begin{corollary} \label{cor:three-qubit-pd-modulus}
  For every $U\in U(2^3)$, Theorem~\ref{thm:pd_modulus_from_qmiep} gives a
  $PD$-equivalent implementation of the form
  \begin{equation}\label{qc:three-qubit-pd-modulus}
  \statementcircuit{
    & \qw & \multigate{2}{U_{PD}} & \qw \\
    & \qw & \ghost{U_{PD}}        & \qw \\
    & \qw & \ghost{U_{PD}}        & \qw
  }
  \quad = \quad
  \statementcircuit{
    & \qw & \qw               & \multigate{2}{\rgate} & \gate{H}          & \ucgrzgate{} & \gate{H}          & \multigate{1}{\rgate} & \ctrl{2}    & \qw               & \qw \\
    & \qw & \multigate{1}{L_D} & \ghost{\rgate}        & \multigate{1}{M_D} & \ucgrzctrl   & \multigate{1}{N_D} & \ghost{\rgate}        & \qw         & \multigate{1}{O_D} & \qw \\
    & \qw & \ghost{L_D}        & \ghost{\rgate}        & \ghost{M_D}        & \ucgrzctrl   & \ghost{N_D}        & \qw                   & \control\qw & \ghost{O_D}        & \qw
  }
  \end{equation}
  with $\costUPD(3)=17$.
\end{corollary}
This saves one \cnot{} gate compared with $\costUD(3)=18$.
The same 2-qubit costs, together with $\costU(2)=3$, also yield
$\costU(3)=19$ in Theorem~\ref{thm:pd-migrated-n-q-circuit},
trivially matching the state-of-the-art $\costUBlockZXZ$ count.

\subsubsection{Mode 1: \oursanalytic{}}
For $n>3$, we obtain a $PD$-equivalent implementation by replacing the terminal
block $O$ in Theorem~\ref{thm:pd-migrated-n-q-circuit} with $O_{PD}$ and discarding
the residual permutation and diagonal factors.
This follows the same approach as the $D$-equivalent synthesis in
Proposition~\ref{prop:diagonal-migrated-synthesis}.
The resulting four $(n-1)$-qubit $PD$-equivalent blocks are synthesized recursively
using the 3-qubit base case in Corollary~\ref{cor:three-qubit-pd-modulus}.

\begin{proposition} \label{prop:pd-modulus-recursive-naive}
  For $n>3$ and any $U \in U(2^n)$, the following circuit identity gives a recursive
  $PD$-equivalent implementation, with the 3-qubit base case given by
  Corollary~\ref{cor:three-qubit-pd-modulus}:
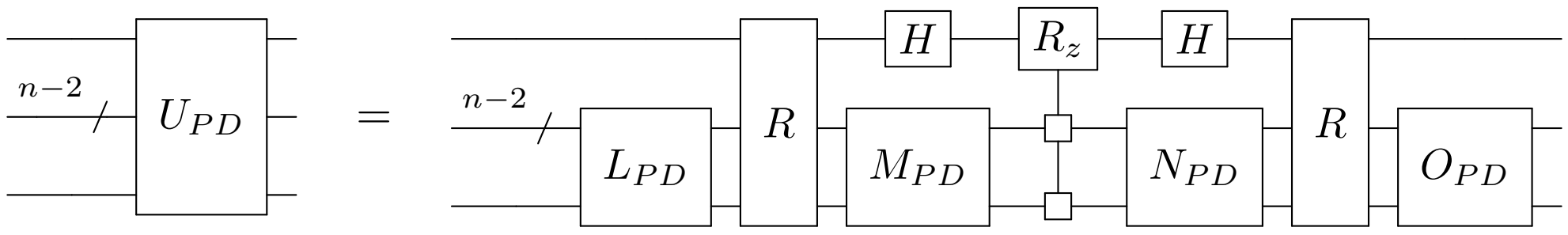
  \begin{equation*}
  \statementcircuit{
    & \qw          & \multigate{2}{U_{PD}} & \qw \\
    & \qcbundle{n-2} & \ghost{U_{PD}}        & \qw \\
    & \qw          & \ghost{U_{PD}}        & \qw
}
  \quad = \quad
  \statementcircuit{
    & \qw          & \qw                   & \multigate{2}{\rgate} & \gate{H}              & \ucgrzgate{} & \gate{H}              & \multigate{2}{\rgate} & \qw                   & \qw \\
    & \qcbundle{n-2} & \multigate{1}{L_{PD}} & \ghost{\rgate}        & \multigate{1}{M_{PD}} & \ucgrzctrl   & \multigate{1}{N_{PD}} & \ghost{\rgate}        & \multigate{1}{O_{PD}} & \qw \\
    & \qw          & \ghost{L_{PD}}        & \ghost{\rgate}        & \ghost{M_{PD}}        & \ucgrzctrl   & \ghost{N_{PD}}        & \ghost{\rgate}        & \ghost{O_{PD}}        & \qw
}
  \end{equation*}
  which gives the following cost relation:
  \begin{equation}\label{eq:cost:UPD_analytic_recursive}
    \costPDanalytic{n}
    =4\costPDanalytic{n-1}+2\costR(n)+\costucgrz(n).
  \end{equation}
\end{proposition}
In this case, the permutation difference acts on the lower 3 qubits
and thus satisfies the multiplexor restriction in Definition~\ref{def:PD-equivalence}.

Combining the preceding results, \oursanalytic{} applies $PD$-migration to the Block-ZXZ decomposition
of an $n$-qubit unitary $U$ ($n\geq 3$), yielding the exact circuit in Theorem~\ref{thm:pd-migrated-n-q-circuit}.
It uses the recursion in Proposition~\ref{prop:pd-modulus-recursive-naive} for the $PD$-equivalent
implementations of $L$, $M$, and $N$ and synthesizes the updated terminal block $O$ exactly.
See Algorithm~\ref{alg:analytic-synthesis} in Appendix~\ref{app:analytic-synthesis} for the full procedure.

Note that the procedure solves the QMIEP only for unitaries in $U(2^2)$, and every such instance admits an analytic solution.
Solving the recurrence (see Appendix~\ref{app:recurrence}) in Eq.~\eqref{eq:cost:UPD_analytic_recursive} and
substituting the result into Eq.~\eqref{eq:cost:U_by_PD_migrate}
yields the following \cnot{} count for \oursanalytic{}.

\begin{corollary}
    For $n\geq3$, every $U\in U(2^n)$ can be implemented
    using arbitrary 1-qubit gates and
    \begin{equation}\label{eq:analytic-unitary-cost}
    \costanalytic{n} = \OurAnalyticCost
    \end{equation}
    \cnot{} gates.
\end{corollary}

\subsubsection{Mode 2: \oursheuristic{}}

\oursheuristic{} also applies $PD$-migration to the Block-ZXZ decomposition as in Theorem~\ref{thm:pd-migrated-n-q-circuit}.
To obtain the $PD$-equivalent implementations required for migration, it applies the construction
in Eqs.~\eqref{qc:pd-modulus-derivation:a}--\eqref{qc:pd-modulus-derivation:h} recursively to each of $L$, $M$, and $N$.
Exact synthesis of an $n$-qubit target requires QMIEP solutions only up to $n-2$ qubits.
See Algorithm~\ref{alg:heuristic-synthesis} in Appendix~\ref{app:heuristic-synthesis} for the full procedure.
This gives the following cost guarantee for \oursheuristic{}.

\begin{corollary}\label{cor:heuristic-unitary-cost}
    For $n\geq3$, a Haar-generic unitary $U\in U(2^n)$ can be implemented,
    with non-zero probability, using arbitrary 1-qubit gates and
    \begin{equation}\label{eq:heuristic-unitary-cost}
    \costheuristic{n} = \OurHeuristicCost
    \end{equation}
    \cnot{} gates.
\end{corollary}

See Appendix~\ref{app:heuristic-unitary-cost-proof} for the proof.

Although the formal guarantees of the corollary are very limited,
in Section~4.2.2 we report empirical evidence supporting QMIEP solvability for all $n$-qubit unitaries.
Thus, the cost bound in Equation~\eqref{eq:heuristic-unitary-cost} is expected to hold universally.

\section{Further Analysis of QMIEP}\label{sec:theory}
In this section, we present further analysis and results on the proposed QMIEP with an antipodal spectrum.
In Section~\ref{sec:5_1:U4}, we provide support for Claim~\ref{claim:two-qubit-case}, which guarantees solvability for all unitaries in the $n=2$ case.
In Section~\ref{sec:general-case}, we provide support for Claim~\ref{claim:general-case} and present numerical methods for finding solutions in the general case $n>2$.
In Section~\ref{sec:further-optimal-d-equivalent}, we present the empirical observation that the QMIEP was solvable with
$P=I$ for most sampled unitaries and explain how this leads to further CNOT reductions.

\subsection{\texorpdfstring{$U(4)$}{U(4)} case} \label{sec:5_1:U4}
In this section, we present a proof of Claim~\ref{claim:two-qubit-case}, along with the necessary technical propositions.

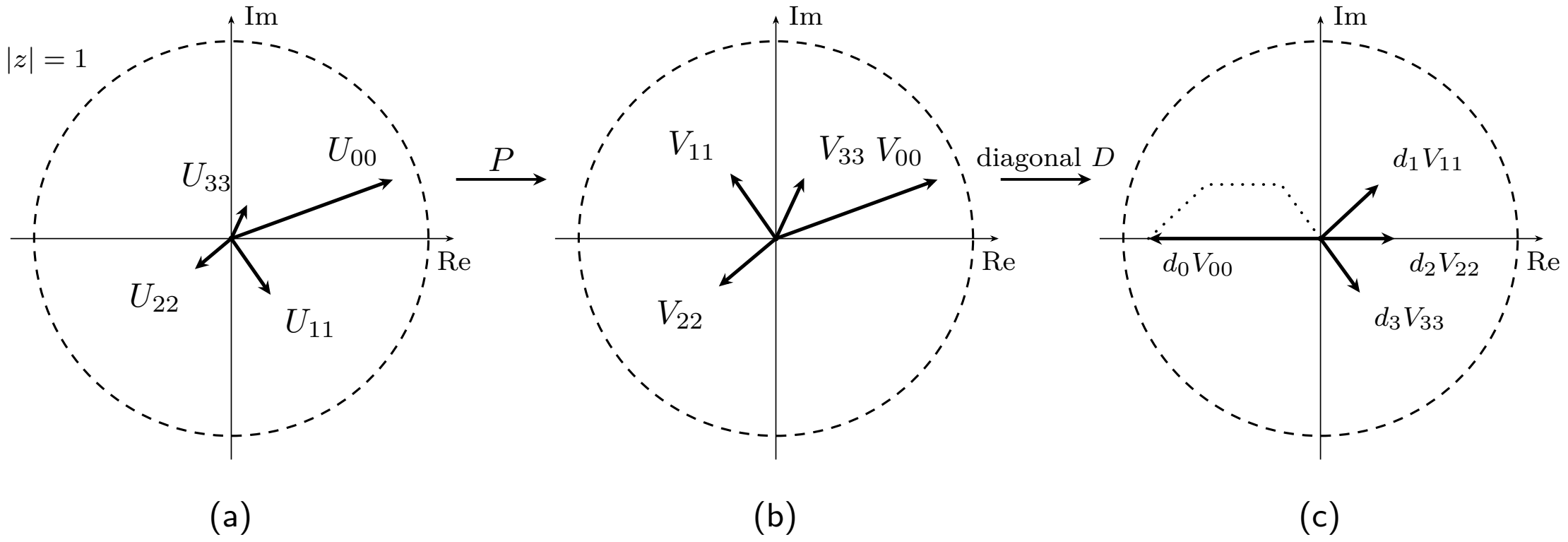
\begin{figure}[t]
    \centering
    \tikzset{
        qmiep panel/.style={x=2.05cm,y=2.05cm,font=\small,text=black},
        qmiep entry/.style={draw=black,solid,line width=1.05pt,
            -{Stealth[length=4pt,width=3.6pt]}},
        qmiep axis/.style={draw=black,solid,line width=.35pt,
            -{Stealth[length=2.5pt]}},
        qmiep disk/.style={draw=black,dash pattern=on 3pt off 2.5pt,
            line width=.65pt},
        qmiep transition/.style={draw=black,solid,line width=.8pt,
            -{Stealth[length=4pt,width=3.6pt]}},
        qmiep closure/.style={draw=black,dash pattern=on 0pt off 3pt,line cap=round,
            line width=.8pt}
    }
    \newcommand{\qmiepunitdisk}{%
        \path[use as bounding box] (-1.3,-1.2) rectangle (1.3,1.25);
        \draw[qmiep axis] (-1.12,0) -- (1.13,0)
            node[below,font=\scriptsize] {$\operatorname{Re}$};
        \draw[qmiep axis] (0,-1.12) -- (0,1.13)
            node[right,font=\scriptsize] {$\operatorname{Im}$};
        \draw[qmiep disk] (0,0) circle[radius=1];
        \fill[black] (0,0) circle[radius=.014];
    }
    \begin{subfigure}[t]{.32\linewidth}
        \centering
        \resizebox{\linewidth}{!}{%
            \begin{tikzpicture}[qmiep panel]
                \qmiepunitdisk
                \node[font=\scriptsize,anchor=south east] at (-.66,.77) {$|z|=1$};
                \draw[qmiep entry] (0,0) -- (0.820151512930,0.298510738295)
                    node[above left] {$U_{00}$};
                \draw[qmiep entry] (0,0) -- (0.199587984253,-0.285041181881)
                    node[below right] {$U_{11}$};
                \draw[qmiep entry] (0,0) -- (-0.184641989159,-0.154933025003)
                    node[below left] {$U_{22}$};
                \draw[qmiep entry] (0,0) -- (0.079829930853,0.171195839177)
                    node[above left] {$U_{33}$};
                \draw[qmiep transition,overlay] (1.14,.30) -- (1.60,.30)
                    node[midway,above,font=\small,inner sep=2pt] {$P$};
            \end{tikzpicture}%
        }
        \caption{
        }
        \label{fig:u4-diagonal-unbalanced}
    \end{subfigure}\hfill
    \begin{subfigure}[t]{.32\linewidth}
        \centering
        \resizebox{\linewidth}{!}{%
            \begin{tikzpicture}[qmiep panel]
                \qmiepunitdisk
                \draw[qmiep entry] (0,0) -- (0.820151512930,0.298510738295)
                    node[above left] {$V_{00}$};
                \draw[qmiep entry] (0,0) -- (-0.231039746661,0.329958953672)
                    node[above left] {$V_{11}$};
                \draw[qmiep entry] (0,0) -- (-0.289408301595,-0.242842399128)
                    node[below left] {$V_{22}$};
                \draw[qmiep entry] (0,0) -- (0.143387039507,0.307494498534)
                    node[above right] {$V_{33}$};
                \draw[qmiep transition,overlay] (1.14,.30) -- (1.60,.30)
                    node[midway,above,font=\scriptsize,inner sep=2pt] {diagonal $D$};
            \end{tikzpicture}%
        }
        \caption{%
        }
        \label{fig:u4-diagonal-balanced}
    \end{subfigure}\hfill
    \begin{subfigure}[t]{.32\linewidth}
        \centering
        \resizebox{\linewidth}{!}{%
            \begin{tikzpicture}[qmiep panel]
                \qmiepunitdisk
                \coordinate (q0) at (0,0);
                \coordinate (q1) at (-0.872787010123,0);
                \coordinate (q2) at (-0.577674798612,0.274155171868);
                \coordinate (q3) at (-0.199879092117,0.274155171868);
                \draw[qmiep closure] (q1) -- (q2) -- (q3) -- (q0);
                \draw[qmiep entry] (0,0) -- (q1)
                    node[below right,font=\scriptsize] {$d_0V_{00}$};
                \draw[qmiep entry] (0,0) -- (0.295112211511,0.274155171868)
                    node[above right,font=\scriptsize] {$d_1V_{11}$};
                \draw[qmiep entry] (0,0) -- (0.377795706495,0)
                    node[below right,font=\scriptsize] {$d_2V_{22}$};
                \draw[qmiep entry] (0,0) -- (0.199879092117,-0.274155171868)
                    node[below right,font=\scriptsize] {$d_3V_{33}$};
            \end{tikzpicture}%
        }
        \caption{%
        }
        \label{fig:u4-diagonal-closure}
    \end{subfigure}
    \caption{%
    Solving the QMIEP for $U\in U(4)$ through permutation and phase adjustments.
    (a) The diagonal entries of $U$ are unbalanced in this example,
    violating Equation~\eqref{eq:u4-diagonal-condition}, so no choice of phases allows the vectors to form a closed polygon (i.e., sum to zero).
    (b) A row permutation yields $V=PU$ with balanced diagonal magnitudes,
    as guaranteed by Proposition~\ref{prop:permutation_u4}.
    (c) A diagonal unitary $D=\operatorname{diag}(d_0,\ldots,d_3)$ adjusts the phases
    so that the vectors $d_iV_{ii}$ form a closed polygon when placed head to tail.
    The dotted segments represent translated copies of these vectors.
    Their zero sum gives $\tr(DPU)=0$,
    so $DPU$ has an antipodal spectrum.
    }
    \label{fig:u4-diagonal-balance}
\end{figure}

\begin{proposition} \label{prop:polygonclosure}
    For $U \in U(4)$, let
    $m \in \argmax_{i\in\{0,1,2,3\}} |U_{ii}|$. Then, 
    \begin{equation}\label{eq:u4-diagonal-condition}
        |U_{mm}| \leq \sum_{i\neq m}|U_{ii}|,
    \end{equation}
    if and only if there exists a diagonal unitary $\diagonal$ such that \(\tr(\diagonal U)=0\).
\end{proposition}

\begin{proof}
    Write the diagonal entries of $U$ as $U_{ii}=a_i e^{i\theta_i}$,
    where $a_i=|U_{ii}|$ and $\theta_i\in\mathbb{R}$ for $i=0,1,2,3$, taking $\theta_i=0$ when $a_i=0$.
    By relabelling the indices, we may assume that $a_0\geq a_1\geq a_2\geq a_3$.

    Suppose the condition~\eqref{eq:u4-diagonal-condition} holds.
    Let $V:=DU$, where
    $D=\operatorname{diag}(d_0,d_1,d_2,d_3)$.
    Set $d_0=e^{i(\pi-\theta_0)}$ and $d_3=e^{-i\theta_3}$.
    It remains to choose $d_1$ and $d_2$ so that
    $V_{11}+V_{22}=a_0-a_3=:r$.

    Write $V_{11}=a_1e^{i\alpha}$ and $V_{22}=a_2e^{-i\beta}$,
    where $\alpha,\beta\in\mathbb{R}$ will be chosen below. The required condition becomes
    $a_1e^{i\alpha}+a_2e^{-i\beta}=r$.
    \begin{itemize}
        \item If $r=0$, all $a_i$ are equal, so $(\alpha,\beta)=(0,\pi)$ suffices.
        \item If $r>0$ and $a_2=0$, then $a_3=0$ and the hypothesis forces $a_0=a_1$,
        so $(\alpha,\beta)=(0,0)$ suffices.
        \item Otherwise, $r,a_1,a_2>0$, and we set
        \begin{align*}
            \alpha &= \arccos\left(\frac{r^2+a_1^2-a_2^2}{2ra_1}\right),
            & \beta &= \arccos\left(\frac{r^2+a_2^2-a_1^2}{2ra_2}\right).
        \end{align*}
        It follows that $r-(a_1-a_2)=(a_0-a_1)+(a_2-a_3)\geq0$,
        and the hypothesis gives $r\leq a_1+a_2$.
        Thus $|a_1-a_2|\leq r\leq a_1+a_2$, so both $\arccos$ arguments lie in $[-1,1]$ and $\alpha,\beta$ are well defined.
    \end{itemize}
    These angles satisfy $a_1\cos\alpha+a_2\cos\beta=r$ and
    $a_1\sin\alpha=a_2\sin\beta$. Choosing $\alpha,\beta$ as above gives $\tr(DU)=0$, as desired.

    Conversely, suppose that $\tr(DU)=0$ for some diagonal unitary $D$, and set $V:=DU$.
    If Equation~\eqref{eq:u4-diagonal-condition} were violated, then $|V_{ii}|=a_i$ and $\tr(V)=0$ would give
    \[
        a_0-a_3
        \leq |V_{00}+V_{33}|
        = |V_{11}+V_{22}|
        \leq a_1+a_2
        < a_0-a_3,
    \]
    a contradiction. Thus, Equation~\eqref{eq:u4-diagonal-condition} holds.
\end{proof}
The proof gives the following analytic expression for the desired diagonal unitary $D$, up to relabelling:
\begin{equation*}
    D=\operatorname{diag}\!\left(
        e^{i(\pi - \theta_0)},\;
        e^{i(\alpha-\theta_1)},\;
        e^{-i(\beta+\theta_2)},\;
        e^{-i\theta_3}
    \right),
\end{equation*}
where $\alpha,\beta$ are chosen as in the proof.

\begin{proposition}\label{prop:trace_cond_u4}
    If $U\in U(4)$ satisfies $\tr(U)=0$, then its spectrum is antipodal.
\end{proposition}
\begin{proof}
    Newton's identities and unitarity give the characteristic polynomial
    $p(t):=\det(tI-U)$ in the form
    (cf.~\cite[Eq.~(24)]{zhang2003geometric})
    \[
        p(t)=t^4-\tr(U)t^3
            +\frac{(\tr(U))^2-\tr(U^2)}{2}t^2
            -\det(U)\overline{\tr(U)}\,t+\det(U).
    \]
    Since $\tr(U)=0$, the polynomial is even: $p(t)=p(-t)$.
    Its roots therefore occur in pairs $\lambda$ and $-\lambda$,
    so the spectrum of $U$ is antipodal.
\end{proof}

Provided that the condition in Equation~\eqref{eq:u4-diagonal-condition} holds, Proposition~\ref{prop:polygonclosure} gives a diagonal unitary $D$ such that $\tr(DU)=0$.
By Proposition~\ref{prop:trace_cond_u4}, $DU$ therefore has an antipodal spectrum.
However, the condition need not hold for every unitary matrix.
The following example supports the necessity of introducing a permutation for the antipodal QMIEP.

\begin{example}[Necessity of a permutation]\label{ex:u4_perm_necc}
    The condition in Equation~\eqref{eq:u4-diagonal-condition} can fail throughout a neighborhood of a certain unitary matrix.
    Consider the permutation matrix
    \[
        V^*=\begin{pmatrix}
            1 & 0 & 0 & 0\\
            0 & 0 & 1 & 0\\
            0 & 0 & 0 & 1\\
            0 & 1 & 0 & 0
        \end{pmatrix}.
    \]
    Since $|V^*_{00}|=1$ and $|V^*_{ii}|=0$ for $i=1,2,3$,
    $V^*$ violates Equation~\eqref{eq:u4-diagonal-condition}.
    However, choosing $P=(V^*)^\dagger$ gives $PV^*=I$, which satisfies the condition.
    The same argument applies to any permutation matrix with exactly one fixed point.

    The same conclusion holds for unitary matrices in a neighborhood of $V^*$.
    Let $V\in U(4)$ satisfy $\|V-V^*\|<\epsilon=1/4$, where $\|\cdot\|$ denotes the operator norm.
    Then $|V_{00}|>1-\epsilon$ and $|V_{ii}|<\epsilon$ for $i=1,2,3$, so
    \[
        |V_{00}|>1-\epsilon=3\epsilon>\sum_{i\neq0}|V_{ii}|.
    \]
    Thus $V$ also violates the condition.
    On the other hand, setting $W:=PV$ gives
    $\|W-I\|=\|P(V-V^*)\|<\epsilon$.
    Hence, for each $i\in\{0,1,2,3\}$,
    \[
        |W_{ii}|<1+\epsilon<3(1-\epsilon)<\sum_{j\neq i}|W_{jj}|,
    \]
    so $W$ satisfies Equation~\eqref{eq:u4-diagonal-condition}.
\end{example}

The following proposition shows that, for every $U\in U(4)$, there exists a permutation matrix $P$ such that $PU$ satisfies Equation~\eqref{eq:u4-diagonal-condition}.
The proof uses the following two facts from matrix analysis.

\begin{fact}[{Jacobi's identity~\cite[Eq. (0.8.4.1)]{horn2013}}]\label{fact:jacobi}
    Let $M$ be the submatrix obtained by deleting the $i$-th row and $j$-th column of a unitary matrix $U$. Then
    $|\det(M)|=|U_{ij}|$.
\end{fact}
\begin{fact}[{\cite[Eq. (0.3.2.1)]{horn2013}}]\label{fact:leibniz}
    The determinant of an $n\times n$ matrix $M$ is given by
    \[
        \det(M)=\sum_{\sigma\in S_n}\operatorname{sgn}(\sigma)\prod_{i=1}^{n}M_{i,\sigma(i)},
    \]
    where $S_n$ is the set of all permutations of $\{1,\ldots,n\}$.
\end{fact}

\begin{proposition} \label{prop:permutation_u4}
    For $U\in U(4)$, there exists a permutation matrix $P$ such that $V=PU$ satisfies
    $|V_{mm}| \leq \sum_{i\neq m}|V_{ii}|$
    for $m\in\argmax_{i\in\{0,1,2,3\}} |V_{ii}|$.
\end{proposition}
\begin{proof}
    Choose $(i_0,j_0)\in\operatorname*{arg\,max}_{i,j}|U_{ij}|$ and set
    $a:=|U_{i_0j_0}|$, so $0<a\leq1$.
    Let $I:=\{0,1,2,3\}\setminus\{i_0\}$ and
    $J:=\{0,1,2,3\}\setminus\{j_0\}$.
    Let $M$ be the $3\times3$ submatrix obtained by deleting the $i_0$-th row and $j_0$-th column of $U$.
    For convenience, we index $M$ using the corresponding row and column indices of $U$;
    thus $M_{ij}=U_{ij}$ for $i\in I$ and $j\in J$.

    For each bijection $\tau:I\to J$, set
    $s_\tau:=\sum_{k\in I}|M_{k,\tau(k)}|$ and
    $p_\tau:=\prod_{k\in I}|M_{k,\tau(k)}|$.
    Suppose, for a contradiction, that $s_\tau<a$ for all six bijections.
    The arithmetic--geometric mean inequality gives
    \begin{equation}\label{eq:arith-geo}
        p_\tau\leq(s_\tau/3)^3<a^3/27.
    \end{equation}
    Using Equation~\eqref{eq:arith-geo}, we obtain
    \[
        a=|\det(M)|\leq\sum_\tau p_\tau
        <\frac{6a^3}{27}\leq\frac{2a}{9}<a,
    \]
    which is a contradiction. Here, the equality follows from Fact~\ref{fact:jacobi},
    and the first inequality follows from Fact~\ref{fact:leibniz} and the triangle inequality.

    Hence some bijection $\tau$ satisfies $s_\tau\geq a$.
    Choosing the permutation matrix
    \begin{equation*}
        P=\ket{j_0}\bra{i_0}+\sum_{k\in I}\ket{\tau(k)}\bra{k}
    \end{equation*}
    gives the desired result.
\end{proof}
The preceding propositions together establish
Claim~\ref{claim:two-qubit-case}.
\begin{proof}[Proof of Claim~\ref{claim:two-qubit-case}]
    Let $U \in U(4)$. By Proposition~\ref{prop:permutation_u4}, there exists a permutation $P$ such that
    $PU$ satisfies the condition of Proposition~\ref{prop:polygonclosure}, which gives a diagonal unitary $D$ with $\tr(DPU)=0$.
    The spectrum of $DPU$ is therefore antipodal by Proposition~\ref{prop:trace_cond_u4}.

\end{proof}
Figure~\ref{fig:u4-diagonal-balance} illustrates the steps of applying a permutation matrix and a diagonal unitary to $U$ to solve the antipodal QMIEP in $U(4)$.

\subsection{General Case}
\label{sec:general-case}

\subsubsection{On Solvability} \label{sec:theory_general_case}
In this subsection, we discuss some solvability results for the problem, including Claim~\ref{claim:general-case}.
The following proposition naturally extends the trace condition in
Proposition~\ref{prop:trace_cond_u4} to the general case $n\geq3$.

\begin{proposition}\label{prop:odd-moment-antipodality}
    Let $U\in U(2^n)$, where $n\geq1$.
    If $\tr(U^k)=0$ for every odd integer $k$ with $1\leq k\leq 2^{n-1}$,
    then the spectrum of $U$ is antipodal.
\end{proposition}
\begin{proof}
    The proof combines the coefficient symmetry of the characteristic polynomial
    with Newton's identities; see Appendix~\ref{app:odd-moment-antipodality}.
\end{proof}

Hence, the problem reduces to finding $P,D$ such that $\tr(W^k)=0$ for $W:=DPU$
and every odd integer $k$ with $1\leq k\leq 2^{n-1}$,
extending the single condition $k=1$ for $n=2$ to a system of moment constraints.
These additional constraints explain why establishing solvability is more difficult in the general case.
Nevertheless, we can establish the following local result:
\begin{proposition}[Local solvability implying Claim~\ref{claim:general-case}]\label{prop:identity-neighborhood-solvable}
    For every $n\geq1$, there exists $\epsilon>0$ such that the QMIEP is solvable
    for every $U\in U(2^n)$ satisfying $\|U-I\|<\epsilon$.
    Consequently, the solvable set has positive Haar measure.
\end{proposition}
\begin{proof}
    The proof applies the implicit function theorem to an antipodality residual near the identity.
    The construction uses a simple spectrum to ensure smooth local eigenphase branches.
    See Appendix~\ref{app:identity-neighborhood} for the full proof.
\end{proof}

Permutations can still be necessary in the general case, as illustrated by extending
Example~\ref{ex:u4_perm_necc} to $1\oplus P_\sigma$,
where $\sigma$ is a cycle of length $2^n-1$ and $P_\sigma$ denotes the corresponding permutation matrix.
Up to a reordering of the basis, every permutation matrix can be written as
$P=\bigoplus_{j=1}^{r}P_{\sigma_j}$, where $\sigma_1,\ldots,\sigma_r$ are its disjoint cycles.
The following proposition characterizes, in terms of these cycle lengths,
when a trace moment cannot be cancelled by diagonal phases.

\begin{proposition}\label{prop:permutation-moment-obstruction}
    Let $P$ be a $2^n\times 2^n$ permutation matrix with cycle lengths
    $\ell_1,\ldots,\ell_r$.
    For an integer $k\geq1$, let $J_k:=\{j\in\{1,\ldots,r\}:\ell_j\mid k\}$
    be the set of indices of cycles whose lengths divide $k$.
    Then $\tr((DP)^k)\neq0$ for every diagonal unitary $D$
    if and only if there exists $j_*\in J_k$ such that
    \begin{equation} \label{eq:general_perm_condi}
        \ell_{j_*}>
        \sum_{j\in J_k\setminus\{j_*\}}\ell_j.
    \end{equation}
\end{proposition}
\begin{proof}
    The proof expresses the trace in terms of the disjoint cycles and applies
    the polygon closure condition; see Appendix~\ref{app:permutation-moment-obstruction}.
\end{proof}

As in Example~\ref{ex:u4_perm_necc}, consider a permutation matrix satisfying Equation~\eqref{eq:general_perm_condi} for some odd $k$.
By continuity, there exists an $\epsilon$-ball around this matrix in $U(2^n)$ consisting entirely of unitaries for which the QMIEP is unsolvable with $P=I$.
This ball has positive Haar measure, supporting the necessity of introducing permutations into the problem.

\begin{proposition}\label{prop:permutation-neighborhood-obstruction}
    Let $k\geq1$ be an integer, and let $P$ be a $2^n\times 2^n$ permutation matrix satisfying Equation~\eqref{eq:general_perm_condi}.
    Then there exists $\epsilon>0$ such that, for every $U\in U(2^n)$ with $\|U-P\|<\epsilon$,
    $\tr((DU)^k)\neq0$ for every diagonal unitary $D$.
\end{proposition}
\begin{proof}
    For fixed $k$, let $f(U):=\min_D |\tr((DU)^k)|$, where $D$ ranges over all diagonal unitaries.
    By continuity of the trace and compactness of the set of diagonal unitaries, the minimum is well defined and $f$ is continuous.
    Proposition~\ref{prop:permutation-moment-obstruction} gives $f(P)>0$, so $f$ remains positive and nonzero in some neighborhood of $P$.
\end{proof}

The solvability guarantees established in this paper are nevertheless limited.
For each permutation matrix $P$, left multiplication by $P^\dagger$ maps a neighborhood of $P$ to a neighborhood of $I$, since $\|P^\dagger U-I\|=\|U-P\|$.
The construction underlying Proposition~\ref{prop:identity-neighborhood-solvable} therefore yields, for every $U$ sufficiently close to $P$, a diagonal unitary $D$ such that $DP^\dagger U$ has an antipodal spectrum.
Thus, each permutation matrix gives a solvable neighborhood of positive Haar measure, but even their union appears to cover only a very small portion of $U(2^n)$.

Nevertheless, our numerical solver succeeded on every tested unitary (Section~\ref{sec:numerical_heuristic}), providing empirical support for universal solvability and for the practical utility of our approach.
Universal solvability therefore remains a conjecture, whose resolution is a natural next challenge for this line of research:
\begin{conjecture}\label{conj:qmiep_generals}
    For every $n\geq1$, a Haar-random $U\sim U(2^n)$ admits a permutation matrix $P$ and a diagonal unitary $D$
    such that $\spec(DPU)$ is antipodal.
\end{conjecture}

\subsubsection{Solving via Gauss--Newton}
\label{sec:numerical_heuristic}
We formulate the search for QMIEP solutions for $U\in U(2^n)$ with $n\geq3$
as a numerical optimization problem.
For a fixed candidate permutation matrix $P$, consider the parameterized unitary
\[
    W(\vec\theta)=D(\vec\theta)PU,
    \qquad
    D(\vec\theta)=\operatorname{diag}
    (1,e^{i\theta_1},\ldots,e^{i\theta_{2^n-1}}).
\]
We fix the first diagonal entry of $D$ to 1 because a global phase does not
affect antipodality.

Write the spectral decomposition of $W(\vec\theta)$ as
\[
    W(\vec\theta)
    =Q\operatorname{diag}(e^{i\alpha_1},\ldots,e^{i\alpha_{2^n}})Q^\dagger,
\]
where $Q$ and $\alpha_i$ depend on $\vec\theta$, and the eigenphases are ordered as
$-\pi<\alpha_1\leq\cdots\leq\alpha_{2^n}\leq\pi$.
We define the residual $r : \mathbb{R}^{2^{n}-1} \rightarrow \mathbb{R}^{2^{n-1}} $ by
\begin{equation}\label{eq:qmiep-pairing-residual}
    \bigl(r(\vec\theta)\bigr)_a
    =\alpha_{a+2^{n-1}}(\vec\theta)-\alpha_a(\vec\theta)-\pi,
    \qquad a=1,\ldots,2^{n-1}.
\end{equation}
By construction, $W(\vec\theta)$ has an antipodal spectrum if and only if
$r(\vec{\theta}) = 0$.

To find $\vec\theta^*$ satisfying $r(\vec\theta^*)=0$, we use a damped
Gauss--Newton method with line search~\cite{nocedal2006}:
\begin{equation}\label{eq:qmiep-gauss-newton-update}
    \begin{aligned}
        p_t&=-\bigl(J_t^\top J_t+\lambda_t I_{2^n-1}\bigr)^{-1}
        J_t^\top r(\vec\theta_t),\\
        \vec\theta_{t+1}&=\vec\theta_t+\eta_t p_t.
    \end{aligned}
\end{equation}
Here, $J_t$ is the Jacobian of $r$ at $\vec\theta_t$,
$\lambda_t$ is the damping parameter, and $\eta_t$ is the line-search step size.

The Jacobian entries can be computed directly from the eigenvectors of $W(\vec\theta_t)$:
\begin{equation}\label{eq:qmiep-jacobian}
    (J_t)_{aj}=|Q_{j+1,a+2^{n-1}}|^2-|Q_{j+1,a}|^2,
    \qquad a=1,\ldots,2^{n-1},\quad j=1,\ldots,2^n-1,
\end{equation}
where $Q$ is evaluated at $\vec\theta_t$.
We compute $p_t$ by solving the corresponding linear system using Cholesky
factorization, avoiding explicit matrix inversion.
We defer further implementation details to Appendix~\ref{app:implementation-details}.

We ran the solver on a Mac mini with a 12-core Apple M4 Pro CPU and 24\,GB of RAM.
The solver succeeded on all Haar-random unitaries
tested at each qubit count from 3 to 10.
This consistent success provides empirical support for the conjectured
universal solvability of the QMIEP from Conjecture~\ref{conj:qmiep_generals}.
Table~\ref{tab:qmiep-solver-runtime} reports the sample counts and mean solver times,
which were only a few seconds even at 10 qubits.

\begin{table}[!t]
    \centering
    \caption{Sample counts and mean QMIEP solver runtimes for Haar-random unitaries
    at each qubit count, using the method described in Section~\ref{sec:numerical_heuristic}.
    We adjusted the sample counts to account for the increasing solver runtime at larger qubit counts.}
    \label{tab:qmiep-solver-runtime}
    \small
    \renewcommand{\arraystretch}{1.15}
    \begin{tabular}{crr}
        \toprule
        \textbf{Qubits} & \textbf{Samples} & \textbf{Mean Solver Time (s)} \\
        \midrule
        3  & 1,000,000 & 0.000048 \\
        4  & 1,000,000 & 0.000153 \\
        5  & 1,000,000 & 0.000672 \\
        6  &   100,000 & 0.004493 \\
        7  &    10,000 & 0.028123 \\
        8  &    10,000 & 0.151093 \\
        9  &     1,000 & 0.674908 \\
        10 &      100 & 4.461843 \\
        \bottomrule
    \end{tabular}
\end{table}

\subsection{\oursheuristicopt{} for Further CNOT Savings}
\label{sec:further-optimal-d-equivalent}

In this section, we report that the QMIEP admitted a solution with $P=I$ for most of the Haar-random
unitaries we sampled.
We then describe the procedure for \oursheuristicopt{}, which exploits such solutions to achieve
further \cnot{} savings, as reflected in the costs reported in
Table~\ref{tab:cnot-count-comparison}.

\begin{table}[!t]
    \centering
    \caption{Instances for which no QMIEP solution with $P=I$ was obtained
    among the indicated numbers of Haar-random unitaries at each qubit count.
    Sample counts were adjusted to account for the increasing solver runtime at larger qubit counts.}
    \label{tab:qmiep-nonidentity-permutation-frequency}
    \small
    \renewcommand{\arraystretch}{1.1}
    \begin{tabular}{crr}
        \toprule
        \textbf{Qubits} & \textbf{Samples} & \textbf{No solution found with $\boldsymbol{P=I}$} \\
        \midrule
        2  & 2,000,000 & 49,747 (2.48735\%) \\
        3  & 1,000,000 &      4 (0.00040\%) \\
        4  & 1,000,000 &      0 (0.00000\%) \\
        5  & 1,000,000 &      0 (0.00000\%) \\
        6  &   100,000 &      0 (0.00000\%) \\
        7  &    10,000 &      0 (0.00000\%) \\
        8  &    10,000 &      0 (0.00000\%) \\
        9  &     1,000 &      0 (0.00000\%) \\
        10 &      100 &      0 (0.00000\%) \\
        \bottomrule
    \end{tabular}
\end{table}

\paragraph{Numerical Observations on the QMIEP with $P=I$}
We sampled Haar-random unitaries at qubit counts $n=2,\ldots,10$ and report the results in
Table~\ref{tab:qmiep-nonidentity-permutation-frequency}.
For $n=2$, the criterion was whether the sampled unitary satisfied the condition in
Equation~\eqref{eq:u4-diagonal-condition},
in which case the QMIEP could be solved without applying an additional permutation.
Of the samples, $\simeq 2.5\%$ violated the criterion, meaning that they strictly required $P \neq
I$ for the QMIEP.

For $n=3,\ldots,10$, we fixed $P=I$ and used the numerical heuristic described in
Section~\ref{sec:numerical_heuristic}
to search for a QMIEP solution for each sampled unitary.
The search failed to find a solution with $P=I$ for four 3-qubit instances (0.00040\%);
it succeeded for all tested instances at $n=4,\ldots,10$.
Since the method is heuristic, failure to find a solution does not establish that no solution
exists with $P=I$.

When the QMIEP is solvable with $P=I$, further \cnot{} savings are possible,
yielding \oursheuristicopt{}, as described below. Appendix~\ref{app:algorithmic-description} gives
the complete algorithmic description.

\paragraph{Base Case: Further Improved 3-qubit $U_D$}
Consider the Block-ZXZ decomposition of $U\in U(2^3)$ in
Equation~\eqref{qc:pd-modulus-derivation:a}, obtained from Equation~\eqref{qc:block-zxz-ucg:block},
\[
U = (I \oplus A)(H \otimes I)(B_1 \oplus B_2)(H \otimes I)(I \oplus C),
\]
where $A,B_1,B_2,C\in U(2^2)$.
Suppose the QMIEP solver finds $P_A=I$ for $A$. Follow the circuit identities
through Equation~\eqref{qc:pd-modulus-derivation:g}. Each 2-qubit subunitary
has a $D$-equivalent implementation using two \cnot{} gates. In particular,
the terminal block leaves only its diagonal residual $D_O$, giving
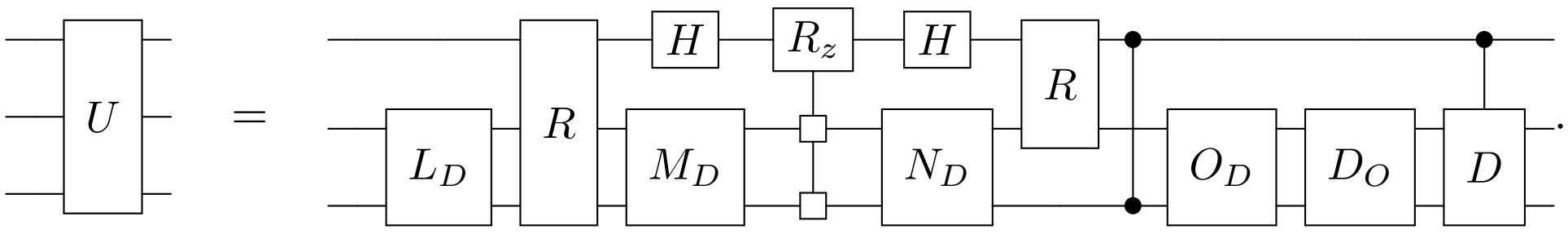
\[
\statementcircuit{
    & \qw & \multigate{2}{U} & \qw \\
    & \qw & \ghost{U} & \qw \\
    & \qw & \ghost{U} & \qw
}
\quad = \quad
\statementcircuit{
    & \qw & \qw & \multigate{2}{\rgate} & \gate{H} & \ucgrzgate{} & \gate{H} & \multigate{1}{\rgate} & \ctrl{2} & \qw & \qw & \ctrl{1} & \qw \\
    & \qw & \multigate{1}{L_D} & \ghost{\rgate} & \multigate{1}{M_D} & \ucgrzctrl & \multigate{1}{N_D} & \ghost{\rgate} & \qw & \multigate{1}{O_D} & \multigate{1}{D_O} & \multigate{1}{D} & \qw \\
    & \qw & \ghost{L_D} & \ghost{\rgate} & \ghost{M_D} & \ucgrzctrl & \ghost{N_D} & \qw & \control\qw & \ghost{O_D} & \ghost{D_O} & \ghost{D} & \qw
}.
\]
Discarding the two final diagonal factors gives a $D$-equivalent implementation.
The rightmost 3-qubit $R$ block of Equation~\eqref{qc:diagonal-migrated-modulus}
is replaced by a 2-qubit $R$ block and a $CZ$, reducing its cost from three
\cnot{} gates to two. Using $\costUD(2)=2$ and Equation~\eqref{eq:ucgrz-r-cost},
we find that the entire circuit uses
\[
\costDheuristicopt{3}
=4\cdot2+\costR(3)+\costR(2)+\costucgrz(3)+1
=17
\]
\cnot{} gates, matching the $PD$-equivalent cost in Corollary~\ref{cor:three-qubit-pd-modulus}.
This is a best-case cost, attained when the QMIEP for $A\in U(2^2)$ is solved
with $P=I$, which occurred in $\simeq 97.5\%$ of the sampled cases.

\paragraph{Recursive Extension to $n$ Qubits}
The principle for $n=3$ extends recursively to larger $n$.
For $n\geq4$, start from Equation~\eqref{qc:pd-modulus-derivation:a} and again
suppose the solver finds a solution with $P_A=I$ for $A\in U(2^{n-1})$. Apply the construction of
Theorem~\ref{thm:pd_modulus_from_qmiep} through Equation~\eqref{qc:pd-modulus-derivation:f}.
At this step, we can use the cheaper $D$-equivalent implementation for $N_D$.

For the terminal block $O'=OD_N$, we recursively construct a $D$-equivalent
implementation $O_D$ such that $O'=D_O O_D$. Since $P_A=I$, the remaining
output residual is diagonal.
The four recursive blocks are thus $L_{PD},M_{PD},N_D,O_D$.
The recursive construction can yield $D$-equivalent implementations with costs lower than or equal
to those of the corresponding $PD$-equivalent implementations.
Thus, these implementations can also be used for $L$ and $M$ (by applying $D$-migrations).
When all four blocks attain the best-case $(n-1)$-qubit cost $\costDheuristicopt{n-1}$,
we obtain the following cost
\begin{equation*}
\begin{aligned}
\costDheuristicopt{n}
&=4\costDheuristicopt{n-1}+\costR(n)+\costR(n-1)+\costucgrz(n)+1\\
&=4\costDheuristicopt{n-1}+5\cdot2^{n-2}-1.
\end{aligned}
\end{equation*}
Solving this recurrence with $\costDheuristicopt{3}=17$ gives
\begin{equation}\label{eq:cost_ud_prime}
\costDheuristicopt{n}=\frac{5}{12}4^n-\frac{5}{4}2^n+\frac{1}{3},
\qquad n\geq3.
\end{equation}

\paragraph{Cost Analysis}

The synthesis procedure \oursheuristicopt{} refines \oursheuristic{} by recursively using improved
$D$-equivalent implementations whenever available.
See Algorithm~\ref{alg:heuristic-opt-synthesis} in Appendix~\ref{app:algorithmic-description} for
the full procedure.
In the best case, the procedure achieves the following \cnot{} count:
\begin{equation}\label{eq:cost_u_prime}
\begin{aligned}
\costheuristicopt{n}
&=\costheuristicopt{n-1}+3\costDheuristicopt{n-1}\\
&\quad+2\costR(n)+\costucgrz(n)\\
&=\OurHeuristicOptCost,\qquad n\geq3.
\end{aligned}
\end{equation}
The last equality follows by substituting Equation~\eqref{eq:cost_ud_prime}
and solving the recurrence with $\costheuristicopt{3}=19$.

The counts in Equations~\eqref{eq:cost_ud_prime} and~\eqref{eq:cost_u_prime}
are attained only when the QMIEP is solved with $P_A=I$ at every recursive step
that requires a $D$-equivalent output.
Otherwise, we revert to the standard Block-ZXZ recursion,
while possibly retaining the improved synthesis methods for the recursive subblocks.
The resulting \cnot{} count thus lies between $\costheuristicopt{n}$ and
$\costheuristic{n}$.

Here, $\costheuristicopt{n}$ is achievable in the best case and does not hold universally.
A formal justification follows from Proposition~\ref{prop:permutation-neighborhood-obstruction},
which gives, for $n\geq 2$, a set of positive Haar measure on which no solution with
$P=I$ exists.
Moreover, since the number of steps requiring a $D$-equivalent output grows
exponentially with $n$ and $\simeq2.5\%$ of the sampled 2-qubit QMIEP instances required
$P\neq I$, we expect the probability of attaining the best-case \cnot{} count
to approach zero as $n$ grows.
However, each failure at the lowest level costs only one additional \cnot{}
gate, and failures at higher levels are empirically negligible
(Table~\ref{tab:qmiep-nonidentity-permutation-frequency}).
We therefore expect most of the savings over \oursheuristic{} to be retained on
average.

\begin{conjecture}\label{conj:high-probability-synthesis}
    For $n\geq5$, the average \cnot{} count of \oursheuristicopt{} over
    Haar-random unitaries $U\in U(2^n)$ is much closer to $\costheuristicopt{n}$
    (Equation~\eqref{eq:cost_u_prime}) than to $\costheuristic{n}$.
\end{conjecture}

This construction also partially addresses the issue raised in
Remark~\ref{rem:diagonal-migration-savings}.
The naive $D$-equivalent construction saves only one \cnot{} gate, since only the
final 2-qubit block in the recursion benefits from $D$-equivalence.
In contrast, when the QMIEP is solved with $P=I$, the savings over exact
synthesis grow exponentially with $n$:
\[
\costheuristicopt{n}-\costDheuristicopt{n}=2^{n-1}-n+1,
\qquad n\geq3.
\]
We believe this observation can serve as a starting point for further studies.

\section{Scalability of the Synthesis}\label{sec:scalability}

\paragraph{Experimental Setup}
We benchmarked synthesis problems on Haar-random unitaries from 4 to 12 qubits on a Mac mini with an Apple M4 Pro chip and 24\,GB of RAM.
Our baseline was Qiskit's \texttt{qs\_decomposition} routine~\cite{qiskit2024}, which implements the state-of-the-art optimized Block-ZXZ decomposition~\cite{krol2024}.
The experiments were conducted up to 12 qubits.
For larger inputs, memory pressure caused the system to swap data to disk, hence resulting in significant latency in running time both for the baseline and ours.

\begin{table}[htbp]
    \centering
    \caption{Mean \cnot{} counts and synthesis times for exact synthesis. Sample counts are per method,
    except that \oursanalytic{} and \oursheuristic{} each used 100 samples at 11 qubits.
    We used smaller samples at 11 and 12 qubits to account for longer synthesis times.
    Times are truncated to hundredths of a second, with shorter times shown as $<0.01$.
    Percentages give the reduction in mean \cnot{} count relative to Qiskit QSD.
    Note that the CNOT count results align with those in Table 1.}
    \label{tab:synthesis-runtime}
    \footnotesize
    \setlength{\tabcolsep}{1.5pt}
    \renewcommand{\arraystretch}{1.22}
    \resizebox{\linewidth}{!}{%
    \begin{tabular}{@{}rr *{4}{@{\hspace{6pt}}rr}@{}}
        \toprule
        \multirow{2}{*}{Qubits} & \multirow{2}{*}{Samples}
            & \multicolumn{2}{c}{Qiskit QSD}
            & \multicolumn{2}{c}{\bfseries\boldmath\oursanalytic{}}
            & \multicolumn{2}{c}{\bfseries\boldmath\oursheuristic{}}
            & \multicolumn{2}{c}{\bfseries\boldmath\oursheuristicopt{}} \\
        \cmidrule(lr){3-4}\cmidrule(lr){5-6}\cmidrule(lr){7-8}\cmidrule(lr){9-10}
            & & CNOTs & Time (s)
              & CNOTs & Time (s)
              & CNOTs & Time (s)
              & CNOTs & Time (s) \\
        \midrule
        4 & 100
            & \benchmarkCNOT{blockzxz-4}{95} & $<0.01$ & \benchmarkCNOT{analytic-4}{92}\,(3.2\%) & $<0.01$
            & \benchmarkCNOT{heuristic-4}{92}\,(3.2\%) & $<0.01$ & \benchmarkCNOT{heuristic-opt-4}{92}\,(3.2\%) & $<0.01$ \\
        5 & 100
            & \benchmarkCNOT{blockzxz-5}{423} & $<0.01$ & \benchmarkCNOT{analytic-5}{408}\,(3.5\%) & $<0.01$
            & \benchmarkCNOT{heuristic-5}{402}\,(5.0\%) & $<0.01$ & \benchmarkCNOT{heuristic-opt-5}{399.13}\,(5.6\%) & $<0.01$ \\
        6 & 100
            & \benchmarkCNOT{blockzxz-6}{1,783} & $<0.01$ & \benchmarkCNOT{analytic-6}{1,720}\,(3.5\%) & $<0.01$
            & \benchmarkCNOT{heuristic-6}{1,678}\,(5.9\%) & $<0.01$ & \benchmarkCNOT{heuristic-opt-6}{1,654.33}\,(7.2\%) & $<0.01$ \\
        7 & 100
            & \benchmarkCNOT{blockzxz-7}{7,319} & 0.04 & \benchmarkCNOT{analytic-7}{7,064}\,(3.5\%) & 0.04
            & \benchmarkCNOT{heuristic-7}{6,854}\,(6.4\%) & 0.05 & \benchmarkCNOT{heuristic-opt-7}{6,726.94}\,(8.1\%) & 0.05 \\
        8 & 100
            & \benchmarkCNOT{blockzxz-8}{29,655} & 0.26 & \benchmarkCNOT{analytic-8}{28,632}\,(3.4\%) & 0.26
            & \benchmarkCNOT{heuristic-8}{27,702}\,(6.6\%) & 0.30 & \benchmarkCNOT{heuristic-opt-8}{27,116.55}\,(8.6\%) & 0.31 \\
        9 & 100 & 119,383 & 1.55 & 115,288\,(3.4\%) & 1.54 & 111,382\,(6.7\%) & 1.83 & 108,868.98\,(8.8\%) & 1.87 \\
        10 & 100 & 479,063 & 8.10 & 462,680\,(3.4\%) & 8.20 & 446,678\,(6.8\%) & 9.94 & 436,264.69\,(8.9\%) & 10.27 \\
        11 & 10 & 1,919,319 & 42.50 & 1,853,784\,(3.4\%) & 41.26
            & 1,789,014\,(6.8\%) & 50.04 & 1,746,613\,(9.0\%) & 52.78 \\
        12 & 10 & 7,683,415 & 223.08 & 7,421,272\,(3.4\%) & 218.98
            & 7,160,662\,(6.8\%) & 270.01 & 6,989,621.20\,(9.0\%) & 278.88 \\
        \bottomrule
    \end{tabular}%
    }
\end{table}

The results are as in Table~\ref{tab:synthesis-runtime}.

\paragraph{CNOT savings from each algorithm}
The merit starts with synthesis with $n \geq 4$, and for $n=3$ the methods trivially match the baseline.
The asymptotic reductions achieved for \oursanalytic{} and \oursheuristic{} relative to optimized Block-ZXZ are $3.4\%$ and $6.8\%$, respectively, as qubit sizes get bigger.
\oursheuristicopt{} achieves further CNOT savings, but as already noted, the
cost is determined in a probabilistic manner depending on the input and whether the derived QMIEP instances are solvable with $P=I$.

\paragraph{Synthesis time}
Across the tested qubit counts, \oursanalytic{} incurred (almost) zero overhead in synthesis time relative to optimized Block-ZXZ
while reducing the asymptotic \cnot{} count by $3.4\%$.
This mode solves only two-qubit QMIEP instances, for which our analytic solution is inexpensive to evaluate.
Synthesis took moderately longer with \oursheuristic{} and \oursheuristicopt{} than with the baseline,
and both methods reduced the \cnot{} count further.
The additional classical synthesis time may be a worthwhile trade-off when compiling programs for near-term quantum computers,
where errors in two-qubit gates can limit circuit fidelity.

\section{Applications} \label{sec:application}

In this section, we show how our proposed methods for ($PD$-equivalent)
unitary synthesis also improve the cost of other quantum programming tasks.

\subsection{Block Encoding}

\begin{table*}[t]
    \caption{\cnot{} counts for block encoding a full-rank
    $2^{n-1}\times2^{n-1}$ matrix using $n-1$ data qubits and one ancilla,
    for $n\geq4$.
    For $n=3$, the construction uses optimal 2-qubit unitary synthesis and matches the existing \cnot{} count.
    For \oursheuristicopt{}, the numerical entries are means over 100 randomly sampled matrices for each $n$.
    The cost formula gives an achievable lower bound.}
    \label{tab:block-encoding-cost-comparison}
    \centering
    \begingroup
        \footnotesize
        \renewcommand{\arraystretch}{1.30}
        \setlength{\tabcolsep}{10pt}
        \noindent\resizebox{\linewidth}{!}{%
        \begin{tabular}{lccccccc}
            \toprule
            \multirow{2}{*}{\textbf{Method}}
                & \multicolumn{6}{c}{\textbf{Qubits, }$\bm{n}$}
                & \multirow{2}{*}{\textbf{\boldmath\cnot{} cost formula}} \\
            \cmidrule(lr){2-7}
                & $\mathbf{4}$ & $\mathbf{5}$ & $\mathbf{6}$
                & $\mathbf{7}$ & $\mathbf{8}$ & $\mathbf{9}$ & \\
            \midrule
            \cite{li2026}\textsuperscript{*}
                & $45$ & $205$ & $877$ & $3629$ & $14765$ & $59565$
                & $\siableBEcost$ \\
            \addlinespace[2pt]
            \textbf{\oursanalytic{}}
                & $\mathbf{44}$ & $\mathbf{198}$ & $\mathbf{846}$
                & $\mathbf{3502}$ & $\mathbf{14254}$ & $\mathbf{57518}$
                & $\bm{\OurAnalyticBECost}$ \\
            \addlinespace[1.5pt]
            \cdashline{1-8}[1pt/2.5pt]
            \textbf{\oursheuristic{}}
                & $\mathbf{44}$ & $\mathbf{196}$ & $\mathbf{828}$
                & $\mathbf{3404}$ & $\mathbf{13804}$ & $\mathbf{55596}$
                & $\bm{\OurHeuristicBECost}$ \\
            \textbf{\oursheuristicopt{}}
                & $\mathbf{44}$ & $\mathbf{195.01}$ & $\mathbf{818.19}$
                & $\mathbf{3346.03}$ & $\mathbf{13524.17}$ & $\mathbf{54368.18}$
                & $\bm{\geq \OurHeuristicOptBECost}$ \\
            \bottomrule
        \end{tabular}
        }%
        \par
        \smallskip
        {\raggedright
        \textsuperscript{*}\,Uses Block-ZXZ~\cite{krol2024} with $D$-migration.
        \par}
    \endgroup
\end{table*}

Improvements in unitary synthesis lead to more efficient block encoding.
Block encoding is the task of embedding a (generally non-unitary) matrix $A$, up to a
normalization factor, as the upper-left block of a larger unitary
implemented by a quantum circuit~\cite{low2019,gilyen2019,low2017,martyn2021}.
For a $d\times d$ matrix, at least $\lceil\log_2 d\rceil$ data qubits are required,
together with additional ancilla qubits for the unitary embedding.
Unitary synthesis is useful for block encoding a full-rank $2^{n-1}\times 2^{n-1}$ matrix
using $n-1$ data qubits and one ancilla qubit. The corresponding synthesis problem is stated as follows.

\begin{problem}[Block Encoding of a Full-Rank Square Matrix]
Given a full-rank matrix $A\in\mathbb{C}^{2^{n-1}\times 2^{n-1}}$
and a normalization factor $\alpha\geq\lVert A\rVert_2$, synthesize an $n$-qubit
quantum circuit that implements a unitary $U_A \in U(2^n)$ satisfying
\[
    (\bra{0}\otimes I_{2^{n-1}})\,U_A\,
    (\ket{0}\otimes I_{2^{n-1}})=\frac{A}{\alpha}.
\]
\end{problem}

The approach to this problem is based on the singular-value decomposition (SVD).
An explicit algebraic construction was given in~\cite{martyn2021}.
Its circuit implementation using Block-ZXZ synthesis and $D$-migration
in~\cite{li2026} achieves the current state-of-the-art \cnot{} count as
reported in Table~\ref{tab:block-encoding-cost-comparison}.
The implementation proceeds as follows.
Write $A=W\Sigma V$, where $W,V\in U(2^{n-1})$ and
$\Sigma=\operatorname{diag}(\sigma_0,\ldots,\sigma_{2^{n-1}-1})$.
For the given normalization factor $\alpha\geq\lVert A\rVert_2$, set
    $D_C={\Sigma}/{\alpha}, D_S=\sqrt{I_{2^{n-1}}-D_C^2}$.
Then, a block encoding of $A$ is given by
\begin{equation}\label{eq:svd-block-encoding}
    U_A=(I_2\otimes W)U_{CS}(I_2\otimes V),\quad
    \text{where }U_{CS}=\begin{pmatrix}D_C&D_S\\-D_S&D_C\end{pmatrix}.
\end{equation}
The unitary $U_{CS}$ can be implemented as a uniformly controlled
$R_y$ gate~\cite[Theorem~10]{shende2006}, with the ancilla as the target and the
$n-1$ data qubits as controls.
Applying the proposed $PD$-migration to this construction yields the following implementation.
\begin{proposition}
    Given a full-rank matrix $A\in\mathbb{C}^{2^{n-1}\times 2^{n-1}}$,
    the following circuit implements a unitary $U_A$ that block encodes $A$:
    \[
    \statementcircuit{
      & \qw            & \gate{S^\dagger} & \gate{H} & \ucgrzgate{} & \gate{H} & \gate{S} & \qw \\
      & \qcbundle{n-1} & \qw             & \gate{L_{PD}} & \ucgrzctrl   & \gate{M} & \qw      & \qw
}
    \]
    The \cnot{} cost is
    \begin{equation}\label{eq:block-encoding-cost}
    \costBE(n) = \costUPD(n-1) + \costU(n-1) + \costucgrz(n).
    \end{equation}
\end{proposition}
\begin{proof}
Equation~\eqref{eq:svd-block-encoding} gives the following circuit:
\[
\statementcircuit{
  & \qw            & \qw      & \gate{R_y} & \qw      & \qw \\
  & \qcbundle{n-1} & \gate{V} & \ucgrzctrl & \gate{W} & \qw
}.
\]
The central uniformly controlled $R_y$ gate can be implemented using $\ucgrz$
via the identity $R_y(\theta)=SHR_z(\theta)HS^\dagger$, yielding
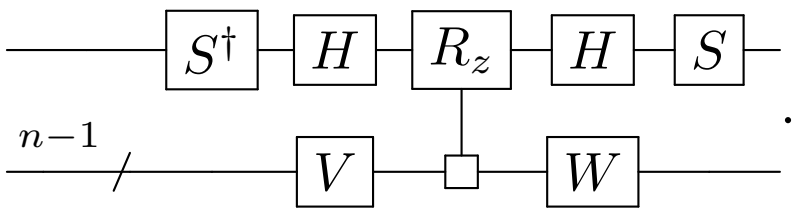
\begin{equation*}
\statementcircuit{
  & \qw            & \gate{S^\dagger} & \gate{H} & \ucgrzgate{} & \gate{H} & \gate{S} & \qw \\
  & \qcbundle{n-1} & \qw             & \gate{V} & \ucgrzctrl   & \gate{W} & \qw      & \qw
}.
\end{equation*}
Applying $PD$-migration to $V$ as in Equation~\eqref{qc:pd-migration},
absorbing the residual $PD$ into $W$, and relabelling the subunitaries
then yields the stated circuit.
\end{proof}

For the \oursanalytic{} and \oursheuristic{} entries in Table~\ref{tab:block-encoding-cost-comparison},
we evaluate Eq.~\eqref{eq:block-encoding-cost} using Eqs.~\eqref{eq:analytic-unitary-cost} and~\eqref{eq:heuristic-unitary-cost}, respectively.
The \oursheuristicopt{} best-case bound uses Eq.~\eqref{eq:cost_ud_prime} for $L_{PD}$ with $P=I$
and Eq.~\eqref{eq:cost_u_prime} for the exact block $M$.

\subsection{Isometry Synthesis}\label{sec:isometry-synthesis}

Another important quantum programming task is isometry synthesis~\cite{iten2016,li2026,malvetti2021}.
The task is to synthesize an $n$-qubit quantum circuit whose unitary matrix has
the columns of a given $m$-to-$n$-qubit isometry, where $m<n$, as its first $2^m$ columns.
The case $m=n-1$ is of particular interest because of its use in generic state preparation
(see Section~\ref{sec:state_prep}). Improved unitary synthesis reduces the cost in this case
(for $m<n-1$, the best-known method in terms of \cnot{} cost is the column-by-column method presented in~\cite{iten2016}).

The target synthesis problem is formulated as follows.
\begin{problem}[$(n-1)$-to-$n$-Qubit Isometry Synthesis]
Given an isometry $J\in\mathbb{C}^{2^n\times 2^{n-1}}$ with
$J^\dagger J=I_{2^{n-1}}$, synthesize an $n$-qubit quantum circuit
that implements a unitary $U_{Iso}\in U(2^n)$ satisfying
\[
    U_{Iso}\bigl(\ket{0}\otimes\ket{i}\bigr)=J\ket{i},
    \qquad i=0,\ldots,2^{n-1}-1.
\]
For the $PD$-equivalent variant, synthesize a unitary
$U_{Iso_{PD}}\in U(2^n)$ such that
\begin{equation}\label{eq:pd-isometry-definition}
 U_{Iso_{PD}}\bigl(\ket{0}\otimes\ket{i}\bigr)
    =e^{i\theta_i}J\ket{\sigma(i)}
\end{equation}
for all $i$ in the same range, with phases $\theta_i\in\mathbb{R}$
and some input-register permutation~$\sigma$.
\end{problem}
The $PD$-equivalent variant defined above will be used as a building block
for state preparation in Section~\ref{sec:state_prep}.

In~\cite{iten2016}, QSD~\cite{shende2006} was adapted to isometry synthesis.
The same basic approach can also be formulated using the Block-ZXZ
decomposition~\cite{devos2014,fuhr2015,devos2016}.
For an $(n-1)$-to-$n$-qubit isometry, the first qubit is initialized to $\ket{0}$,
so the input-side block can be chosen as an uncontrolled unitary.
This gives a unitary extension of the isometry of the form
\begin{equation}\label{eq:isometry-qsd-decomposition}
    U_{Iso}=(A_1\oplus A_2)(H\otimes I_{2^{n-1}})
    (I_{2^{n-1}}\oplus B)(H\otimes I_{2^{n-1}})(I_2\otimes C),
\end{equation}
where $A_1,A_2,B,C\in U(2^{n-1})$, as in the Block-ZXZ decomposition.
In~\cite{li2026}, $D$-migration and \cnot{} absorption from~\cite{krol2024}
were used to yield a state-of-the-art \cnot{} count for isometry synthesis.
Using $PD$-migration further reduces this count, giving the following circuit:
\begin{proposition}\label{prop:pd-isometry-synthesis}
    Let $n\geq3$ and let $J\in\mathbb{C}^{2^n\times 2^{n-1}}$ be an isometry.
    The following circuit implements $J$:
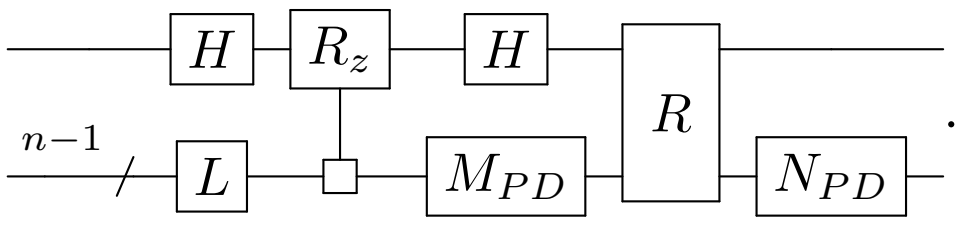
    \[
    \statementcircuit{
      & \qw            & \gate{H} & \ucgrzgate{} & \gate{H}      & \multigate{1}{\rgate} & \qw           & \qw \\
      & \qcbundle{n-1} & \gate{L} & \ucgrzctrl   & \gate{M_{PD}} & \ghost{\rgate}        & \gate{N_{PD}} & \qw
}.
\]
    Replacing $L$ by a $PD$-equivalent implementation $L_{PD}$ satisfying $L_{PD}P_LD_L=L$ yields a
    $PD$-equivalent implementation of $J$. 
    
    The \cnot{} costs are
    \begin{align}
        \costIso(n) &= 2\costUPD(n-1) + \costU(n-1)+ \costucgrz(n) + \costR(n),
        \label{eq:isometry-cost}\\
        \costIsoPD(n)&= 3\costUPD(n-1) + \costucgrz(n) + \costR(n).
        \label{eq:pd-isometry-cost}
\end{align}
\end{proposition}
\begin{proof}
\begingroup
\setlength{\abovedisplayskip}{6pt}
\setlength{\belowdisplayskip}{6pt}
\setlength{\abovedisplayshortskip}{4pt}
\setlength{\belowdisplayshortskip}{4pt}
Equation~\eqref{eq:isometry-qsd-decomposition} gives the following isometry circuit:
\[
\statementcircuit{
  & \qw            & \gate{H}         & \ctrl{1}         & \gate{H} & \mux\qwx[1]            & \qw \\
  & \qw            & \multigate{1}{C} & \multigate{1}{B} & \qw      & \multigate{1}{A_1,A_2} & \qw \\
  & \qcbundle{n-2} & \ghost{C}        & \ghost{B}        & \qw      & \ghost{A_1,A_2}        & \qw
}.
\]
Decomposing the final multiplexor $A_1\oplus A_2$ as in
Equation~\eqref{eq:multiplexor} and then decomposing the resulting
controlled-diagonal gate using Propositions~\ref{prop:shende_theorem_7}
and~\ref{prop:ucgrz} yields the following circuit after absorbing
adjacent unitaries into $V$ and $N$:
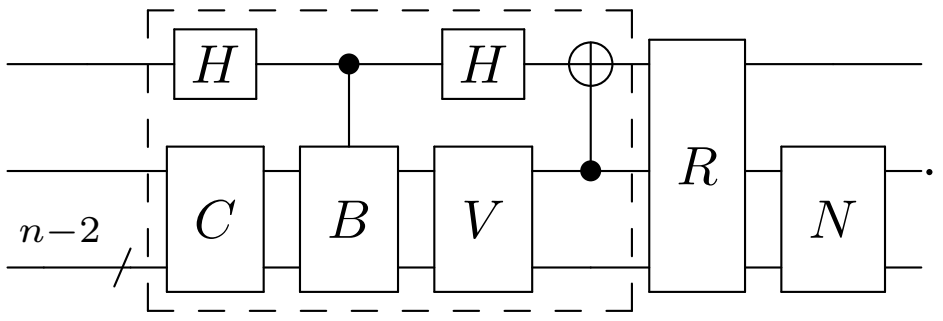
\[
\statementcircuit{
  & \qw            & \gate{H}         & \ctrl{1}         & \gate{H}         & \targ     & \multigate{2}{\rgate} & \qw              & \qw \\
  & \qw            & \multigate{1}{C} & \multigate{1}{B} & \multigate{1}{V} & \ctrl{-1} & \ghost{\rgate}        & \multigate{1}{N} & \qw \\
  & \qcbundle{n-2} & \ghost{C}        & \ghost{B}        & \ghost{V}        & \qw \gategroup{1}{3}{3}{6}{.7em}{--} & \ghost{\rgate}        & \ghost{N}        & \qw
}.
\]
\begin{samepage}

As in Equation~\eqref{qc:block-zxz-cz}, the dashed subcircuit can be
rewritten as a single multiplexor sandwiched between Hadamard gates.
Decomposing this multiplexor yields the following circuit for some
unitaries $L,M\in U(2^{n-1})$:
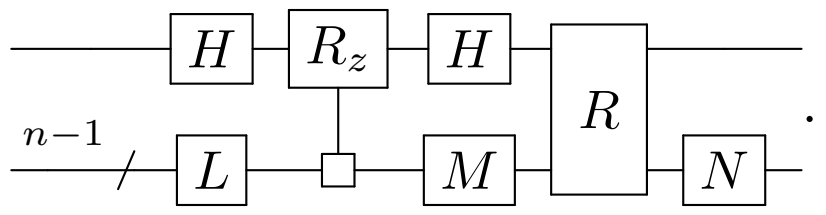
\[
\statementcircuit{
  & \qw            & \gate{H} & \ucgrzgate{} & \gate{H} & \multigate{1}{\rgate} & \qw      & \qw \\
  & \qcbundle{n-1} & \gate{L} & \ucgrzctrl   & \gate{M} & \ghost{\rgate}        & \gate{N} & \qw
}.
\]
Applying $PD$-migration from right to left, starting with $N$,
using Proposition~\ref{prop:pd-thru-r} gives the stated circuit.
Finally, writing $L=L_{PD}P_LD_L$ and discarding the input-side factors
$P_L$ and $D_L$ yields a $PD$-equivalent implementation of the isometry.
\end{samepage}
\endgroup
\end{proof}

For the \oursanalytic{} and \oursheuristic{} entries in Table~\ref{tab:isometry-cost-comparison},
we evaluate Eq.~\eqref{eq:isometry-cost} using Eqs.~\eqref{eq:analytic-unitary-cost} and~\eqref{eq:heuristic-unitary-cost}, respectively.
The \oursheuristicopt{} best-case bound uses Eq.~\eqref{eq:cost_ud_prime} for both $PD$-equivalent blocks
and Eq.~\eqref{eq:cost_u_prime} for the exact block.

\begin{table*}[t]
    \caption{\cnot{} counts for exact synthesis of an $(n-1)$-to-$n$-qubit
    isometry for $n\geq4$.
    It is known that, for $n=2$, exact and $D$-equivalent implementations each use two \cnot{} gates~\cite{iten2016,shende2004c}.
    For $n=3$, the construction uses optimal 2-qubit unitary synthesis and matches the existing \cnot{} count.
    For \oursheuristicopt{}, the numerical entries are means over 100 randomly sampled isometries for each $n$.
    The cost formula gives an achievable lower bound.}
    \label{tab:isometry-cost-comparison}
    \centering
    \begingroup
        \footnotesize
        \newcommand{\isometrymethodcell}[1]{\parbox[c][2.2\baselineskip][c]{9em}{\raggedright #1}}
        \renewcommand{\arraystretch}{1.30}
        \setlength{\tabcolsep}{10pt}
        \noindent\resizebox{\linewidth}{!}{%
        \begin{tabular}{l@{\hspace{12pt}}ccccccc}
            \toprule
            \multirow{2}{*}{\textbf{Method}}
                & \multicolumn{6}{c}{\textbf{Output qubits, }$\bm{n}$}
                & \multirow{2}{*}{\textbf{\boldmath\cnot{} cost formula}} \\
            \cmidrule(lr){2-7}
                & $\mathbf{4}$ & $\mathbf{5}$ & $\mathbf{6}$
                & $\mathbf{7}$ & $\mathbf{8}$ & $\mathbf{9}$ & \\
            \midrule
            \isometrymethodcell{\cite{iten2016}\textsuperscript{*}}
                & $73$ & $329$ & $1393$ & $5729$ & $23233$ & $93569$
                & $\ItenIsoCost$ \\
            \isometrymethodcell{\cite{li2026}\textsuperscript{**}}
                & $70$ & $314$ & $1330$ & $5474$ & $22210$ & $89474$
                & $\LiIsoCost$ \\
            \isometrymethodcell{\textbf{\oursanalytic{}}}
                & $\mathbf{68}$ & $\mathbf{303}$ & $\mathbf{1283}$
                & $\mathbf{5283}$ & $\mathbf{21443}$ & $\mathbf{86403}$
                & $\bm{\OurAnalyticIsoCost}$ \\
            \addlinespace[1.5pt]
            \cdashline{1-8}[1pt/2.5pt]
            \isometrymethodcell{\textbf{\oursheuristic{}}}
                & $\mathbf{68}$ & $\mathbf{299}$ & $\mathbf{1253}$
                & $\mathbf{5129}$ & $\mathbf{20753}$ & $\mathbf{83489}$
                & $\bm{\OurHeuristicIsoCost}$ \\
            \isometrymethodcell{\textbf{\oursheuristicopt{}}}
                & $\mathbf{68}$ & $\mathbf{297.04}$ & $\mathbf{1236.34}$
                & $\mathbf{5036.35}$ & $\mathbf{20320.01}$ & $\mathbf{81617.37}$
                & $\bm{\geq \OurHeuristicOptIsoCost}$ \\
            \bottomrule
        \end{tabular}
        }%
        \par
        \smallskip
        {\raggedright
        \textsuperscript{*}\,Based on QSD~\cite{shende2006}.\\
        \textsuperscript{**}\,Based on Block-ZXZ~\cite{krol2024} with \cnot{} absorption and $D$-migration.
        \par}
    \endgroup
\end{table*}

\begin{table*}[t]
    \caption{\cnot{} counts for $n$-qubit state preparation.
    Optimal constructions are known for \mbox{$n=2,3$}~\cite{shende2004c,znidaric2008}.
    Our method matches the result in~\cite{li2026} for $n=4,5$, which relies on 2-qubit unitaries.
    The improvement begins at $n=6$, when the construction first uses unitary blocks on more than 2 qubits.
    For \oursheuristicopt{}, the numerical entries are means over 100 randomly sampled states for each $n$.}
    \label{tab:state-preparation-cost-comparison}
    \centering
    \begingroup
        \footnotesize
        \renewcommand{\arraystretch}{1.30}
        \setlength{\tabcolsep}{0pt}
        \begin{tabular}{@{}m{.23\linewidth}%
            *{7}{>{\centering\arraybackslash}m{.11\linewidth}}@{}}
            \toprule
            \multirow{2}{*}{\textbf{Method}}
                & \multicolumn{7}{c}{\textbf{Qubits, }$\bm{n}$} \\
            \cmidrule(lr){2-8}
                & $\mathbf{6}$ & $\mathbf{7}$ & $\mathbf{8}$ & $\mathbf{9}$
                & $\mathbf{10}$ & $\mathbf{11}$ & $\mathbf{12}$ \\
            \midrule
            \cite{araujo2024}\textsuperscript{*}
                & $46$ & $100$ & $213$ & $441$ & $914$ & $1862$ & $3789$ \\
            \cite{li2026}\textsuperscript{**}
                & $42$ & $93$ & $199$ & $418$ & $867$ & $1774$ & $3612$ \\
            \addlinespace[2pt]
            \textbf{\oursanalytic{}}
                & $\mathbf{41}$ & $\mathbf{90}$ & $\mathbf{192}$ & $\mathbf{403}$
                & $\mathbf{836}$ & $\mathbf{1711}$ & $\mathbf{3484}$ \\
            \addlinespace[1.5pt]
            \cdashline{1-8}[1pt/2.5pt]
            \textbf{\oursheuristic{}}
                & $\mathbf{41}$ & $\mathbf{90}$ & $\mathbf{190}$ & $\mathbf{397}$
                & $\mathbf{818}$ & $\mathbf{1669}$ & $\mathbf{3386}$ \\
            \textbf{\oursheuristicopt{}}
                & $\mathbf{40.05}$ & $\mathbf{89.02}$ & $\mathbf{185.11}$ & $\mathbf{390.16}$
                & $\mathbf{797.22}$ & $\mathbf{1634.54}$ & $\mathbf{3301.11}$ \\
            \bottomrule
        \end{tabular}
        \par
        \smallskip
        {\raggedright
        \textsuperscript{*}\,Based on QSD~\cite{shende2006}.\\
        \textsuperscript{**}\,Based on Block-ZXZ~\cite{krol2024} with $D$-migration.
        \par}
    \endgroup
\end{table*}

\subsection{State Preparation} \label{sec:state_prep}
State preparation plays an essential role in many quantum applications.
The corresponding synthesis problem is stated as follows.

\begin{problem}[State Preparation]
Given an $n$-qubit state $\ket{\psi}$, synthesize an $n$-qubit quantum circuit implementing a unitary $U_{SP}$ such that $U_{SP}\ket{0}=\ket{\psi}$.
\end{problem}

Many techniques have been proposed for this problem~\cite{mottonen2005,shende2006,plesch2011,sun2023,araujo2024,li2026}.
State preparation using unitaries derived from the Schmidt decomposition was first considered
in~\cite{plesch2011} and was extended to incorporate isometry synthesis in~\cite{iten2016}.
An explicit quantum circuit construction using QSD-based unitary synthesis was proposed and implemented in~\cite{araujo2024}.
The current state-of-the-art result~\cite{li2026} applies $D$-equivalent
unitary synthesis based on the optimized Block-ZXZ decomposition~\cite{krol2024}.
We improve this result using the proposed $PD$-equivalent unitary and isometry synthesis methods.

The basic construction proceeds as follows.
The Schmidt decomposition expresses an $n$-qubit state $\ket{\psi}$ as
\[
    \ket{\psi}=\sum_{i=0}^{2^m-1}\lambda_i\ket{e_i}_A\otimes\ket{f_i}_B,
\]
where $\{\ket{e_i}_A\}$ and $\{\ket{f_i}_B\}$ are orthonormal sets on registers
$A$ (the first $m=\lfloor n/2\rfloor$ qubits) and $B$ (the remaining qubits), respectively.
Choose $U$ and $V$ such that $U\ket{i}=\ket{e_i}_A$ and $V\ket{i}=\ket{f_i}_B$.
Here, $U$ is an $m$-qubit unitary, while $V$ is an $m$-qubit unitary for $n=2m$
and an $m$-to-$(m+1)$-qubit isometry for $n=2m+1$.
Applying their inverse circuits, followed by a \cnot{} staircase, reduces the problem to $m$-qubit state preparation.
In this approach, combining our improved unitary and isometry synthesis methods with suitable $PD$-equivalent implementations
further reduces the \cnot{} cost compared with the state of the art, as follows:

\begin{proposition}
    Let $n\geq4$ and $m=\lfloor n/2\rfloor$.
    The following circuit implements $n$-qubit state preparation:
    \[
    \statementcircuit{
      \push{\makebox[3.5em][r]{$q_1$}}       & \qw & \multigate{3}{\text{\normalfont\shortstack{$m$-qubit\\state preparation}}} & \ctrl{4} & \qw      & \qw      & \multigate{3}{U_D} & \qw \\
      \push{\makebox[3.5em][r]{$q_2$}}       & \qw & \ghost{\text{\normalfont\shortstack{$m$-qubit\\state preparation}}}       & \qw      & \ctrl{4} & \qw      & \ghost{U_D}        & \qw \\
                                           & \push{\vdots} & \nghost{\text{\normalfont\shortstack{$m$-qubit\\state preparation}}} & & & & \nghost{U_D} & \\
      \push{\makebox[3.5em][r]{$q_m$}}       & \qw & \ghost{\text{\normalfont\shortstack{$m$-qubit\\state preparation}}}       & \qw      & \qw      & \ctrl{4} & \ghost{U_D}        & \qw \\
      \push{\makebox[3.5em][r]{$q_{m+1}$}}   & \qw & \qw & \targ & \qw   & \qw   & \multigate{4}{V_{PD}} & \qw \\
      \push{\makebox[3.5em][r]{$q_{m+2}$}}   & \qw & \qw & \qw   & \targ & \qw   & \ghost{V_{PD}}        & \qw \\
                                           & \push{\vdots} & & & & & \nghost{V_{PD}} & \\
      \push{\makebox[3.5em][r]{$q_{2m}$}}    & \qw & \qw & \qw   & \qw   & \targ & \ghost{V_{PD}}        & \qw \\
      \push{\makebox[3.5em][r]{%
        \shortstack[r]{$q_{2m+1}$\\[-0.2ex]
          {\scriptsize\normalfont odd $n$ only}}}}
                                           & \qw & \qw & \qw   & \qw   & \qw   & \ghost{V_{PD}}        & \qw
}
    \]
    For $n=2m$, $V_{PD}$ is an $m$-qubit $PD$-equivalent unitary, and the bottom wire is omitted.
    For $n=2m+1$, replace $V_{PD}$ by $V_{Iso_{PD}}$, a unitary extension of an $m$-to-$(m+1)$-qubit $PD$-equivalent isometry as defined in Equation~\eqref{eq:pd-isometry-definition} (up to wire ordering), with $q_{2m+1}$ initialized to $\ket{0}$.
    The \cnot{} cost satisfies the recurrence
    \begin{equation}\label{eq:state-preparation-cost}
        \costSP(n)=\costSP(m)+m+\costUD(m)
        +\begin{cases}
            \costUPD(m), & n=2m,\\[2pt]
            \costIsoPD(m+1), & n=2m+1.
        \end{cases}
    \end{equation}
\end{proposition}

\begin{proof}
Consider the inverse circuit and write the target state in Schmidt form as
$\ket{\psi}=\sum_i\lambda_i\ket{e_i}_A\ket{f_i}_B$.
For $n=2m$, write $V_{PD}\ket{j}=e^{i\theta_j}\ket{f_{\sigma(j)}}_B$;
for $n=2m+1$, write
$V_{Iso_{PD}}(\ket{j}\otimes\ket{0})=e^{i\theta_j}\ket{f_{\sigma(j)}}_B$.
Applying the inverse circuit on register $B$ and relabelling the summation index gives
\[
    \sum_j\lambda_{\sigma(j)}e^{-i\theta_j}
    \ket{e_{\sigma(j)}}_A\ket{j}_B,
\]
with the extra qubit in $\ket{0}$ when $n$ is odd.
Choose the column ordering of $U$ so that $U\ket{j}=\ket{e_{\sigma(j)}}_A$
and apply $U_D^\dagger$, where $D U_D^\dagger=U^\dagger$.
The state then becomes $\sum_j c_j\ket{j}_A\ket{j}_B$,
where $c_j=\lambda_{\sigma(j)}e^{i\theta'_j}$.
The phases $\theta'_j\in\mathbb{R}$ include both $-\theta_j$
and the additional phases introduced by $U_D^\dagger$.
The \cnot{} staircase then reduces the problem to preparing the $m$-qubit state $\sum_j c_j\ket{j}_A$,
which can be prepared recursively.
\end{proof}
Note that using $PD$-equivalent implementations for both $U$ and $V$ generally fails,
since different permutations of the computational basis prevent the \cnot{} staircase from disentangling the registers.

It is known that 2-qubit state preparation can be implemented with one \cnot{} gate~\cite{shende2004c},
and 3-qubit state preparation with three \cnot{} gates~\cite{znidaric2008}.
Using these base cases and the unitary-synthesis costs $\costanalytic{n}$ and
$\costheuristic{n}$ from Equations~\eqref{eq:analytic-unitary-cost}
and~\eqref{eq:heuristic-unitary-cost}, respectively, in the recurrence
in Equation~\eqref{eq:state-preparation-cost}, with the $PD$-equivalent isometry
cost from Equation~\eqref{eq:pd-isometry-cost}, yields the counts for each mode of unitary synthesis in
Table~\ref{tab:state-preparation-cost-comparison}.

\section{Related Work}\label{sec:related}

\paragraph{Other Quantum Circuit Synthesis Objectives}
Another important synthesis objective is $T$-count minimization.
Alongside \cnot{}-count minimization, it plays a major role in many quantum
circuit synthesis problems, including unitary synthesis~\cite{
gosset2014,heyfron2019,amy2019,kissinger2020,
kliuchnikov2013exact,ross2016,hao2026,
low2024,gosset2026,tan2026,sanders2019,bausch2022,
paetznick2014,bocharov2015rus,gidney2018,amy2013,amy2014,selinger2013}.
Logical $T$ gates require substantial additional resources in many fault-tolerant
architectures, particularly surface-code implementations based on magic-state
distillation~\cite{bravyi2005}. This overhead motivates $T$-count minimization.
Our results show that imposing a separable structure on the target diagonal
unitary reduces the \cnot{} count. This structure may also lower the $T$ cost
of synthesizing the controlled diagonal.
An interesting direction for future work
is to investigate whether this approach can be combined with existing
$T$-gate optimization techniques.

Another axis of synthesis concerns the number of ancillary qubits.
In this work, we focus on ancilla-free synthesis.
When additional qubits are available, they can be used to reduce circuit depth,
giving rise to a space--time trade-off~\cite{sun2023,zhang2022depth,rosenthal2026,low2024,gosset2026,tan2026,giles2013,kliuchnikov2013multi,amy2014,selinger2013}.
This introduces a Pareto trade-off between qubit count and circuit depth:
a shallower circuit that uses ancillas is not necessarily preferable to a
deeper ancilla-free circuit. The comparison depends on the cost of qubit
interactions and the relative costs of space and time, which vary substantially
across hardware platforms. An important direction for future work is to assess
how the cost improvements achieved by our ancilla-free approach compare with
those of ancilla-assisted methods. 
Our approach, which exploits a separable spectrum in conjunction with $PD$-migration,
may also be combined with existing ancilla-assisted methods.

\paragraph{Other Approaches to Unitary Synthesis}

In addition to the comparable work discussed in Section~\ref{sec:introduction},
several studies~\cite{younis2021qfast,davis2020qsearch,rakyta2022approaching,wierichs2025brickwall}
optimize parameterized quantum circuit (PQC) ansatzes,
typically composed of \cnot{} gates and 1-qubit rotations.
These approaches can be understood in terms of a trade-off between
\cnot{} count and classical synthesis effort.
Methods based on analytic matrix decomposition require a moderate amount of classical computation
but typically yield higher \cnot{} counts.
By contrast, PQC-based approaches can, in principle, find
\cnot{}-optimal circuits, but the cost of parameter optimization limits their scalability.
For example, even for 5-qubit unitaries, the method in Ref.~\cite{wierichs2025brickwall} has a reported average synthesis time of 1466.67 seconds.

From this perspective, either \oursheuristic{} or \oursheuristicopt{} can be viewed as a balanced choice.
Each incorporates numerical optimization (Section~\ref{sec:general-case}) for QMIEP subproblems,
reducing the \cnot{} count at the cost of additional but affordable classical computation.

\paragraph{Multiplicative Inverse Eigenvalue Problem}
Early work on the MIEP addressed its formulation and solvability~\cite{downing1956,HADELER196965,FRIEDLAND1975127}.
Subsequent studies considered variants with restricted diagonal multipliers~\cite{chu1998iep}.
Numerical approaches to inverse eigenvalue problems include Newton-type methods~\cite{friedland1987numerical},
least-squares formulations~\cite{chen1996least}, and gradient-based methods~\cite{chu1990projected}.
Our QMIEP imposes unitary constraints, allows permutations, and prescribes only the antipodal structure of the target spectrum.
These methods may help refine the QMIEP algorithms in Section~\ref{sec:numerical_heuristic},
alongside the search for analytic solutions and a proof of universal solvability (as stated through Conjecture~\ref{conj:qmiep_generals}).

\section{Conclusion}\label{sec:conclusion}
In this work, we developed a new technique for unitary synthesis that reduces the \cnot{} count.
We introduced $PD$-migration and formulated the quantum
multiplicative inverse eigenvalue problem (QMIEP) to support it.
By modifying a unitary's spectrum through permutation and diagonal factors,
QMIEP enables $PD$-equivalent implementations with fewer \cnot{} gates;
$PD$-migration then absorbs these factors to obtain an exact implementation.
Our fully analytic construction (\oursanalytic{}) achieves an asymptotic $3.4\%$ reduction in
\cnot{} count over the state of the art.
The numerical heuristic (\oursheuristic{}) increases this reduction to $6.8\%$, although its universality remains conjectural.
The optimized heuristic (\oursheuristicopt{}) can reduce the \cnot{} count further when a QMIEP subproblem admits a solution with $P=I$.

\paragraph{Future Work}
Several directions remain open.
\begin{itemize}
    \item A natural next step is to resolve
    Conjectures~\ref{conj:qmiep_generals} and~\ref{conj:high-probability-synthesis}.
    Both concern QMIEP, and studying them may yield solution methods beyond
    the numerical approach in Section~\ref{sec:numerical_heuristic}.
    Although the numerical overhead may be moderate, faster solvers would
    still improve the scalability of unitary synthesis.

    \item Another interesting direction is to consider other versions of QMIEP.
    Throughout this paper, we have formulated QMIEP with an antipodal target spectrum,
    since it yields a separable spectrum that is cheaper to implement during synthesis.
    Beyond separability, other spectral structures (represented by diagonal unitaries)
    could be considered as QMIEP targets.
    For example, a certain class of diagonal unitaries with inexpensive implementations
    (and hence inexpensive controlled versions) can be characterized~\cite{welch2014}.
    Designing such classes of diagonal unitaries as QMIEP targets could lead to
    cheaper implementations in unitary synthesis and potentially facilitate analytic solutions.
\end{itemize}

\section*{Acknowledgments}

Acknowledgments will be added upon final publication.

\section*{Author contributions}

Kang took the lead in conceiving the study, formulating the research problem, deriving the main solutions and results, and writing the manuscript.
Oh led discussions on the technical side of the manuscript and its presentation,
validation,
and supervised the project.

\paragraph{AI Usage Disclosure}

We developed all the core ideas behind the synthesis method, including the construction of $PD$-equivalent implementations and $PD$-migration.
We carried out most of the technical derivations, including those for the circuit identities.

We used large language models, primarily ChatGPT Astra~6 and earlier versions of ChatGPT, for parts of the technical work.
These models pointed us to references on the classical MIEP that inspired us to formulate its quantum counterpart, the QMIEP.
They also suggested several key references, including a book we consulted on eigenvalue perturbations~\cite{kato}.
We wrote the proof of Proposition~\ref{prop:permutation_u4} with guidance from the models on applying Facts~\ref{fact:jacobi} and~\ref{fact:leibniz}.

AI also played an important role in the proof of Corollary~\ref{cor:heuristic-unitary-cost}.
Our initial inductive argument relied on each subunitary admitting the required $PD$-equivalent implementation with positive probability.
The models pointed out that, for rigor, we needed to convey this argument in a Haar-measure theoretic way.
Through these discussions, we developed the measure-theoretic proof in Appendix~\ref{app:heuristic-unitary-cost-proof} with some pointers to important references.

We used the models interactively to generate code for the software and to prototype several numerical approaches.
We selected the damped Newtonian approach and developed it further, as described in Section~\ref{sec:numerical_heuristic}.

We also used these frontier AI models to seek formal proofs of Conjectures~\ref{conj:qmiep_generals} and~\ref{conj:high-probability-synthesis}.
With the very limited token budget available to us, we did not obtain a complete proof of either conjecture.
We believe both conjectures are simple enough for current models to prove with a sufficiently large token budget.
We plan to continue working toward proofs of both conjectures regardless of how much AI assistance we use.

We relied mainly on ChatGPT and partly on Claude to enhance the manuscript overall.
This included polishing sentences, correcting grammatical errors and typos, and preparing figures and circuit diagrams with TikZ and \texttt{qcircuit}.

We reviewed and validated the entire manuscript and take responsibility for its presentation.

\section*{Data availability}

Software implementations of the synthesis algorithms and code 
to reproduce all numerical results in this paper are available at \artifactlink.

\bibliographystyle{quantum}
\bibliography{references}

@preamble{ "\MakeDoiLinksExplicit" }

@article{shende2004,
  author  = {Shende, Vivek V. and Markov, Igor L. and Bullock, Stephen S.},
  title   = {Minimal universal two-qubit controlled-{NOT}-based circuits},
  journal = {Phys. Rev. A},
  volume  = {69},
  number  = {6},
  pages   = {062321},
  year    = {2004},
  doi     = {10.1103/PhysRevA.69.062321},
  eprint  = {quant-ph/0308033},
  archivePrefix = {arXiv}
}

@article{shende2004b,
  author  = {Shende, Vivek V. and Bullock, Stephen S. and Markov, Igor L.},
  title   = {Recognizing small-circuit structure in two-qubit operators},
  journal = {Phys. Rev. A},
  volume  = {70},
  number  = {1},
  pages   = {012310},
  year    = {2004},
  doi     = {10.1103/PhysRevA.70.012310},
  eprint  = {quant-ph/0308045},
  archivePrefix = {arXiv}
}

@article{shende2004c,
  author  = {Shende, Vivek V. and Markov, Igor L. and Bullock, Stephen S.},
  title   = {Finding small two-qubit circuits},
  journal = {Proc. SPIE},
  volume  = {5436},
  pages   = {348},
  year    = {2004},
  doi     = {10.1117/12.542381}
}

@inproceedings{aaronson2007,
  author        = {Aaronson, Scott and Kuperberg, Greg},
  title         = {Quantum versus Classical Proofs and Advice},
  booktitle     = {Twenty-Second Annual IEEE Conference on Computational Complexity (CCC'07)},
  pages         = {115--128},
  year          = {2007},
  doi           = {10.1109/CCC.2007.27},
  eprint        = {quant-ph/0604056},
  archivePrefix = {arXiv}
}

@article{shende2006,
  author  = {Shende, Vivek V. and Bullock, Stephen S. and Markov, Igor L.},
  title   = {Synthesis of quantum-logic circuits},
  journal = {IEEE Trans. Comput.-Aided Des. Integr. Circuits Syst.},
  volume  = {25},
  number  = {6},
  pages   = {1000--1010},
  year    = {2006},
  doi     = {10.1109/TCAD.2005.855930},
  eprint  = {quant-ph/0406176},
  archivePrefix = {arXiv}
}

@article{devos2016,
  author  = {De Vos, Alexis and De Baerdemacker, Stijn},
  title   = {Block-{ZXZ} synthesis of an arbitrary quantum circuit},
  journal = {Phys. Rev. A},
  volume  = {94},
  number  = {5},
  pages   = {052317},
  year    = {2016},
  doi     = {10.1103/PhysRevA.94.052317},
  eprint  = {1512.07240},
  archivePrefix = {arXiv}
}

@misc{krol2024,
  author        = {Krol, Anna M. and Al-Ars, Zaid},
  title         = {Beyond Quantum {S}hannon: Circuit Construction for General $n$-Qubit Gates Based on Block {ZXZ}-Decomposition},
  year          = {2024},
  eprint        = {2403.13692},
  archivePrefix = {arXiv},
  primaryClass  = {quant-ph},
  doi           = {10.48550/arXiv.2403.13692}
}

@misc{qiskit2024,
  author        = {Javadi-Abhari, Ali and Treinish, Matthew and Krsulich, Kevin and
                   Wood, Christopher J. and Lishman, Jake and Gacon, Julien and
                   Martiel, Simon and Nation, Paul D. and Bishop, Lev S. and
                   Cross, Andrew W. and Johnson, Blake R. and Gambetta, Jay M.},
  title         = {Quantum computing with {Q}iskit},
  year          = {2024},
  doi           = {10.48550/arXiv.2405.08810},
  eprint        = {2405.08810},
  archivePrefix = {arXiv},
  primaryClass  = {quant-ph}
}

@inproceedings{younis2021qfast,
  author        = {Younis, Ed and Sen, Koushik and Yelick, Katherine and Iancu, Costin},
  title         = {{QFAST}: Conflating Search and Numerical Optimization for Scalable Quantum Circuit Synthesis},
  booktitle     = {2021 IEEE International Conference on Quantum Computing and Engineering (QCE)},
  pages         = {232--243},
  year          = {2021},
  doi           = {10.1109/QCE52317.2021.00041},
  eprint        = {2103.07093},
  archivePrefix = {arXiv}
}

@inproceedings{davis2020qsearch,
  author        = {Davis, Marc G. and Smith, Ethan and Tudor, Ana and Sen, Koushik and Siddiqi, Irfan and Iancu, Costin},
  title         = {Towards Optimal Topology Aware Quantum Circuit Synthesis},
  booktitle     = {2020 IEEE International Conference on Quantum Computing and Engineering (QCE)},
  pages         = {223--234},
  year          = {2020},
  doi           = {10.1109/QCE49297.2020.00036}
}

@article{rakyta2022approaching,
  author        = {Rakyta, P\'eter and Zimbor\'as, Zolt\'an},
  title         = {Approaching the theoretical limit in quantum gate decomposition},
  journal       = {Quantum},
  volume        = {6},
  pages         = {710},
  year          = {2022},
  doi           = {10.22331/q-2022-05-11-710},
  eprint        = {2109.06770},
  archivePrefix = {arXiv}
}

@misc{wierichs2025brickwall,
  author        = {Wierichs, David and Kottmann, Korbinian and Killoran, Nathan},
  title         = {Unitary synthesis with optimal brick wall circuits},
  year          = {2025},
  eprint        = {2511.16736},
  archivePrefix = {arXiv},
  primaryClass  = {quant-ph},
  doi           = {10.48550/arXiv.2511.16736}
}

@misc{wierichs2025,
  author        = {Wierichs, David and West, Maxwell and Forestano, Roy T. and Cerezo, M. and Killoran, Nathan},
  title         = {Recursive {Cartan} Decompositions for Unitary Synthesis},
  year          = {2025},
  eprint        = {2503.19014},
  archivePrefix = {arXiv},
  primaryClass  = {quant-ph},
  doi           = {10.48550/arXiv.2503.19014}
}

@article{mottonen2004,
  author  = {M\"ott\"onen, Mikko and Vartiainen, Juha J. and Bergholm, Ville and Salomaa, Martti M.},
  title   = {Quantum Circuits for General Multiqubit Gates},
  journal = {Phys. Rev. Lett.},
  volume  = {93},
  pages   = {130502},
  year    = {2004},
  doi     = {10.1103/PhysRevLett.93.130502},
  eprint  = {quant-ph/0404089},
  archivePrefix = {arXiv}
}

@article{mottonen2005,
  author  = {M\"ott\"onen, Mikko and Vartiainen, Juha J. and Bergholm, Ville and Salomaa, Martti M.},
  title   = {Transformation of quantum states using uniformly controlled rotations},
  journal = {Quantum Inf. Comput.},
  volume  = {5},
  number  = {6},
  pages   = {467--473},
  year    = {2005},
  doi     = {10.26421/QIC5.6-5},
  eprint  = {quant-ph/0407010},
  archivePrefix = {arXiv}
}

@article{bergholm2005,
  author  = {Bergholm, Ville and Vartiainen, Juha J. and M\"ott\"onen, Mikko and Salomaa, Martti M.},
  title   = {Quantum circuits with uniformly controlled one-qubit gates},
  journal = {Phys. Rev. A},
  volume  = {71},
  number  = {5},
  pages   = {052330},
  year    = {2005},
  doi     = {10.1103/PhysRevA.71.052330},
  eprint  = {quant-ph/0410066},
  archivePrefix = {arXiv}
}

@article{barenco1995,
  author  = {Barenco, Adriano and Bennett, Charles H. and Cleve, Richard and DiVincenzo, David P. and Margolus, Norman and Shor, Peter and Sleator, Tycho and Smolin, John A. and Weinfurter, Harald},
  title   = {Elementary gates for quantum computation},
  journal = {Phys. Rev. A},
  volume  = {52},
  number  = {5},
  pages   = {3457--3467},
  year    = {1995},
  doi     = {10.1103/PhysRevA.52.3457},
  eprint  = {quant-ph/9503016},
  archivePrefix = {arXiv}
}

@misc{knill1995,
  author        = {Knill, Emanuel},
  title         = {Approximation by Quantum Circuits},
  year          = {1995},
  eprint        = {quant-ph/9508006},
  archivePrefix = {arXiv},
  doi           = {10.48550/arXiv.quant-ph/9508006}
}

@misc{tucci1999,
  author        = {Tucci, Robert R.},
  title         = {A Rudimentary Quantum Compiler (2nd ed.)},
  year          = {1999},
  eprint        = {quant-ph/9902062},
  archivePrefix = {arXiv},
  doi           = {10.48550/arXiv.quant-ph/9902062}
}

@article{cybenko2001,
  author  = {Cybenko, George},
  title   = {Reducing quantum computations to elementary unitary operations},
  journal = {Comput. Sci. Eng.},
  volume  = {3},
  number  = {2},
  pages   = {27--32},
  year    = {2001},
  doi     = {10.1109/5992.908999}
}

@misc{aho2003,
  author        = {Aho, Alfred V. and Svore, Krysta M.},
  title         = {Compiling Quantum Circuits using the {Palindrome} Transform},
  year          = {2003},
  eprint        = {quant-ph/0311008},
  archivePrefix = {arXiv},
  doi           = {10.48550/arXiv.quant-ph/0311008}
}

@article{vartiainen2004,
  author  = {Vartiainen, Juha J. and M\"ott\"onen, Mikko and Salomaa, Martti M.},
  title   = {Efficient decomposition of quantum gates},
  journal = {Phys. Rev. Lett.},
  volume  = {92},
  number  = {17},
  pages   = {177902},
  year    = {2004},
  doi     = {10.1103/PhysRevLett.92.177902},
  eprint  = {quant-ph/0312218},
  archivePrefix = {arXiv}
}

@incollection{mottonen2006,
  author        = {M\"ott\"onen, Mikko and Vartiainen, Juha J.},
  title         = {Decompositions of general quantum gates},
  booktitle     = {Trends in Quantum Computing Research},
  editor        = {Shannon, S.},
  publisher     = {Nova Science Publishers},
  address       = {New York},
  chapter       = {7},
  pages         = {149},
  year          = {2006},
  eprint        = {quant-ph/0504100},
  archivePrefix = {arXiv},
  doi           = {10.48550/arXiv.quant-ph/0504100}
}

@article{krol2022,
  author  = {Krol, Anna M. and Sarkar, Aritra and Ashraf, Imran and Al-Ars, Zaid and Bertels, Koen},
  title   = {Efficient Decomposition of Unitary Matrices in Quantum Circuit Compilers},
  journal = {Appl. Sci.},
  volume  = {12},
  number  = {2},
  pages   = {759},
  year    = {2022},
  doi     = {10.3390/app12020759},
  eprint  = {2101.02993},
  archivePrefix = {arXiv}
}

@article{mansky2023,
  author  = {Mansky, Maximilian Balthasar and Londo{\~n}o Castillo, Santiago and Ramos Puigvert, Victor and Linnhoff-Popien, Claudia},
  title   = {Near-optimal quantum circuit construction via {Cartan} decomposition},
  journal = {Phys. Rev. A},
  volume  = {108},
  number  = {5},
  pages   = {052607},
  year    = {2023},
  doi     = {10.1103/PhysRevA.108.052607},
  eprint  = {2212.12934},
  archivePrefix = {arXiv}
}

@article{sun2023,
  author  = {Sun, Xiaoming and Tian, Guojing and Yang, Shuai and Yuan, Pei and Zhang, Shengyu},
  title   = {Asymptotically Optimal Circuit Depth for Quantum State Preparation and General Unitary Synthesis},
  journal = {IEEE Trans. Comput.-Aided Des. Integr. Circuits Syst.},
  volume  = {42},
  number  = {10},
  pages   = {3301--3314},
  year    = {2023},
  doi     = {10.1109/TCAD.2023.3244885},
  eprint  = {2108.06150},
  archivePrefix = {arXiv}
}

@article{araujo2024,
  author  = {Araujo, Israel F. and Blank, Carsten and Araujo, Ismael C. S. and da Silva, Adenilton J.},
  title   = {Low-Rank Quantum State Preparation},
  journal = {IEEE Trans. Comput.-Aided Des. Integr. Circuits Syst.},
  volume  = {43},
  number  = {1},
  pages   = {161--170},
  year    = {2024},
  doi     = {10.1109/TCAD.2023.3297972},
  eprint  = {2111.03132},
  archivePrefix = {arXiv}
}

@article{znidaric2008,
  author    = {{\v{Z}}nidari{\v{c}}, Marko and Giraud, Olivier and Georgeot, Bertrand},
  title     = {Optimal number of controlled-{NOT} gates to generate a three-qubit state},
  journal   = {Phys. Rev. A},
  volume    = {77},
  number    = {3},
  pages     = {032320},
  numpages  = {5},
  year      = {2008},
  month     = mar,
  publisher = {American Physical Society},
  doi       = {10.1103/PhysRevA.77.032320},
  url       = {https://link.aps.org/doi/10.1103/PhysRevA.77.032320}
}

@article{plesch2011,
  title = {Quantum-state preparation with universal gate decompositions},
  author = {Plesch, Martin and Brukner, \ifmmode \check{C}\else \v{C}\fi{}aslav},
  journal = {Phys. Rev. A},
  volume = {83},
  issue = {3},
  pages = {032302},
  numpages = {5},
  year = {2011},
  month = {Mar},
  publisher = {American Physical Society},
  doi = {10.1103/PhysRevA.83.032302},
  url = {https://link.aps.org/doi/10.1103/PhysRevA.83.032302}
}

@article{iten2016,
  title = {Quantum circuits for isometries},
  author = {Iten, Raban and Colbeck, Roger and Kukuljan, Ivan and Home, Jonathan and Christandl, Matthias},
  journal = {Phys. Rev. A},
  volume = {93},
  issue = {3},
  pages = {032318},
  numpages = {19},
  year = {2016},
  month = {Mar},
  publisher = {American Physical Society},
  doi = {10.1103/PhysRevA.93.032318},
  url = {https://link.aps.org/doi/10.1103/PhysRevA.93.032318}
}

@article{malvetti2021,
  title = {Quantum {C}ircuits for {S}parse {I}sometries},
  author = {Malvetti, Emanuel and Iten, Raban and Colbeck, Roger},
  journal = {Quantum},
  volume = {5},
  pages = {412},
  year = {2021},
  month = mar,
  doi = {10.22331/q-2021-03-15-412},
  url = {https://doi.org/10.22331/q-2021-03-15-412}
}

@book{nielsen2010,
  author    = {Nielsen, Michael A. and Chuang, Isaac L.},
  title     = {Quantum Computation and Quantum Information},
  edition   = {10th Anniversary},
  publisher = {Cambridge University Press},
  address   = {Cambridge},
  year      = {2010},
  doi       = {10.1017/CBO9780511976667}
}

@book{horn2013,
  author    = {Horn, Roger A. and Johnson, Charles R.},
  title     = {Matrix Analysis},
  edition   = {2nd},
  publisher = {Cambridge University Press},
  address   = {Cambridge},
  year      = {2013},
  doi       = {10.1017/CBO9781139020411},
  isbn      = {9780521839402}
}

@article{vidal2004,
  author  = {Vidal, Guifr\'e and Dawson, Christopher M.},
  title   = {Universal quantum circuit for two-qubit transformations with three controlled-{NOT} gates},
  journal = {Phys. Rev. A},
  volume  = {69},
  number  = {1},
  pages   = {010301},
  year    = {2004},
  doi     = {10.1103/PhysRevA.69.010301},
  eprint  = {quant-ph/0307177},
  archivePrefix = {arXiv}
}

@article{vatan2004,
  author  = {Vatan, Farrokh and Williams, Colin},
  title   = {Optimal quantum circuits for general two-qubit gates},
  journal = {Phys. Rev. A},
  volume  = {69},
  number  = {3},
  pages   = {032315},
  year    = {2004},
  doi     = {10.1103/PhysRevA.69.032315},
  eprint  = {quant-ph/0308006},
  archivePrefix = {arXiv}
}

@article{zhang2003geometric,
  author  = {Zhang, Jun and Vala, Jiri and Sastry, Shankar and Whaley, K. Birgitta},
  title   = {Geometric theory of nonlocal two-qubit operations},
  journal = {Phys. Rev. A},
  volume  = {67},
  number  = {4},
  pages   = {042313},
  year    = {2003},
  doi     = {10.1103/PhysRevA.67.042313},
  eprint  = {quant-ph/0209120},
  archivePrefix = {arXiv}
}

@article{feynman1982,
  author  = {Feynman, Richard P.},
  title   = {Simulating physics with computers},
  journal = {Int. J. Theor. Phys.},
  volume  = {21},
  number  = {6--7},
  pages   = {467--488},
  year    = {1982},
  doi     = {10.1007/BF02650179}
}

@article{lloyd1996,
  author  = {Lloyd, Seth},
  title   = {Universal quantum simulators},
  journal = {Science},
  volume  = {273},
  number  = {5278},
  pages   = {1073--1078},
  year    = {1996},
  doi     = {10.1126/science.273.5278.1073}
}

@article{berry2015,
  author  = {Berry, Dominic W. and Childs, Andrew M. and Cleve, Richard and Kothari, Robin and Somma, Rolando D.},
  title   = {Simulating {H}amiltonian dynamics with a truncated {T}aylor series},
  journal = {Phys. Rev. Lett.},
  volume  = {114},
  number  = {9},
  pages   = {090502},
  year    = {2015},
  doi     = {10.1103/PhysRevLett.114.090502},
  eprint  = {1412.4687},
  archivePrefix = {arXiv}
}

@article{low2017,
  author  = {Low, Guang Hao and Chuang, Isaac L.},
  title   = {Optimal {H}amiltonian simulation by quantum signal processing},
  journal = {Phys. Rev. Lett.},
  volume  = {118},
  number  = {1},
  pages   = {010501},
  year    = {2017},
  doi     = {10.1103/PhysRevLett.118.010501},
  eprint  = {1606.02685},
  archivePrefix = {arXiv}
}

@article{low2019,
  author  = {Low, Guang Hao and Chuang, Isaac L.},
  title   = {{H}amiltonian simulation by qubitization},
  journal = {Quantum},
  volume  = {3},
  pages   = {163},
  year    = {2019},
  doi     = {10.22331/q-2019-07-12-163},
  eprint  = {1610.06546},
  archivePrefix = {arXiv}
}

@inproceedings{gilyen2019,
  author        = {Gily\'en, Andr\'as and Su, Yuan and Low, Guang Hao and Wiebe, Nathan},
  title         = {Quantum singular value transformation and beyond: exponential improvements for quantum matrix arithmetics},
  booktitle     = {Proc. 51st Annu. ACM SIGACT Symp. Theory of Computing (STOC)},
  pages         = {193--204},
  year          = {2019},
  doi           = {10.1145/3313276.3316366},
  eprint        = {1806.01838},
  archivePrefix = {arXiv}
}

@article{martyn2021,
  author  = {Martyn, John M. and Rossi, Zane M. and Tan, Andrew K. and Chuang, Isaac L.},
  title   = {Grand unification of quantum algorithms},
  journal = {PRX Quantum},
  volume  = {2},
  number  = {4},
  pages   = {040203},
  year    = {2021},
  doi     = {10.1103/PRXQuantum.2.040203},
  eprint  = {2105.02859},
  archivePrefix = {arXiv}
}

@article{shor1997,
  author  = {Shor, Peter W.},
  title   = {Polynomial-time algorithms for prime factorization and discrete logarithms on a quantum computer},
  journal = {SIAM J. Comput.},
  volume  = {26},
  number  = {5},
  pages   = {1484--1509},
  year    = {1997},
  doi     = {10.1137/S0097539795293172},
  eprint  = {quant-ph/9508027},
  archivePrefix = {arXiv}
}

@inproceedings{grover1996,
  author        = {Grover, Lov K.},
  title         = {A fast quantum mechanical algorithm for database search},
  booktitle     = {Proc. 28th Annu. ACM Symp. Theory of Computing (STOC)},
  pages         = {212--219},
  year          = {1996},
  doi           = {10.1145/237814.237866},
  eprint        = {quant-ph/9605043},
  archivePrefix = {arXiv}
}

@article{peruzzo2014,
  author  = {Peruzzo, Alberto and McClean, Jarrod and Shadbolt, Peter and Yung, Man-Hong and Zhou, Xiao-Qi and Love, Peter J. and Aspuru-Guzik, Al\'an and O'Brien, Jeremy L.},
  title   = {A variational eigenvalue solver on a photonic quantum processor},
  journal = {Nat. Commun.},
  volume  = {5},
  pages   = {4213},
  year    = {2014},
  doi     = {10.1038/ncomms5213},
  eprint  = {1304.3061},
  archivePrefix = {arXiv}
}

@misc{farhi2014,
  author        = {Farhi, Edward and Goldstone, Jeffrey and Gutmann, Sam},
  title         = {A quantum approximate optimization algorithm},
  year          = {2014},
  eprint        = {1411.4028},
  archivePrefix = {arXiv},
  primaryClass  = {quant-ph},
  doi           = {10.48550/arXiv.1411.4028}
}

@article{cerezo2021,
  author  = {Cerezo, M. and Arrasmith, Andrew and Babbush, Ryan and Benjamin, Simon C. and Endo, Suguru and Fujii, Keisuke and McClean, Jarrod R. and Mitarai, Kosuke and Yuan, Xiao and Cincio, Lukasz and Coles, Patrick J.},
  title   = {Variational quantum algorithms},
  journal = {Nat. Rev. Phys.},
  volume  = {3},
  number  = {9},
  pages   = {625--644},
  year    = {2021},
  doi     = {10.1038/s42254-021-00348-9},
  eprint  = {2012.09265},
  archivePrefix = {arXiv}
}

@article{abbas2024,
  author  = {Abbas, Amira and Ambainis, Andris and Augustino, Brandon and B\"artschi, Andreas and Buhrman, Harry and Coffrin, Carleton and others},
  title   = {Challenges and opportunities in quantum optimization},
  journal = {Nat. Rev. Phys.},
  volume  = {6},
  number  = {12},
  pages   = {718--735},
  year    = {2024},
  doi     = {10.1038/s42254-024-00770-9},
  eprint  = {2312.02279},
  archivePrefix = {arXiv}
}

@article{downing1956,
  author  = {Downing, Jr., A. C. and Householder, A. S.},
  title   = {Some Inverse Characteristic Value Problems},
  journal = {Journal of the ACM},
  volume  = {3},
  number  = {3},
  pages   = {203--207},
  year    = {1956},
  doi     = {10.1145/320831.320839}
}

@article{HADELER196965,
title = {Multiplikative inverse eigenwertprobleme},
journal = {Linear Algebra and its Applications},
volume = {2},
number = {1},
pages = {65-86},
year = {1969},
issn = {0024-3795},
doi = {10.1016/0024-3795(69)90008-1},
url = {https://www.sciencedirect.com/science/article/pii/0024379569900081},
author = {K.P. Hadeler}
}

@article{FRIEDLAND1975127,
title = {On inverse multiplicative eigenvalue problems for matrices},
journal = {Linear Algebra and its Applications},
volume = {12},
number = {2},
pages = {127-137},
year = {1975},
issn = {0024-3795},
doi = {10.1016/0024-3795(75)90061-0},
url = {https://www.sciencedirect.com/science/article/pii/0024379575900610},
author = {Shmuel Friedland}
}

@article{chu1998iep,
  author  = {Chu, Moody T.},
  title   = {Inverse Eigenvalue Problems},
  journal = {SIAM Review},
  volume  = {40},
  number  = {1},
  pages   = {1--39},
  year    = {1998},
  doi     = {10.1137/S0036144596303984}
}

@article{friedland1987numerical,
  author  = {Friedland, S. and Nocedal, J. and Overton, M. L.},
  title   = {The Formulation and Analysis of Numerical Methods for Inverse Eigenvalue Problems},
  journal = {SIAM Journal on Numerical Analysis},
  volume  = {24},
  number  = {3},
  pages   = {634--667},
  year    = {1987},
  doi     = {10.1137/0724043}
}

@article{chen1996least,
  author  = {Chen, Xuzhou and Chu, Moody T.},
  title   = {On the Least Squares Solution of Inverse Eigenvalue Problems},
  journal = {SIAM Journal on Numerical Analysis},
  volume  = {33},
  number  = {6},
  pages   = {2417--2430},
  year    = {1996},
  doi     = {10.1137/S0036142994264742}
}

@article{chu1990projected,
  author  = {Chu, Moody T. and Driessel, Kenneth R.},
  title   = {The Projected Gradient Method for Least Squares Matrix Approximations with Spectral Constraints},
  journal = {SIAM Journal on Numerical Analysis},
  volume  = {27},
  number  = {4},
  pages   = {1050--1060},
  year    = {1990},
  doi     = {10.1137/0727062}
}

@article{li2026,
  author  = {Li, Zexian and Zhang, Guofeng and Zhang, Xiao-Ming},
  title   = {Reducing {\cnot{}} Counts for State Preparation and Block Encoding via Diagonal Matrix Migration},
  journal = {IEEE Trans. Comput.-Aided Des. Integr. Circuits Syst.},
  volume  = {},
  number  = {},
  pages   = {1--1},
  year    = {2026},
  doi     = {10.1109/TCAD.2026.3707423}
}

@article{fuhr2015,
  author  = {F\"uhr, Hartmut and Rzeszotnik, Ziemowit},
  title   = {On biunimodular vectors for unitary matrices},
  journal = {Linear Algebra Appl.},
  volume  = {484},
  pages   = {86--129},
  year    = {2015},
  doi     = {10.1016/j.laa.2015.06.019}
}

@article{devos2014,
  author  = {De Vos, Alexis and De Baerdemacker, Stijn},
  title   = {Scaling a Unitary Matrix},
  journal = {Open Syst. Inf. Dyn.},
  volume  = {21},
  number  = {4},
  pages   = {1450013},
  year    = {2014},
  doi     = {10.1142/S1230161214500139}
}

@article{ivashkov2024,
  author  = {Ivashkov, Petr and Uchehara, Gideon and Jiang, Liang and Wang, Derek S. and Seif, Alireza},
  title   = {High-Fidelity, Multiqubit Generalized Measurements with Dynamic Circuits},
  journal = {PRX Quantum},
  volume  = {5},
  number  = {3},
  pages   = {030315},
  year    = {2024},
  doi     = {10.1103/PRXQuantum.5.030315}
}

@article{davies1970,
  author  = {Davies, E. B. and Lewis, J. T.},
  title   = {An operational approach to quantum probability},
  journal = {Commun. Math. Phys.},
  volume  = {17},
  pages   = {239--260},
  year    = {1970},
  doi     = {10.1007/BF01647093}
}

@article{koh2026,
  author  = {Koh, Jin Ming and Xue, Wen-Tan and Tai, Tommy and Koh, Dax Enshan and Lee, Ching Hua},
  title   = {Interacting non-{Hermitian} edge and cluster bursts on a digital quantum processor},
  journal = {Nat. Commun.},
  volume  = {17},
  pages   = {9120},
  year    = {2026},
  doi     = {10.1038/s41467-026-75904-x}
}

@article{gosset2014,
  author  = {Gosset, David and Kliuchnikov, Vadym and Mosca, Michele and Russo, Vincent},
  title   = {An algorithm for the {T}-count},
  journal = {Quantum Inf. Comput.},
  volume  = {14},
  number  = {15--16},
  pages   = {1261--1276},
  year    = {2014},
  doi     = {10.26421/QIC14.15-16-1},
  eprint  = {1308.4134},
  archivePrefix = {arXiv}
}

@article{heyfron2019,
  author  = {Heyfron, Luke E. and Campbell, Earl T.},
  title   = {An efficient quantum compiler that reduces {T} count},
  journal = {Quantum Sci. Technol.},
  volume  = {4},
  number  = {1},
  pages   = {015004},
  year    = {2019},
  doi     = {10.1088/2058-9565/aad604},
  eprint  = {1712.01557},
  archivePrefix = {arXiv}
}

@article{amy2013,
  author  = {Amy, Matthew and Maslov, Dmitri and Mosca, Michele and Roetteler, Martin},
  title   = {A Meet-in-the-Middle Algorithm for Fast Synthesis of Depth-Optimal Quantum Circuits},
  journal = {IEEE Trans. Comput.-Aided Des. Integr. Circuits Syst.},
  volume  = {32},
  number  = {6},
  pages   = {818--830},
  year    = {2013},
  doi     = {10.1109/TCAD.2013.2244643},
  eprint  = {1206.0758},
  archivePrefix = {arXiv}
}

@misc{ibmfez2025,
  author       = {{IBM Quantum}},
  title        = {{IBM Fez} calibration snapshot},
  year         = {2025},
  howpublished = {Qiskit IBM Runtime 0.36.0, \url{https://github.com/Qiskit/qiskit-ibm-runtime/blob/10ebe5af87aabd930c02a87504ec0aef948d5acb/qiskit_ibm_runtime/fake_provider/backends/fez/props_fez.json}},
  url          = {https://github.com/Qiskit/qiskit-ibm-runtime/blob/10ebe5af87aabd930c02a87504ec0aef948d5acb/qiskit_ibm_runtime/fake_provider/backends/fez/props_fez.json},
  note         = {Calibration snapshot dated January 23, 2025}
}

@misc{zhang2026,
  author        = {Zhang, Xiao-Ming and Rattew, Arthur G. and Wu, Bujiao and Styliaris, Georgios and Sun, Xiaoming and Koczor, B{\'a}lint and Yuan, Xiao},
  title         = {From Bits to Qubits: The Theory and Practice of Quantum Data Encoding},
  year          = {2026},
  eprint        = {2609.08058},
  archivePrefix = {arXiv},
  primaryClass  = {quant-ph},
  doi           = {10.48550/arXiv.2609.08058}
}

@article{kliuchnikov2013exact,
  doi           = {10.26421/QIC13.7-8-4},
  author        = {Kliuchnikov, Vadym and Maslov, Dmitri and Mosca, Michele},
  title         = {Fast and efficient exact synthesis of single qubit unitaries generated by {Clifford} and {T} gates},
  journal       = {Quantum Inf. Comput.},
  volume        = {13},
  number        = {7--8},
  pages         = {607--630},
  year          = {2013},
  eprint        = {1206.5236},
  archivePrefix = {arXiv},
  url           = {https://arxiv.org/abs/1206.5236}
}

@article{ross2016,
  doi           = {10.26421/QIC16.11-12-1},
  author        = {Ross, Neil J. and Selinger, Peter},
  title         = {Optimal ancilla-free {Clifford+T} approximation of z-rotations},
  journal       = {Quantum Inf. Comput.},
  volume        = {16},
  number        = {11--12},
  pages         = {901--953},
  year          = {2016},
  eprint        = {1403.2975},
  archivePrefix = {arXiv},
  url           = {https://arxiv.org/abs/1403.2975}
}

@inproceedings{hao2026,
  author        = {Hao, Tianyi and Xu, Amanda and Tannu, Swamit},
  title         = {Reducing {T} Gates with Unitary Synthesis},
  booktitle     = {Proceedings of the 31st ACM International Conference on Architectural Support for Programming Languages and Operating Systems, Volume 2},
  year          = {2026},
  publisher     = {ACM},
  doi           = {10.1145/3779212.3790210},
  eprint        = {2503.15843},
  archivePrefix = {arXiv},
  url           = {https://arxiv.org/abs/2503.15843}
}

@article{amy2019,
  author        = {Amy, Matthew and Mosca, Michele},
  title         = {{T}-Count Optimization and {Reed--Muller} Codes},
  journal       = {IEEE Trans. Inf. Theory},
  volume        = {65},
  number        = {8},
  pages         = {4771--4784},
  year          = {2019},
  doi           = {10.1109/TIT.2019.2906374},
  eprint        = {1601.07363},
  archivePrefix = {arXiv}
}

@article{kissinger2020,
  author        = {Kissinger, Aleks and van de Wetering, John},
  title         = {Reducing the number of non-{Clifford} gates in quantum circuits},
  journal       = {Phys. Rev. A},
  volume        = {102},
  pages         = {022406},
  year          = {2020},
  doi           = {10.1103/PhysRevA.102.022406},
  eprint        = {1903.10477},
  archivePrefix = {arXiv}
}

@article{amy2014,
  author        = {Amy, Matthew and Maslov, Dmitri and Mosca, Michele},
  title         = {Polynomial-Time {T}-Depth Optimization of {Clifford+T} Circuits via Matroid Partitioning},
  journal       = {IEEE Trans. Comput.-Aided Des. Integr. Circuits Syst.},
  volume        = {33},
  number        = {10},
  pages         = {1476--1489},
  year          = {2014},
  doi           = {10.1109/TCAD.2014.2341953},
  eprint        = {1303.2042},
  archivePrefix = {arXiv}
}

@article{low2024,
  author        = {Low, Guang Hao and Kliuchnikov, Vadym and Schaeffer, Luke},
  title         = {Trading {T} gates for dirty qubits in state preparation and unitary synthesis},
  journal       = {Quantum},
  volume        = {8},
  pages         = {1375},
  year          = {2024},
  doi           = {10.22331/q-2024-06-17-1375},
  eprint        = {1812.00954},
  archivePrefix = {arXiv}
}

@article{gosset2026,
  author        = {Gosset, David and Kothari, Robin and Wu, Kewen},
  title         = {Quantum state preparation with optimal {T}-count},
  journal       = {Quantum},
  volume        = {10},
  pages         = {2168},
  year          = {2026},
  doi           = {10.22331/q-2026-07-22-2168},
  eprint        = {2411.04790},
  archivePrefix = {arXiv}
}

@article{tan2026,
  author        = {Tan, Xinyu},
  title         = {Unitary Synthesis with Fewer {T} Gates},
  journal       = {PRX Quantum},
  volume        = {7},
  number        = {3},
  pages         = {033058},
  year          = {2026},
  doi           = {10.1103/pxhd-9s9q},
  url           = {https://journals.aps.org/prxquantum/abstract/10.1103/pxhd-9s9q}
}

@article{sanders2019,
  author        = {Sanders, Yuval R. and Low, Guang Hao and Scherer, Artur and Berry, Dominic W.},
  title         = {Black-Box Quantum State Preparation without Arithmetic},
  journal       = {Phys. Rev. Lett.},
  volume        = {122},
  pages         = {020502},
  year          = {2019},
  doi           = {10.1103/PhysRevLett.122.020502},
  eprint        = {1807.03206},
  archivePrefix = {arXiv}
}

@article{bausch2022,
  author        = {Bausch, Johannes},
  title         = {Fast Black-Box Quantum State Preparation},
  journal       = {Quantum},
  volume        = {6},
  pages         = {773},
  year          = {2022},
  doi           = {10.22331/q-2022-08-04-773},
  eprint        = {2009.10709},
  archivePrefix = {arXiv}
}

@article{paetznick2014,
  doi           = {10.26421/QIC14.15-16-2},
  author        = {Paetznick, Adam and Svore, Krysta M.},
  title         = {Repeat-Until-Success: Non-deterministic decomposition of single-qubit unitaries},
  journal       = {Quantum Inf. Comput.},
  volume        = {14},
  number        = {15--16},
  pages         = {1277--1301},
  year          = {2014},
  eprint        = {1311.1074},
  archivePrefix = {arXiv},
  url           = {https://arxiv.org/abs/1311.1074}
}

@article{bocharov2015rus,
  author        = {Bocharov, Alex and Roetteler, Martin and Svore, Krysta M.},
  title         = {Efficient Synthesis of Universal Repeat-Until-Success Quantum Circuits},
  journal       = {Phys. Rev. Lett.},
  volume        = {114},
  pages         = {080502},
  year          = {2015},
  doi           = {10.1103/PhysRevLett.114.080502},
  eprint        = {1404.5320},
  archivePrefix = {arXiv}
}

@article{selinger2013,
  author        = {Selinger, Peter},
  title         = {Quantum circuits of {T}-depth one},
  journal       = {Phys. Rev. A},
  volume        = {87},
  pages         = {042302},
  year          = {2013},
  doi           = {10.1103/PhysRevA.87.042302},
  eprint        = {1210.0974},
  archivePrefix = {arXiv}
}

@article{gidney2018,
  author        = {Gidney, Craig},
  title         = {Halving the cost of quantum addition},
  journal       = {Quantum},
  volume        = {2},
  pages         = {74},
  year          = {2018},
  doi           = {10.22331/q-2018-06-18-74},
  eprint        = {1709.06648},
  archivePrefix = {arXiv}
}

@article{bravyi2005,
  author        = {Bravyi, Sergey and Kitaev, Alexei},
  title         = {Universal quantum computation with ideal {Clifford} gates and noisy ancillas},
  journal       = {Phys. Rev. A},
  volume        = {71},
  number        = {2},
  pages         = {022316},
  year          = {2005},
  doi           = {10.1103/PhysRevA.71.022316},
  eprint        = {quant-ph/0403025},
  archivePrefix = {arXiv}
}

@article{giles2013,
  author        = {Giles, Brett and Selinger, Peter},
  title         = {Exact synthesis of multiqubit {Clifford+T} circuits},
  journal       = {Phys. Rev. A},
  volume        = {87},
  pages         = {032332},
  year          = {2013},
  doi           = {10.1103/PhysRevA.87.032332},
  eprint        = {1212.0506},
  archivePrefix = {arXiv}
}

@misc{kliuchnikov2013multi,
  author        = {Kliuchnikov, Vadym},
  title         = {Synthesis of unitaries with {Clifford+T} circuits},
  year          = {2013},
  eprint        = {1306.3200},
  archivePrefix = {arXiv},
  primaryClass  = {quant-ph},
  doi           = {10.48550/arXiv.1306.3200}
}

@article{zhang2022depth,
  author        = {Zhang, Xiao-Ming and Li, Tongyang and Yuan, Xiao},
  title         = {Quantum State Preparation with Optimal Circuit Depth: Implementations and Applications},
  journal       = {Phys. Rev. Lett.},
  volume        = {129},
  pages         = {230504},
  year          = {2022},
  doi           = {10.1103/PhysRevLett.129.230504},
  eprint        = {2201.11495},
  archivePrefix = {arXiv}
}

@article{rosenthal2026,
  author        = {Rosenthal, Gregory},
  title         = {Query and Depth Upper Bounds for Quantum Unitaries via {Grover} Search},
  journal       = {Quantum},
  volume        = {10},
  pages         = {2144},
  year          = {2026},
  doi           = {10.22331/q-2026-06-30-2144},
  eprint        = {2111.07992},
  archivePrefix = {arXiv}
}

@book{nocedal2006,
  author    = {Nocedal, Jorge and Wright, Stephen J.},
  title     = {Numerical Optimization},
  edition   = {2nd},
  publisher = {Springer},
  address   = {New York, NY},
  series    = {Springer Series in Operations Research and Financial Engineering},
  year      = {2006},
  doi       = {10.1007/978-0-387-40065-5}
}

@article{welch2014,
  author  = {Welch, Jonathan and Greenbaum, Daniel and Mostame, Sarah and Aspuru-Guzik, Al\'an},
  title   = {Efficient quantum circuits for diagonal unitaries without ancillas},
  journal = {New J. Phys.},
  volume  = {16},
  number  = {3},
  pages   = {033040},
  year    = {2014},
  doi     = {10.1088/1367-2630/16/3/033040},
  eprint  = {1306.3991},
  archivePrefix = {arXiv}
}

@book{kato,
  author = {Kato, Tosio},
  title = {Perturbation Theory for Linear Operators},
  publisher = {Springer},
  year = {1995},
  series = {Classics in Mathematics},
  doi = {10.1007/978-3-642-66282-9}
}

@book{lee2013,
  author    = {Lee, John M.},
  title     = {Introduction to Smooth Manifolds},
  edition   = {2nd},
  series    = {Graduate Texts in Mathematics},
  volume    = {218},
  publisher = {Springer},
  year      = {2013},
  doi       = {10.1007/978-1-4419-9982-5}
}

@article{winkelmann2002,
  author  = {Winkelmann, J{\"o}rg},
  title   = {Generic subgroups of {Lie} groups},
  journal = {Topology},
  volume  = {41},
  number  = {1},
  pages   = {163--181},
  year    = {2002},
  doi     = {10.1016/S0040-9383(00)00029-X}
}

\appendix

\section{Algorithmic Description of Ours}
\label{app:algorithmic-description}

\subsection{Algorithm for \oursanalytic{}}
\label{app:analytic-synthesis}

\begin{algorithm}[H]
\caption{\oursanalytic{} with exact, $D$-equivalent, or $PD$-equivalent output}
\label{alg:analytic-synthesis}
\small
\begin{algorithmic}[1]
\Require $U\in U(2^n)$ with $n\geq3$ for the initial call, and the output $\mathit{target}\in\{\mathrm{exact},\mathrm{D},\mathrm{PD}\}$
\Ensure Return $\mathcal C$, $(\mathcal C,P,D)$, or $(\mathcal C,D)$ according to the output target.
Here $\mathcal C$ is the synthesized quantum circuit, $D$ is a diagonal unitary, and $P=P_0\oplus P_1$ is a multiplexed permutation.
\begin{itemize}
    \item For $\mathit{target}=\mathrm{exact}$, $\mathcal C$ implements $U$ using $\costanalytic{n}$ \cnot{} gates.
    \item For $\mathit{target}=\mathrm{PD}$, $\mathcal C$ implements $U_{PD}=P^\dagger D^\dagger U$ using $\costPDanalytic{n}$ \cnot{} gates.
    \item For $\mathit{target}=\mathrm{D}$, $\mathcal C$ implements $U_D=D^\dagger U$ using $\costanalytic{n}-1$ \cnot{} gates.
\end{itemize}
\Function{SynthesizeAnalytic}{$U,\mathit{target}$}
    \If{$n=2$}
        \If{$\mathit{target}=\mathrm{exact}$}
            \State \Return a three-\cnot{} circuit $\mathcal C$ for $U$~\cite{vidal2004,vatan2004,shende2004}
        \ElsIf{$\mathit{target}=\mathrm{D}$}
            \State Obtain $U=D U_D$ and a two-\cnot{} circuit $\mathcal C$ for $U_D$~\cite{shende2004}
            \State \Return $(\mathcal C,D)$
        \Else \Comment{$\mathit{target}=\mathrm{PD}$}
            \State Obtain the results for $\mathit{target}=\mathrm{D}$ with $P=I_4$
            \State \Return $(\mathcal C,P,D)$
        \EndIf
    \EndIf
    \If{$n=3$ and $\mathit{target}=\mathrm{PD}$}
        \State Use Corollary~\ref{cor:three-qubit-pd-modulus} to obtain $P,D$ and a circuit $\mathcal C$ for $U_{PD}$ of the form in Eq.~\eqref{qc:three-qubit-pd-modulus}
        \State \Return $(\mathcal C,P,D)$
    \EndIf
    \State \label{line:analytic-decompose}Decompose $U$ into $[L,\rgate(\vec\phi_L),H_t,M,\ucgrz,H_t,N,\rgate(\vec\phi_R),O]$ via Eq.~\eqref{qc:block-zxz-cx-absorbed}
    \For{each step $X\to Y$ along $L\to M\to N\to O$}\label{line:analytic-migration}
        \State $(\mathcal C_X,P_X,D_X)\gets\textsc{SynthesizeAnalytic}(X,\mathrm{PD})$
        \State Propagate $D_XP_X$ toward $Y$, reordering rotation parameters (Prop.~\ref{prop:pd-thru-r})
        \State $Y\gets YD_XP_X$
    \EndFor
    \State Form $\mathcal C$ with $\mathcal C_L,\mathcal C_M,\mathcal C_N$ in place of $L,M,N$ and $O$ as its terminal block
    \State Expand $R$ and $\ucgrz$ using Prop.~\ref{prop:ucgrz}
    \If{$\mathit{target}=\mathrm{exact}$}
        \State Replace $O$ in $\mathcal C$ with $\textsc{SynthesizeAnalytic}(O,\mathrm{exact})$
        \State \Return $\mathcal C$
    \ElsIf{$\mathit{target}=\mathrm{D}$}
        \State $(\mathcal C_O,D_O)\gets\textsc{SynthesizeAnalytic}(O,\mathrm{D})$
        \State Replace $O$ in $\mathcal C$ with $\mathcal C_O$
        \State \Return $(\mathcal C,I_2\otimes D_O)$
    \Else \Comment{$\mathit{target}=\mathrm{PD}$}
        \State $(\mathcal C_O,P_O,D_O)\gets\textsc{SynthesizeAnalytic}(O,\mathrm{PD})$
        \State Replace $O$ in $\mathcal C$ with $\mathcal C_O$
        \State \Return $(\mathcal C,I_2\otimes P_O,I_2\otimes D_O)$
    \EndIf \label{line:analytic-output-end}
\EndFunction
\end{algorithmic}
\end{algorithm}

\subsection{Algorithm for \oursheuristic{}}
\label{app:heuristic-synthesis}

\begin{algorithm}[H]
\caption{\oursheuristic{} with exact, $D$-equivalent, or $PD$-equivalent output}
\label{alg:heuristic-synthesis}
\small
\begin{algorithmic}[1]
\Require $U\in U(2^n)$, $n\geq2$, and the output $\mathit{target}\in\{\mathrm{exact},\mathrm{D},\mathrm{PD}\}$
\Ensure On success, return $\mathcal C$, $(\mathcal C,P,D)$, or $(\mathcal C,D)$ according to the output target.
Here $\mathcal C$ is the synthesized quantum circuit, $D$ is a diagonal unitary, and $P=P_0\oplus P_1$ is a multiplexed permutation.
\begin{itemize}
    \item For $\mathit{target}=\mathrm{exact}$, $\mathcal C$ implements $U$ using $\costheuristic{n}$ \cnot{} gates.
    \item For $\mathit{target}=\mathrm{PD}$, $\mathcal C$ implements $U_{PD}=P^\dagger D^\dagger U$ using $\costPDheuristic{n}$ \cnot{} gates.
    \item For $\mathit{target}=\mathrm{D}$, $\mathcal C$ implements $U_D=D^\dagger U$ using $\costheuristic{n}-1$ \cnot{} gates.
\end{itemize}
\Function{SynthesizeHeuristic}{$U,\mathit{target}$}
    \If{$n\leq3$}
        \State \Return $\textsc{SynthesizeAnalytic}(U,\mathit{target})$
    \EndIf
    \If{$\mathit{target}\neq\mathrm{PD}$}
        \State Follow lines~\ref{line:analytic-decompose}--\ref{line:analytic-output-end} of Algorithm~\ref{alg:analytic-synthesis}, recursing with \textsc{SynthesizeHeuristic}
    \Else
        \State Decompose $U$ into $[I\oplus C,H_t,B_1\oplus B_2,H_t,I\oplus A]$ via Eq.~\eqref{qc:pd-modulus-derivation:a}
        \State Find a QMIEP solution $(P_A,D_A)$ for the $(n-1)$-qubit unitary $A$ (Sec.~\ref{sec:numerical_heuristic})
        \State Rewrite the circuit in the form of Eq.~\eqref{qc:pd-modulus-derivation:d}
        \For{each step $X\to Y$ along $L\to M\to N$}\label{line:heuristic-migration}
            \State $(\mathcal C_X,P_X,D_X)\gets\textsc{SynthesizeHeuristic}(X,\mathrm{PD})$
            \State Propagate $D_XP_X$ toward $Y$, reordering rotation parameters (Prop.~\ref{prop:pd-thru-r})
            \State $Y\gets YD_XP_X$
        \EndFor
        \State \label{line:heuristic-synthesize-n}$(\mathcal C_N,D_N)\gets\textsc{SynthesizeHeuristic}(N,\mathrm{D})$
        \State $(\mathcal C_O,P_O,D_O)\gets\textsc{SynthesizeHeuristic}(OD_N,\mathrm{PD})$
        \State Assemble $\mathcal C$ as in Eq.~\eqref{qc:pd-modulus-from-qmiep}, expanding $R$, $\ucgrz$, and $CZ$
        \State $P\gets P_O\oplus(P_A^\dagger P_O)$ and $D\gets D_O\oplus(P_A^\dagger D_A^\dagger D_O P_A)$
        \State \Return $(\mathcal C,P,D)$
    \EndIf
\EndFunction
\end{algorithmic}
\end{algorithm}

\subsection{Algorithm for \oursheuristicopt{}}
\label{app:heuristic-opt-synthesis}

\begin{algorithm}[H]
\caption{\oursheuristicopt{} with exact, $D$-equivalent, or $PD$-equivalent output}
\label{alg:heuristic-opt-synthesis}
\small
\begin{algorithmic}[1]
\Require $U\in U(2^n)$, $n\geq2$, and the output $\mathit{target}\in\{\mathrm{exact},\mathrm{D},\mathrm{PD}\}$
\Ensure On success, return $\mathcal C$, $(\mathcal C,P,D)$, or $(\mathcal C,D)$ according to the output target.
Here $\mathcal C$ is the synthesized quantum circuit, $D$ is a diagonal unitary, and $P=P_0\oplus P_1$ is a multiplexed permutation.
Let $c$ denote the number of \cnot{} gates in $\mathcal C$.
\begin{itemize}
    \item For $\mathit{target}=\mathrm{exact}$, $\mathcal C$ implements $U$, with
    $\costheuristicopt{n}\leq c\leq\costheuristic{n}$.
    \item For $\mathit{target}=\mathrm{PD}$, $\mathcal C$ implements $U_{PD}=P^\dagger D^\dagger U$, with
    $\costDheuristicopt{n}\leq c\leq\costPDheuristic{n}$.
    If the returned permutation is the identity, $\mathcal C$ implements $U_D=D^\dagger U$.
    \item For $\mathit{target}=\mathrm{D}$, $\mathcal C$ implements $U_D=D^\dagger U$, with
    $\costDheuristicopt{n}\leq c\leq\costheuristic{n}-1$.
\end{itemize}
\Function{SynthesizeHeuristicOpt}{$U,\mathit{target}$}
    \If{$n=2$}
        \State \Return $\textsc{SynthesizeAnalytic}(U,\mathit{target})$
    \EndIf
    \State Decompose $U$ into $[I\oplus C,H_t,B_1\oplus B_2,H_t,I\oplus A]$ via Eq.~\eqref{qc:pd-modulus-derivation:a}
    \If{$\mathit{target}\neq\mathrm{exact}$}
        \State Find a QMIEP solution $(P_A,D_A)$ for $A$, trying $P_A=I$ first
    \EndIf
    \If{$\mathit{target}=\mathrm{exact}$ or $(\mathit{target}=\mathrm{D}$ and $P_A\ne I)$}
        \State Rewrite the factors of Eq.~\eqref{qc:pd-modulus-derivation:a} in the form of Eq.~\eqref{qc:block-zxz-cx-absorbed}
        \State Follow lines~\ref{line:analytic-migration}--\ref{line:analytic-output-end} of Algorithm~\ref{alg:analytic-synthesis}, recursing with \textsc{SynthesizeHeuristicOpt}
    \Else
        \State Rewrite the circuit in the form of Eq.~\eqref{qc:pd-modulus-derivation:d}
        \State Follow lines~\ref{line:heuristic-migration}--\ref{line:heuristic-synthesize-n} of Algorithm~\ref{alg:heuristic-synthesis}, recursing with \textsc{SynthesizeHeuristicOpt}
        \If{$\mathit{target}=\mathrm{PD}$}
            \State $(\mathcal C_O,P_O,D_O)\gets\textsc{SynthesizeHeuristicOpt}(OD_N,\mathrm{PD})$
        \Else \Comment{$\mathit{target}=\mathrm{D}$ and $P_A=I$}
            \State $(\mathcal C_O,D_O)\gets\textsc{SynthesizeHeuristicOpt}(OD_N,\mathrm{D})$
        \EndIf
        \State Assemble $\mathcal C$ as in Eq.~\eqref{qc:pd-modulus-from-qmiep}, expanding $R$, $\ucgrz$, and $CZ$
        \If{$\mathit{target}=\mathrm{D}$}
            \State \Return $(\mathcal C,D_O\oplus(D_A^\dagger D_O))$
        \EndIf
        \State $P\gets P_O\oplus(P_A^\dagger P_O)$ and $D\gets D_O\oplus(P_A^\dagger D_A^\dagger D_O P_A)$
        \State \Return $(\mathcal C,P,D)$
    \EndIf
\EndFunction
\end{algorithmic}
\end{algorithm}

\section{Cost Recurrence Solving}
\label{app:recurrence}

As a representative example, we solve the recurrence for the $PD$-equivalent synthesis cost $a_n=\costPDanalytic{n}$.
Corollary~\ref{cor:three-qubit-pd-modulus} and
Eqs.~\eqref{eq:cost:UPD_analytic_recursive} and~\eqref{eq:ucgrz-r-cost} give
\[
    a_3=17, \qquad a_n=4a_{n-1}+3\cdot 2^{n-1}-2 \quad (n\geq 4).
\]
Dividing by $4^n$ and summing the resulting differences gives
\begin{align*}
    \frac{a_n}{4^n}-\frac{17}{64}
    &=\sum_{k=4}^{n}\left(\frac{a_k}{4^k}-\frac{a_{k-1}}{4^{k-1}}\right)
      =\sum_{k=4}^{n}\frac{3\cdot 2^{k-1}-2}{4^k}\\
    &=\frac32\underbrace{\sum_{k=4}^{n}2^{-k}}_{1/8-2^{-n}}
      -2\underbrace{\sum_{k=4}^{n}4^{-k}}_{(1/64-4^{-n})/3}.
\end{align*}
Hence, for $n\geq 3$,
\[
    a_n=\underbrace{\left(\frac{17}{64}+\frac{3}{16}-\frac{1}{96}\right)}_{85/192}4^n
        -\frac32\,2^n+\frac23.
\]
Substituting the formula for $a_{n-1}$ into Eq.~\eqref{eq:cost:U_by_PD_migrate} and using $\costanalytic{3}=19$
gives the closed form for $\costanalytic{n}$ in Eq.~\eqref{eq:analytic-unitary-cost}.
The other cost recurrences in this paper can be solved similarly.

\section{Proof of Corollary~\ref{cor:heuristic-unitary-cost}}
\label{app:heuristic-unitary-cost-proof}

We use the following measure-theoretic fact.

\begin{fact}\label{fact:analytic-null-preimages}
Let $X$ and $Y$ be connected Lie groups with Haar measures $\lambda$ and $\mu$, respectively, and let $F\colon X\to Y$ be real analytic.
If $\mu(F(X))>0$, then every Borel set $T\subseteq Y$ satisfies
\[
    \mu(T)=0 \quad\Longrightarrow\quad \lambda\bigl(F^{-1}(T)\bigr)=0.
\]
\end{fact}
\begin{proof}
Haar measures on Lie groups are induced by smooth positive volume forms.
If $\operatorname{rank}(dF_x)<\dim Y$ for every $x\in X$, Sard's theorem~\cite[Thm.~6.10]{lee2013} would give $\mu(F(X))=0$, contrary to the hypothesis.
Thus there is a point $x_*\in X$ with
\[
    \operatorname{rank}(dF_{x_*})=\dim Y.
\]

Since $X$ and $Y$ are connected and $F$ is real analytic with full rank at $x_*$, the result from~\cite[Lem.~2]{winkelmann2002} applies directly to $F$ with respect to the Haar measures $\lambda$ and $\mu$.
Hence, for every Borel set $T\subseteq Y$,
\[
    \mu(T)=0
    \ \Longrightarrow\ \lambda\bigl(F^{-1}(T)\bigr)=0.
\]
\end{proof}

\begin{proof}[Proof of Corollary~\ref{cor:heuristic-unitary-cost}]
Fix $n\geq4$, set $d=2^{n-1}$, and draw $U$ from the Haar probability measure $\mu$ on $U(2d)$.
In the Block-ZXZ decomposition~\eqref{qc:pd-modulus-derivation:a}, let $A(U)\in U(d)$ denote the target unitary of the final controlled gate.
For each $Q\in U(d)$, this decomposition gives
\[
    \mu\bigl(\{U\in U(2d): A((I_d\oplus Q)U)=QA(U)\}\bigr)=1.
\]

For each Borel set $E\subseteq U(d)$, define the pushforward measure $\nu$ by
\[
    \nu(E):=\mu\bigl(\{V\in U(2d): A(V)\in E\}\bigr).
\]
Fix $Q\in U(d)$ and write $g=I_d\oplus Q$. Let $S=\{V\in U(2d):A(V)\in QE\}$, so that $g^{-1}S=\{V\in U(2d):A(gV)\in QE\}$. Then
\[
\begin{aligned}
  \nu(QE) &= \Pr\bigl(A(U)\in QE\bigr)\\
          &= \Pr\bigl(A(gU)\in QE\bigr)
          && \text{(by left invariance, $\mu(S)=\mu(g^{-1}S)$)}\\
          &= \Pr\bigl(QA(U)\in QE\bigr)
          && \text{($A(gU)=QA(U)$ on a set of $\mu$-measure one)}\\
          &= \Pr\bigl(A(U)\in E\bigr)
          && \text{(since $Q^{-1}(QE)=E$)}\\
          &= \nu(E).
\end{aligned}
\]
This proves the left invariance of $\nu$. By uniqueness, $\nu$ is the Haar probability measure on $U(d)$,
meaning that $A(U)$ is haar generic.

For each permutation $P$, let
\[
  E_P=\{U\in U(2d): \spec(DP A(U))\text{ is antipodal for some diagonal unitary }D\}.
\]
Since $A(U)$ is Haar distributed, Claim~\ref{claim:general-case} gives $\mu(\bigcup_P E_P)>0$.
This union is finite, so we may fix a permutation $P$ with $\mu(E_P)>0$.

Note with $D$ and $P$ denoting the QMIEP factors, the output residual in Eq.~\eqref{qc:pd-modulus-derivation:d} is $(DP)^\dagger$.
Let $X=U(d)^4\times U(1)^k$ and write $x=(L,M,N,O,\vec{z})$, where $\vec{z}$ collects the diagonal entries of $D$ and the phases of the rotations. The circuit evaluation map $F_{\mathcal C_P}\colon X\to U(2d)$ is
\[
\begin{aligned}
F_{\mathcal C_P}(x)
={}&(I_d\oplus (DP)^\dagger)(I_2\otimes O)\,\mathrm{CZ}_{1,n}(R_R\otimes I_2)\\
 &{}\cdot(H\otimes N)\,\ucgrz\,(H\otimes M)\,R_L(I_2\otimes L).
\end{aligned}
\]
By construction, $E_P\subseteq F_{\mathcal C_P}(X)$, so
\[
    \mu\bigl(F_{\mathcal C_P}(X)\bigr)\geq\mu(E_P)>0.
\]

Let $T\subseteq U(2d)$ consist of the targets admitting a $PD$-equivalent implementation at the stated cost.
There are finitely many \cnot{} patterns at this cost, each with a compact space of gate and residual parameters. Circuit evaluation is continuous, so $T$ is compact and hence Borel.

The circuit requires output types $PD,PD,D,PD$ for $L,M,N,O$, respectively. Under product Haar measure $\lambda$ on $X$, these blocks are independent.
Given the preceding synthesis choices, the next block $B$ absorbs a fixed residual $V$ (either $D$ or $P$ or both), so $BV$ remains Haar distributed.
If each required type is attainable on a set of positive Haar measure in $U(d)$, successive integration gives a positive-measure set of parameters $x$ with $F_{\mathcal C_P}(x)\in T$.

Each entry of $F_{\mathcal C_P}$ is polynomial in the real and imaginary parts of the circuit parameters, so the map is real analytic.
Since $\mu(F_{\mathcal C_P}(X))>0$, Fact~\ref{fact:analytic-null-preimages} applies and gives, by contraposition,
\[
    \lambda\bigl(F_{\mathcal C_P}^{-1}(T)\bigr)>0
    \quad\Longrightarrow\quad \mu(T)>0.
\]

Starting from the universal 3-qubit constructions, the required $PD$-equivalent blocks attain cost $\costPDheuristic{n-1}$ with positive probability.
Applying the same measure argument to the circuit of Theorem~\ref{thm:pd-migrated-n-q-circuit}, with its terminal block synthesized recursively, yields an exact $n$-qubit implementation at cost $\costheuristic{n}$ with positive probability.
Solving the coupled recurrences in Eqs.~\eqref{eq:cost:UPD_by_QMIEP} and~\eqref{eq:cost:U_by_PD_migrate} with $\costUD(m)=\costU(m)-1$ and initial values $\costUPD(3)=17$ and $\costU(3)=19$ gives the closed form in Eq.~\eqref{eq:heuristic-unitary-cost}.
\end{proof}

\section{Proofs of Solvability Results}
\label{app:solvability-proofs}

\subsection{Proof of Proposition~\ref{prop:odd-moment-antipodality}}
\label{app:odd-moment-antipodality}

\begin{proof}
    Write the characteristic polynomial of $U$ as
    \[
        f(t):=\det(tI-U)=\sum_{k=0}^{2^n}a_k t^k.
    \]
    Since $2^n$ is even, $a_0=\det(-U)=\det(U)$.
    By unitarity, for $t\neq0$,
    \[
        \begin{aligned}
            f(t)&=\det(U)\det(tI)\det\!\left(U^\dagger-\frac{1}{t}I\right)=a_0t^{2^n}\det\!\left(\frac{1}{t}I-U^\dagger\right)\\
                &=a_0t^{2^n}\overline{f(1/\overline t)} =a_0\sum_{k=0}^{2^n}\overline{a_k}\,t^{2^n-k}.
        \end{aligned}
    \]
    Comparing coefficients gives
    \[
        a_{2^n-k}=a_0\overline{a_k},
        \qquad 0\leq k\leq 2^n.
    \]
    Since $|a_0|=1$, we have $a_{2^n-k}=0$ if and only if $a_k=0$.
    Thus, to show that $f$ is even, it suffices to show that $a_{2^n-k}=0$
    for every odd integer $k$ with $1\leq k\leq 2^{n-1}$.

    Newton's identities~\cite[Eq.~(2.4.18)]{horn2013} give
    \[
        k a_{2^n-k}
        =-\sum_{j=1}^{k}a_{2^n-k+j}\,\tr(U^j),
        \qquad 1\leq k\leq 2^n.
    \]
    We prove this by induction on odd $k\leq 2^{n-1}$.
    For $k=1$, the identity gives $a_{2^n-1}=-\tr(U)=0$.
    Now let $k>1$ be odd and assume $a_{2^n-\ell}=0$ for every
    odd $\ell<k$.
    In each summand, if $j$ is odd, then $\tr(U^j)=0$ by hypothesis.
    If $j$ is even, then $k-j$ is an odd integer smaller than $k$, so
    $a_{2^n-k+j}=a_{2^n-(k-j)}=0$ by the induction hypothesis.
    Thus $a_{2^n-k}=0$.

    Together, these equalities imply that all odd-degree coefficients of $f$ vanish,
    so $f(t)=f(-t)$, and hence the spectrum of $U$ is antipodal.
\end{proof}

\subsection{Proof of Proposition~\ref{prop:identity-neighborhood-solvable}} \label{app:identity-neighborhood}

\begin{proof}
Let $m=2^{n-1}$, and choose pairwise distinct
$\alpha_0,\ldots,\alpha_{m-1}\in[0,\pi)$.
For $t=(t_0,\ldots,t_{m-1})\in\mathbb{R}^m$, define
\[
    D(t)=\diagonal\bigl(e^{i(\alpha_0+t_0)},\ldots,e^{i(\alpha_{m-1}+t_{m-1})},
    -e^{i\alpha_0},\ldots,-e^{i\alpha_{m-1}}\bigr).
\]
Since $D(0)$ has $2m$ distinct eigenvalues, on a sufficiently small
neighborhood of $D(0)$ we can choose well-defined local real-valued eigenphase
branches $\phi_j(V)$ such that $e^{i\phi_j(V)}$ are the eigenvalues
of $V$ and
\[
    \phi_i(D(0))=\alpha_i,\qquad
    \phi_{m+i}(D(0))=\alpha_i+\pi
    \quad(0\leq i<m).
\]
Such local branches are not guaranteed in general because eigenvalues may be degenerate.
Each branch is uniquely specified by requiring its values to remain
in a sufficiently small real interval around its value at $D(0)$.
For sufficiently small neighborhoods $B\subset U(2m)$ of $I$ and
$T\subset\mathbb{R}^m$ of $0$, define the antipodality residual
\[
    F:B\times T\longrightarrow\mathbb{R}^m,\qquad
    F(U,t)=\bigl(\phi_i(D(t)U)-\phi_{m+i}(D(t)U)+\pi\bigr)_{i=0}^{m-1}.
\]
Thus, $F(U,t)=0$ implies that $D(t)U$ has an antipodal spectrum, with
$\phi_{m+i}(D(t)U)=\phi_i(D(t)U)+\pi$ for each $i$.
We seek to apply the implicit function theorem near $(U,t)=(I,0)$
to solve $F(U,t)=0$ for $t$.
The neighborhood $B$ admits smooth local coordinates in
$\mathbb{R}^{(2m)^2}=\mathbb{R}^{4^n}$, with $I$ corresponding to $0$.
In these coordinates, $F$ is a map from an open subset of
$\mathbb{R}^{4^n+m}$ to $\mathbb{R}^m$.
We verify the conditions for the implicit function theorem as follows.
\begin{itemize}
    \item (Vanishing residual) By the chosen eigenphase values,
    \[
        F_i(I,0)=\alpha_i-(\alpha_i+\pi)+\pi=0
        \quad(0\leq i<m).
    \]
    Hence, $F(I,0)=0$.

    \item (Smoothness) Write $F=H\circ\Lambda\circ G$, where
    \begin{align*}
        G(U,t)&=D(t)U,\\
        \Lambda(V)&=(\phi_0(V),\ldots,\phi_{2m-1}(V)),\\
        H(\theta)&=(\theta_i-\theta_{m+i}+\pi)_{i=0}^{m-1}.
    \end{align*}
    The maps $G$ and $H$ are smooth.
    For matrices with simple eigenvalues, the local eigenvalue branches depend
    smoothly on the matrix~\cite{kato}.
    Since $G(U,t)$ has simple eigenvalues in the chosen neighborhood,
    the corresponding local eigenphase map $\Lambda$ is smooth, and hence so is $F$.

    \item (Nonsingularity) Since $F(I,t)=t$ for $t\in T$, where the eigenphase branches are well defined,
    the Jacobian with respect to $t$ satisfies $D_tF(I,0)=I_m$ and is therefore nonsingular.
\end{itemize}
Hence, the implicit function theorem applies, and the proposition follows.
\end{proof}

\subsection{Proof of Proposition~\ref{prop:permutation-moment-obstruction}}
\label{app:permutation-moment-obstruction}

\begin{proof}
    Let $D=\diagonal(z_1,\ldots,z_{2^n})$, with $|z_i|=1$, and let $C_j$
    be the set of indices belonging to the $j$th cycle. Only cycles with $\ell_j\mid k$
    contribute to the trace, with each such cycle traversed $k/\ell_j$ times. Thus,
    \[
        \tr((DP)^k)
        =\sum_{j\in J_k}\ell_j\left(\prod_{i\in C_j}z_i\right)^{k/\ell_j}.
    \]
    For fixed $k$, each factor $\left(\prod_{i\in C_j}z_i\right)^{k/\ell_j}$
    can be treated as an independent arbitrary unit phase, since the $z_i$
    are freely chosen and the cycles are disjoint.
    The claim follows from the polygon closure condition.
\end{proof}

\section{Implementation Details}
\label{app:implementation-details}

In this section, we report some technical details of applying the
Newtonian method described in Section~\ref{sec:numerical_heuristic}.
Among the possible approaches to solving this numerical problem,
we choose the Newtonian method because the Jacobian can be derived naturally,
which may improve scalability.

\paragraph{Derivation of the Jacobian}
For a differentiable unitary family $U(\vec x)$, let $\lambda(\vec x)$ be a simple
eigenvalue with normalized eigenvector $\ket{q_\lambda(\vec x)}$. Then
\[
    \frac{\partial\lambda}{\partial x_i}
    =\bra{q_\lambda}\frac{\partial U}{\partial x_i}\ket{q_\lambda}.
\]
For $W=D(\vec\theta)PU$, let $\lambda_a=e^{i\alpha_a}$ and let $\ket{q_a}$
be the corresponding eigenvector, the $a$th column of $Q$. Thus,
\[
    \begin{aligned}
        i\lambda_a\frac{\partial\alpha_a}{\partial\theta_i}
        &=\frac{\partial\lambda_a}{\partial\theta_i}
        =\bra{q_a}\frac{\partial W}{\partial\theta_i}\ket{q_a}\\
        &=i\lambda_a|\braket{i+1|q_a}|^2
        =i\lambda_a|Q_{i+1,a}|^2.
    \end{aligned}
\]
(Note that, under our parameterization of $W$, the $(i+1)$st column of
$D(\vec\theta)$ corresponds to the $i$th parameter, $\theta_i$.)
Hence $\partial\alpha_a/\partial\theta_i=|Q_{i+1,a}|^2$, which gives
Equation~\eqref{eq:qmiep-jacobian} by differentiating the residual in
Equation~\eqref{eq:qmiep-pairing-residual}.

\begin{remark}
The Jacobian above assumes that the current iterate $W(\vec\theta_t)$
has a simple spectrum, i.e., no repeated eigenvalues.
For Haar-random $U$ and fixed initial phases and permutation, this condition
holds almost surely at initialization.
Moreover, spectral simplicity is preserved under sufficiently small phase updates.
The observed success of our numerical method is consistent with this local stability.
\end{remark}

\paragraph{Hyperparameters}
The performance of this heuristic algorithm depends on several hyperparameters,
which were tuned through empirical evaluation.
We fix the damping parameter in Equation~\eqref{eq:qmiep-gauss-newton-update}
to $\lambda_t=10^{-8}$ for all $t$ to stabilize the linear solve when $J_t$
is ill-conditioned.
At each iteration, we test $\eta_t=1,\frac12,\ldots,2^{-11}$ in order.
We accept the first step that gives a sufficient decrease in the squared
residual norm according to the Armijo condition
\begin{equation}\label{eq:qmiep-armijo}
    \bigl\|r(\vec\theta_t+\eta_t p_t)\bigr\|^2
    \leq \bigl\|r(\vec\theta_t)\bigr\|^2
    +c\eta_t\,r(\vec\theta_t)^\top J_t p_t,
\end{equation}
where $c=2\times10^{-4}$.

\paragraph{Final Algorithm}
The following is an algorithmic description of the numerical process.

\begin{algorithm}
\caption{Gauss--Newton search for a QMIEP solution}
\label{alg:qmiep-search}
\begin{algorithmic}[1]
\Require $U\in U(2^n)$ with $n\geq 3$, output tolerance $\epsilon>0$, iteration budget $T$ per stage
\Ensure A verified pair $(P,D)$, or \textsc{failure} if the search is exhausted
\For{each permutation $P$, starting with $P=I$}
    \State $\vec\theta\gets\vec 0$
    \For{$t=0,\ldots,T-1$ until the stage stopping criterion is met}\label{line:qmiep-stage}
        \State $\vec\theta_t\gets\vec\theta$
        \State Diagonalize $D(\vec\theta_t)PU=Q\operatorname{diag}(e^{i\alpha_1},\ldots,e^{i\alpha_{2^n}})Q^\dagger$
            with $\alpha_1\leq\cdots\leq\alpha_{2^n}$.
        \State Set $(r_t)_a\gets\alpha_{a+2^{n-1}}-\alpha_a-\pi$ for $a=1,\ldots,2^{n-1}$
            and compute $J_t$ from $Q$ by Equation~\eqref{eq:qmiep-jacobian}.
        \State Obtain a descent direction $p_t$ from Equation~\eqref{eq:qmiep-gauss-newton-update} or \textbf{break}.
        \For{$\eta_t=1,\frac12,\ldots,2^{-11}$ in order}
            \If{the Armijo condition~\eqref{eq:qmiep-armijo} holds}
                \State $\vec\theta\gets\vec\theta_t+\eta_t p_t$ and \textbf{break}
            \EndIf
        \EndFor
        \State \textbf{if} no step was accepted \textbf{then break}
    \EndFor
    \State Recompute $r(\vec\theta)$ from the eigenphases of $D(\vec\theta)PU$.
    \State \textbf{if} $\|r(\vec\theta)\|_\infty\leq\epsilon$ \textbf{then} \Return $(P,D(\vec\theta))$\label{line:qmiep-verify}
\EndFor
\State \Return \textsc{failure}
\end{algorithmic}
\end{algorithm}

We used $\epsilon = 10^{-10}$ when rerunning the numerical experiments throughout the paper.

\end{document}